\documentclass[12pt]{article} 
\usepackage{multirow} 
\usepackage{epsfig} 
\usepackage{amsfonts} 
\usepackage{amssymb}
\usepackage{amsmath}  
\usepackage{amsbsy}  
\usepackage{wasysym}
\usepackage{slashed}
\usepackage{cite}
\usepackage{axodraw}
\usepackage{relsize}
\catcode`\@=11 
\@addtoreset{equation}{section} 
\def\theequation{\thesection.\arabic{equation}} 
 
\@twosidetrue 

\catcode`\@=11  
\def\section{\@startsection{section}{1}{\z@}{3.5ex plus 1ex minus 
.2ex}{2.3ex plus .2ex}{\large\bf}}

\def\thesection{\arabic{section}} 
\def\thesubsection{\arabic{section}.\arabic{subsection}} 
\def\thesubsubsection{\arabic{section}.\arabic{subsection}.\arabic{subsubsection}} 
 
\def\appendix{\setcounter{section}{0} 
 \def\thesection{\Alph{section}} 
 \def\theequation{\Alph{section}.\arabic{equation}} 
\def\thesubsection{\Alph{section}.\arabic{subsection}} 
\def\thesubsubsection{\Alph{section}.\arabic{subsection}.\arabic{subsubsection}} 
 
\def\section{\@startsection{section}{1}{\z@}{3.5ex plus 1ex minus 
   .2ex}{2.3ex plus .2ex}{\large\bf}} }

\newcount\minutes 
\newcount\scratch 
 
\def\timestamp{%
\scratch=\time 
\divide\scratch by 60 
\edef\hours{\the\scratch} 
\multiply\scratch by 60 
\minutes=\time 
\advance\minutes by -\scratch 
---$\,$\hours:\null 
\ifnum\minutes< 10 0\fi 
\the\minutes} 

\def\sla#1{\ifmmode%
\setbox0=\hbox{$#1$}%
\setbox1=\hbox to\wd0{\hss$/$\hss}\else%
\setbox0=\hbox{#1}%
\setbox1=\hbox to\wd0{\hss/\hss}\fi%
#1\hskip-\wd0\box1 }

\def\be{\begin{equation}}
\def\ee{\end{equation}}
\newcommand{\bea}{\begin{eqnarray}}
\newcommand{\eea}{\end{eqnarray}}

\def\beq{\begin{equation}} 
\def\eeq{\end{equation}} 
 
\def\beqn{\begin{eqnarray}} 
\def\eeqn{\end{eqnarray}}

\def\({\left(} 
\def\){\right)} 
\def \as   {\ifmmode \alpha_s \else $\alpha_s$ \fi}

\newcommand\figfont{ }
\begin{document} 
%
%
%
%
%
%
%
%
%
%
%
%
%
%
%
%
\begin{titlepage} 
\nopagebreak 
{\flushright{ 
        \begin{minipage}{5cm}
         FTUV-11-0130 \\ 
         KA-TP-10-2011\\
         LPN11-20 \\
         SFB-CPP-11-22 \\
        \end{minipage}        } 
 
} 
\vfill 
\begin{center} 
{\LARGE \bf 
 \baselineskip 0.5cm 
Towards $pp\to VVjj$ at NLO QCD:  Bosonic\\[2mm] 
contributions to triple vector
boson  \\[2mm]
production plus jet \\[2mm] 
} 
\vskip 0.5cm  
{\large   
Francisco Campanario 
}   
\vskip .2cm
{ {\it Institut f\"ur Theoretische Physik, 
    Universit\"at Karlsruhe, KIT,\\ 76128 Karlsruhe, 
    Germany}
    }\\
 \vskip 1.3cm     
\end{center} 
 
\nopagebreak 
\begin{abstract}
In this work, some of the NLO QCD corrections for $ pp\to VVjj + X$ are presented.
A program in Mathematica based on
 the structure of FeynCalc which automatically simplifies a set of
 amplitudes up to the hexagon level of rank 5 has been created for this purpose.
%
We focus on two different topologies. The first involves all the virtual contributions needed for quadruple 
electroweak vector boson production, i.e. $pp\to VVVV + X$.
In the second, the remaining ``bosonic'' corrections to electroweak triple vector boson production with an additional
jet~($pp\to VVV j + X$) are computed.  We show the
factorization formula of the infrared
divergences of the bosonic contributions for VVVV and VVVj production with $V \in (W,Z,\gamma)$.
Stability issues associated with the evaluation of the hexagons up to
rank 5 are studied.
The CPU time of the FORTRAN subroutines rounds the 2
milliseconds and seems to be competitive with other more sophisticated
methods. Additionally, in Appendix A the master equations to obtain the tensor
coefficients up to the hexagon level in the external momenta convention
are presented including the ones needed for small
Gram determinants.
\end{abstract} 
\vfill 
\today \timestamp 
\hfill 
\vfill 
\end{titlepage} 
\newpage               
\section{Introduction}
\label{sec:intro}
The incoming data from the CERN Large Hadron Collider (LHC) will require
precise theoretical predictions for a variety of signal and background
processes. 
Processes with two vector bosons are of vital
importance since they constitute background signals to Higgs and top physics
as
well as to physics Beyond the Standard
Model~(BSM)~(for an overview see e.g. Ref.~\cite{Campbell:2006wx}). 
A tremendous effort has been carried out by the scientific community to
compute these processes accurately.
Next-to-Leading Order (NLO) QCD diboson production was computed in Refs.~%
\cite{Ohnemus:1991kk,Ohnemus:1991gb,Campbell:1999ah}, and 
with an additional jet in Refs.~\cite{
Dittmaier:2007th,
Campbell:2007ev,
Campanario:2009um,
Dittmaier:2009un,
Binoth:2009wk,
Campanario:2010hp,Campanario:2010hv}.
Electroweak~(EW) diboson production in association with two
jets was computed 
in Refs.~\cite{Oleari:2003tc,Jager:2006zc,Bozzi:2007ur}. Recently, the NLO QCD corrections to
$W^+W^+jj+X$~\cite{Melia:2010bm} and $W^+W^-jj+X$~\cite{Melia:2011dw} production were
computed using the generalized D-dimensional unitarity framework for the calculation of the
 one loop virtual amplitudes. 
%
%
An impressive progress in the calculation of multi-parton one-loop
amplitudes has been
 achieved not only applying new techniques%
~\cite{
Bern:1990cu,
Bern:1993mq,
Bern:1994zx,
Bern:1994cg,
Bern:1994fz,
Bern:1997sc,
Britto:2004nc,
Bern:2005hs,
Forde:2005hh,
Berger:2006ci,
Ossola:2006us,
Anastasiou:2006gt,
Ossola:2007bb,
Badger:2007si,
Ossola:2008xq,
Ellis:2007br,
Ossola:2007ax,
Giele:2008ve,
Mastrolia:2008jb,
Giele:2008bc,
Ellis:2008qc,
NigelGlover:2008ur,
Britto:2008zz,
Lazopoulos:2008ex,
Nagy:2006xy,
Britto:2007tt,
Moretti:2008jj,
Britto:2008vq,
Britto:2008sw,
Berger:2008sj,
vanHameren:2009dr}, but
also based on traditional 
methods%
~\cite{
Passarino:1978jh
,vanOldenborgh:1989wn
,Ezawa:1990dh
,Davydychev:1991va
,Bern:1992em
,Bern:1993kr
,Tarasov:1996br
,Fleischer:1999hq
,Binoth:1999sp
,Denner:2002ii
,Duplancic:2003tv
,Giele:2004iy
,Giele:2004ub
,Binoth:2005ff
,Ellis:2005zh
,Denner:2005nn
,Ellis:2006ss
,Binoth:2006hk
,Diakonidis:2008dt
,Diakonidis:2008ij
,Ossola:2008zzb
,Gong:2008ww
,vanHameren:2009vq
,Belanger:2003sd
,delAguila:2004nf
,vanHameren:2005ed,
Binoth:2008uq}. 
In this paper, relying on traditional techniques, we compute some of the
virtual 
corrections  needed for
$pp \to VV jj + X$ production.  We focus on some of the topologies that appear, namely QCD ``bosonic''
one-loop corrections to the diagrams contributing to $pp
\to VVVV + X $ and $pp
\to VVVj + X$ production. The strategy followed is to collect Feynman
diagrams with a given topology in groups which can be easily checked and
reused in other processes. Thus, our aim is not only
to provide results for $VVjj+X$ production, including a second calculation of $W^+W^-jj+X$ production using a different
method, but also to provide a set of routines that can be used for many
other interesting processes, such as $ W \gamma \gamma j +X$ at
NLO QCD~\cite{Campanario:2011}. It is known that in
the calculation of one-loop multi-leg amplitudes, the presence of small Gram 
and Cayley determinants might yield unstable results. In this paper, we show 
that the use of higher precision in the numerical determination of the tensor
integrals~(similarly to Ref.~\cite{Reiter:2009dk}) together with the use of
Ward Identities to identify the unstable
points, and to a minor extent, following Ref.~\cite{Denner:2002ii}, the use of special routines for the
determination of small Gram determinants for the $C$ and $D$ functions, solve this problem.

The paper is organized as follows; in Section~\ref{sec:calc}, the method
used to perform the calculation, the different
contributions computed and the tests performed to
guarantee the correctness of the results are described. In
Section~\ref{sec:ins}, 
a discussion of the instabilities and the timing of the amplitudes computed are shown. The conclusions are
 given in Section~\ref{sec:concl}. In Appendix A, the tensor reduction
 routines used are presented. In Appendix B, numbers for the
contributions involving the hexagons are given, including proofs of   
the factorization of the infrared divergences and other additional
numerical tests. Finally, in Appendix C, we give the color factors
needed for testing the amplitudes. 
\section{Calculational details}
\label{sec:calc}
\begin{figure}[h!]
\begin{center}
\includegraphics[scale=0.85]{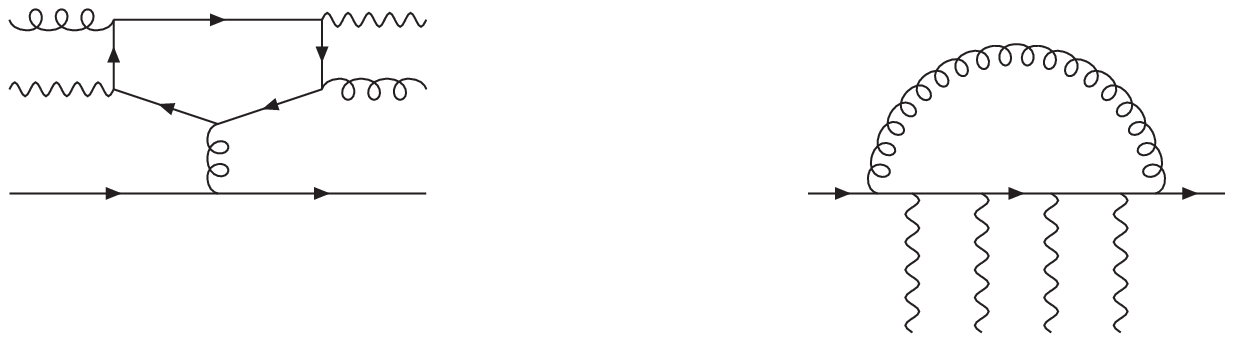}
\end{center}
\caption[]{\label{fig:fer}
{\figfont ``One loop fermion'' contributions~(left) and ``bosonic'' contributions~(right).}  }
\end{figure}

\begin{figure}[h!]
\begin{center}
\includegraphics[scale=0.85]{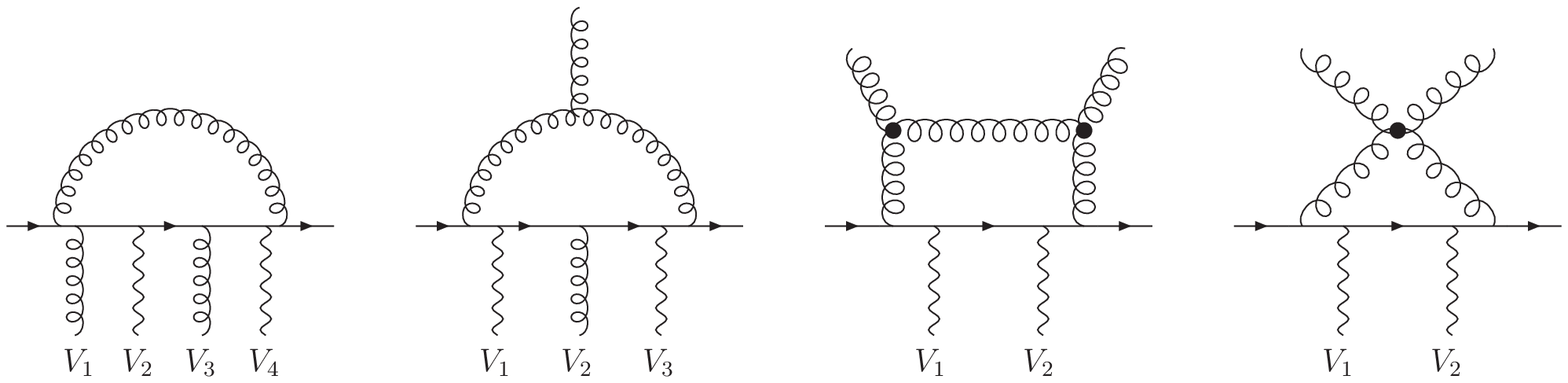}
\end{center}
\caption[]{\label{fig:bos}
{\figfont Different ``bosonic'' topologies. $V_i$ stands for vector bosons
  emitted from the quark line.}  }
\end{figure}

%
%
In the calculation of the NLO QCD virtual corrections to $ pp \to VVjj + X$,
different sub-processes contribute; processes involving two quark pairs,
e.g., $ u u \to d d W^+ W^+$ and processes with one quark pair, e.g., $u
\bar{\mbox{u}}\to W^+ W^- gg$~\footnote{Note that the NLO QCD corrections
  for $pp\to W^+W^+jj+X$ production involve only sub-processes with two quark pairs.}. The latter can be separated into ``one loop
fermion''
corrections and ``bosonic'' 
corrections, Fig.~\ref{fig:fer}. 
%
%
Among the ``bosonic'' contributions, different
topologies appear, Fig.~\ref{fig:bos}. We will focus on 
diagrams with the first two topologies, i.e.,  one loop QED-like
corrections and diagrams formed with one triple gluon
vertex. The remaining ones will be discussed in future publications.
%
%
%

To compute these amplitudes, a function in Mathematica~\cite{Wolfran}
using FeynCalc~\cite{Mertig:1990an} 
which  automatically simplifies a set of amplitudes up to Hexagons of rank
5 has been created. Throughout the
calculation, the quarks are considered to be massless, the anticommuting prescription for
$\gamma_5$ is used and we work in the Feynman Gauge. The result is given in terms of tensor integrals following the
Passarino-Veltman convention~\cite{Passarino:1978jh},
Appendix~\ref{appendixa}, and written automatically in to FORTRAN
routines.  Nevertheless, the amplitudes can be evaluated also in
Mathematica with unlimited precision which is used for testing purposes. To achieve that, the scalar integrals, the tensor
reduction formalism to extract the tensor coefficient integrals, and also the  helicity
method described in Refs.~\cite{Hagiwara:1985yu,Hagiwara:1988pp} to
compute the spinor products describing the quark lines of
Figure~\ref{fig:bos} have been implemented at the FORTRAN and
Mathematica level. For the determination of the tensor integrals up to
the box level, we have implemented the Passarino-Veltman 
tensor reduction formalism~\cite{Passarino:1978jh}. 
We have also applied the LU decomposition method to avoid the
explicit calculation of inverse Gram matrices by solving a system of
linear equations which is a more stable procedure close to singular
points. Finally, for singular Gram determinants, special
tensor reduction routines following Ref.~\cite{Denner:2005nn} have been
implemented, however, the external momenta convention~(Passarino-like) was used.  
The impact of these methods is discussed in detail in
Section~\ref{sec:ins}. For pentagons, in addition to the Passarino-Veltman formalism, the method proposed by Denner and
Dittmaier~\cite{Denner:2005nn}, applied also to hexagons, has been
implemented. For that, the recursion relations of
Ref.~\cite{Denner:2005nn} in terms of the Passarino-Veltman external
momenta convention have been re-derived. This last method is used for the numerical
implementation at the FORTRAN level in Section~\ref{sec:ins}.
All the recursion relations used can be found in Appendix~\ref{appendixa}.

In the following, a more detailed description of the method used is given. To extract the rational terms, two simplifications are done at the
Mathematica level. To illustrate them, the
amplitude of a simple vertex diagram is used,

\begin{minipage}[h!]{\linewidth}
\begin{minipage}[h!]{0.37\linewidth}
\vspace{5pt} 
\hspace*{10pt}
\SetScale{0.7}
\begin{picture}(70,80)(0,0)
\ArrowLine(20,50)(70,50)
\ArrowLine(70,50)(120,50)
\ArrowLine(120,50)(170,50)
\ArrowLine(170,50)(220,50)
\GlueArc(120,50)(40,0,180){5}{10}
\Photon(120,50)(120,10){5}{3} 
\Vertex(120,50){4}

\LongArrowArc(120,60)(20,0,180)
\Text(85,48)[]{$l$}

\Text(-5,36)[]{I~=}
\Text(35,48)[]{$p_1$}
\Text(35,40)[]{$\rightarrow$}

\Text(130,48)[]{$p_3$}
\Text(130,40)[]{$\leftarrow$}
\Text(70,10)[]{$\mu_2$}
\Text(102,20)[]{$p_2$}
\Text(93,20)[]{$\uparrow$}
\end{picture} 
\end{minipage}
\begin{minipage}[h!]{0.57\linewidth}
\[\propto \mathlarger{\int} \mathlarger{\frac{d^d \, l}{(2\pi)^d}} \frac{1}{l^2}\,\bar{v} (p_3) \gamma^\alpha
\frac{(\slashed{l}+\slashed{p}_1+\slashed{p}_2)}{(l+p_1+p_2)^2} \gamma^{\mu_2} \frac{(\slashed{l}+
\slashed{p}_1)}{(l+p_1)^2}\gamma_\alpha u (p_1).  \]
\end{minipage}
\end{minipage}
%
First, a simple Dirac re-ordering manipulation is applied to explicitly contract 
repeated indices and to obtain all the terms proportional to the space dimension, $D$, coming from 
$\gamma_\mu\gamma^\mu = D$ contractions, i.e., \\
{\small
\begin{equation}
I  \propto \mathlarger{\int} \mathlarger{\frac{d^d \, l}{(2\pi)^d}} \,\bar{v} (p_3) \gamma^\alpha
(\slashed{l}+\slashed{p}_1+\slashed{p}_2) \gamma^{\mu_2} (\slashed{l}+
\slashed{p}_1)\gamma_\alpha u (p_1) =\underbrace{-
\int \frac{d^d \, l}{(2\pi)^d} \,
\frac{\bar{v} (p_3) (\slashed{l}) \gamma^{\mu_2} (\slashed{l})\overbrace{\gamma^\alpha \gamma_\alpha}^D u
(p_1)}{(l+p_1+p_2)^2(l+p_1)^2l^2}}_{II}  + \, \, \, \ldots 
\end{equation}
}%
\noindent 
Second, all Dirac's structure containing the 
loop momenta is pulled to the right, such that resulting terms like
$\slashed{l}\slashed{l}=l^2$ are canceled against one of the integral
denominators, 
\begin{eqnarray}
\label{eq.1.2}
II &= &
D \, \bar{v} (p_3)  \gamma^{\mu_2}  u (p_1)  \int
\frac{d^d \, l}{(2\pi)^d}
  \frac{l^2}{l^2(l+p_1)^2(l+p_1+p_2)^2} +\ldots \nonumber \\
& \propto &D\,  \bar{v} (p_3)  \gamma^{\mu_2}  u (p_1) I_{2}(p_2)+\ldots,
\end{eqnarray}
where $I_{2}(p_2) \equiv B_0(p_2)$ is the scalar two point function
defined correspondingly to Eq.~(\ref{A:1}).
In this way we avoid possible additional terms proportional to $D$
coming from the Lorentz structure of the tensor integral. More
concretely, the same term
$II$ could result in terms proportional to $D^2$:
\begin{eqnarray}
II &=& -
D \, \bar{v} (p_3) \gamma_{\nu_1}\gamma^{\mu_2} \gamma_{\nu_2}  u (p_1)  \int
\frac{d^d \, l}{(2\pi)^d}
  \frac{l^{\nu_1}l^{\nu_2}}{l^2(l+p_1)^2(l+p_1+p_2)^2} \nonumber\\
&\propto &- D \, \bar{v} (p_3) \gamma_{\nu_1}\gamma^{\mu_2} \gamma_{\nu_2}  u (p_1)
 \times I^{(3)}_{00}(p_1,p_2) g^{\nu_1\nu_2}+\ldots \nonumber\\ & =&- (2-D)\,D\,     
\bar{v}(p_2) \gamma^{\mu_2} u (p_1) I^{(3)}_{00}(p_1,p_2)  +\ldots ,
\end{eqnarray}
with $I^{3}_{00}(p_1,p_2) \equiv C_{00}(p_1,p_2)$, a three point tensor
coefficient integral of the tensor integral $I^{\mu_1\mu_2}_3(p_1,p_2)$,
defined by Eq.~(\ref{ref:tens1}), and obtained recursively from  Eq.~(\ref{masPV00}). Further simplifications are obtained using the Dirac equation of
motion and rewriting the pair $(l\cdot p_i)$ as a difference of
two propagators, e.g., $(l\cdot p_1)= (l+p_1)^2-l^2$, which are canceled against
the denominators. 
The remaining rational terms stem from ultraviolet
divergent tensor
coefficients, which are treated independently within the tensor reduction routines. As an example, the finite
contribution of the ultraviolet divergent tensor coefficient
$C_{00}$, for massless propagators, following Eq.~(\ref{masPV00}) is obtained by,
\be
\label{eq.c00}
C^{(\mbox{fin})}_{00}=\frac{1}{2 D-4} \left( B_0 + \sum_{n=1}^2 (r_n-r_{n-1})
  I^3_n \right) \bigg|_{\mbox{fin}}=\frac{1}{4}
\left( B^{(\mbox{fin})}_0 +  \sum_{n=1}^{2}(r_n-r_{n-1}) I^{3,(\mbox{fin})}_n \right) +
\frac{ B_0^{(UV)}}{4}, 
\ee
where we have taken $D= 4 -2\, \epsilon$ and series expand in $\epsilon$
all the scalar and tensor coefficient integrals, e.g.,
\be
\label{eq.b0}
B_0= B_0^{(\mbox{fin})} + \frac{1}{\epsilon} B_0^{(UV)},
\ee
with $B_0^{(UV)}=1$. After these steps, the amplitudes, ${\cal M}_v$, in terms of scalar and tensor
coefficient integrals can be written by, 
\be
\label{DR}
{\cal M}_v
={\cal M}^{D=4}_v  + (D-4) {\cal M}^{DR}_v,
\ee
where ${\cal M}^{D=4}_v$ is the amplitude that one would obtain performing the Dirac algebra manipulation 
in four dimensions, $D=4$, and ${\cal M}^{DR}_v$ contains
the rational terms and vanish in Dimensional Reduction ($DR$).  To
get this expansion, we have only rewritten the space dimension D as $D= \bar{d}
+ 4$ such that for example Eq.~(\ref{eq.1.2}) is given by,
\be
D\,  \bar{v} (p_3)  \gamma^{\mu_2}  u (p_1) B_0(p_2) = \underbrace{4 \,  \bar{v}
(p_3)  \gamma^{\mu_2}  u (p_1) B_0(p_2)}_{{\cal M}^{D=4}_v } + \bar{d} \,  \underbrace{\bar{v}
(p_3)  \gamma^{\mu_2}  u (p_1) B_0(p_2)}_{{\cal M}^{DR}_v }.
\ee
Both ${\cal M}^{D=4}_v$  and ${\cal M}^{DR}_v$ are decomposed in terms of,
\be
\label{eq:helmethod}
{\cal M}^{(D=4,DR)} = \sum_{i,j,\tau} \mbox{SM}_{i,\tau}~\mbox{F1}_j,
\ee
where SM$_{i,\tau}$ is a basis of Standard Matrix elements corresponding to
spinor products describing the quark line of Figure~\ref{fig:bos} which are computed
following the helicity method~\cite{Hagiwara:1985yu,Hagiwara:1988pp}  
with a defined helicity, $\tau$. F1$_j$ are complex
functions which are further decomposed into dependent and
independent loop integral parts,
\be
\label{eq:F1}
\mbox{F1}_j= \sum_{k} \mbox{F}_k T_k\Big(  \epsilon(p_n) \cdot p_m
  ;\epsilon(p_i) \cdot \epsilon(p_j) \Big).
\ee
$T_k$ is a monomial function at
most for each polarization vector $ \epsilon(p_x)$. $F_k$ contains
kinematic variables~($p_i\cdot p_j$), the scalar integrals ($B_0\equiv
 I_2, C_0\equiv I_3,D_0\equiv I_4$)\footnote{Note that throughout the
   paper to name the different n-point function integrals, $I_n$, the
   alphabetically-ordered labeling frequently found in the literature is
   also used.}, and the tensor integral coefficients
 $(B_{ij},C_{ij},D_{ij},E_{ij},F_{ij})$. The latter obtained
numerically from the scalar integral basis following the recursion relations
of Appendix A. To illustrate this notation, the following
example is given,
\be
\bar{v} (p_4) \slashed{p}_2 P_+ u (p_1)\, C_0(p_1,p_2)\, \epsilon(p_3) \cdot
\epsilon(p_2)  = \underbrace{\bar{v} (p_4) \slashed{p}_2 P_+ u
  (p_1)}_{SM_{1,+}} \overbrace{\underbrace{C_0(p_1,p_2)}_{F_1}
  \underbrace{\epsilon(p_3) \cdot \epsilon(p_2)}_{T_1}}^{F1_1}= SM_{1,+}
F1_1.
\ee
$P_+$ is the positive helicity projector, $P_+=\frac{1+\gamma_5}{2}$. We note here that this parametrization is convenient for example
to evaluate the amplitude for different polarization vectors or for
performing gauge tests since some of the functions will remain unchanged.
Finally, the full result is obtained from ${\cal M}^{D=4}_v$ and
${\cal M}^{DR}_v$ using the finite and the coefficients of the $1/\epsilon^n$ poles of
the scalar and tensor coefficient integrals (see e.g. Eq.~(\ref{eq.b0})). 
From ${\cal M}^{D=4}_v$, we get
\be
\label{D4terms}
 {\cal M}^{D=4}_v= \widetilde{{\cal M}}_v
 +\frac{{\cal M}^1_v
}{\epsilon} +\frac{{\cal M}^2_v
}{\epsilon^2},
\ee
where $ \widetilde{{\cal M}}_v $ is the finite contribution obtained
using the finite pieces of the scalar and tensor coefficient integrals including the finite contributions from rational terms
arising in ultraviolet tensor coefficient integrals, Eqs.~(\ref{eq.c00}) and (\ref{masPV00}).
${\cal M}^1_v $ and ${\cal M}^2_v $ are obtained from the $1/\epsilon$
and $1/\epsilon^2$ pole contributions, respectively. 
Similarly, from $(D-4) {\cal M}^{DR}$, one gets,
\be
\label{rational}
(D-4) {\cal M}^{DR}_v= \widetilde{{\cal N}_v} 
+\frac{{\cal N}_v^1
}{\epsilon},
\ee
where $\widetilde{{\cal N}}_{v}$ and ${\cal N}^1_{v}$ are obtained from the
$1/\epsilon$  and $ 1/\epsilon^2 $ poles, respectively. Indeed, ${\cal N}^1_{v}$  is analytically zero
since rational terms in one-loop QCD amplitudes, omitting wave
renormalization graphs (WRF) are of UV origin~\cite{Bredenstein:2008zb}. Eq.~(\ref{rational}) will allow us to
check this statement numerically.
Additionally, up to the pentagon level, the divergent part is computed in Dimensional
Regularization analytically in Mathematica. A library containing all the divergent tensor coefficients for massless
particles including pentagons of rank 4 has been created with this purpose. This allows us to obtain also
the terms of Eq.~(\ref{rational}) and the divergent parts of
Eq.~(\ref{D4terms}) analytically up to this level. 

It is known that the IR divergences depend on the kinematics of the external particles
involved in the process, i.e., whether they are massless/massive
on-shell/off-shell particles. Nevertheless, since we reconstruct the
divergences numerically, to compute generally the
amplitude of a particular diagram for a given helicity, $\tau$, it is
sufficient to consider off-shell vector bosons, $V_i$ in
Figure~\ref{fig:bos}. This basically means that  all $p_i^2$ terms are kept
and the transversality property
of the vector on-shell bosons ($\epsilon(p_x)\cdot p_x=0$) is not
applied analytically, allowing us, the latter, to perform gauge tests
($\epsilon(p_x)\to p_x$) for on-shell massive particles. We note here that the polarization vectors can be
understood as generic effective currents, $\epsilon^\mu(p_x) \to J_{\mbox{eff}}^\mu(p_x)$,
which can include the leptonic decay of the vector bosons or 
physics BSM.
%
%
Thus, denoted by ${\cal M}_{V_1V_2V_3V_4}$ the amplitude of the first
diagram of Figure~\ref{fig:bos}  would be given by,
\be
{\cal 
M
}_{V_1V_2V_3V_4,\tau}
=g^{V_1f}_{\tau}g^{V_2f}_{\tau}g^{V_3f}_{\tau}g^{V_4f}_{\tau}\frac{g_0^2}{(4 \pi)^2} {\cal
  C}^{V_1 V_2 V_3 V_4}_{ij}  {\cal
  M}^{ij}_{\tau},
\ee
where $g_0$ is the strong unrenormalized coupling, ${\cal M}_{\tau}^{ij}$ is the corresponding diagram with
off-shell vector bosons (or effective currents) with color indices $ij$
and given in terms of Eq.~(\ref{DR}) which is computed only once and for
all vector bosons, $V_i\in (\gamma,W,Z,g)$. ${\cal C}^{V_1 V_2 V_3 V_4}_{ij}$ is a color diagram dependent factor,
e.g., ${\cal C}^{\gamma \gamma g \gamma}_{ij}=(T_a)_{ij} (C_F-1/2 C_A)$
and from now on the color sub-indices will be omitted. $g^{V_i f}_{\tau}$
is the
coupling following the notation of Ref.~\cite{Hagiwara:1988pp} with e.g.,
$g^{\gamma f}_{\pm}=|\mbox{e}| Q_f$,  $g^{W
  f}_{-}=|\mbox{e}|/(\sqrt{2}\sin \theta_w)$, $g^{Z
  f}_{-}=|\mbox{e}| (T_{3f}-Q_f\sin^2 \theta_w)/(\sin \theta_w\cos \theta_w)$,
where $\theta_W$ is the weak mixing angle, $T_{3f}$ is the third
component of the isospin of the (left-handed) fermions, and $|\mbox{e}|$
the electric charge.

This procedure is advantageous for several reasons.  First, when considering off-shell
vector bosons, for example, diagrams with the same QED-like topology as the first
one of Fig.~\ref{fig:bos} will give us,
already, all the corrections needed for EW
quadruple vector boson production, $pp \to VVVV + X$ and part of the $pp \to VVVj +
X$ and $pp \to VVjj +
X$  contributions, independently whether we are
considering massless or massive particles or the bosons emitted from the
quark line are gluons. ${\cal C}^{V_1 V_2 V_3 V_4}_{ij}$ will associate the correct color factor to each diagram depending
whether one is considering $pp \to VVVV +X $, $pp \to VVVj+X$ or  $pp \to
VVjj+X$ and  $g^{V_i f}_{\tau}$ the corresponding couplings.
Also, the leptonic decay of the vector bosons and
new physic effects can be immediately incorporated by using the
appropriate effective currents. Second, cross term related diagrams are obtained from the same
analytical amplitude just by permuting the
vector boson momenta and polarization, and thus, e.g., from the 24 cross related
diagrams to the first one of Fig.~\ref{fig:bos}, only the permutation
depicted has to be computed. Moreover, additional checks can be
implemented, as for example the
factorization of the IR divergences against the born amplitude for
different processes starting from the same analytical structure. In the
following, we will refer to the first topology of Fig.~\ref{fig:bos} as contributions
to $pp \to VVVV+X$ and, to the second as ``bosonic'' contributions to  $pp \to VVVj+X$ since they appear
for the first time in these processes.

\subsection{Contributions to  $ p p \to VVVV+X$} 
\begin{figure}[!h]
\begin{center}
\includegraphics[scale=0.9]{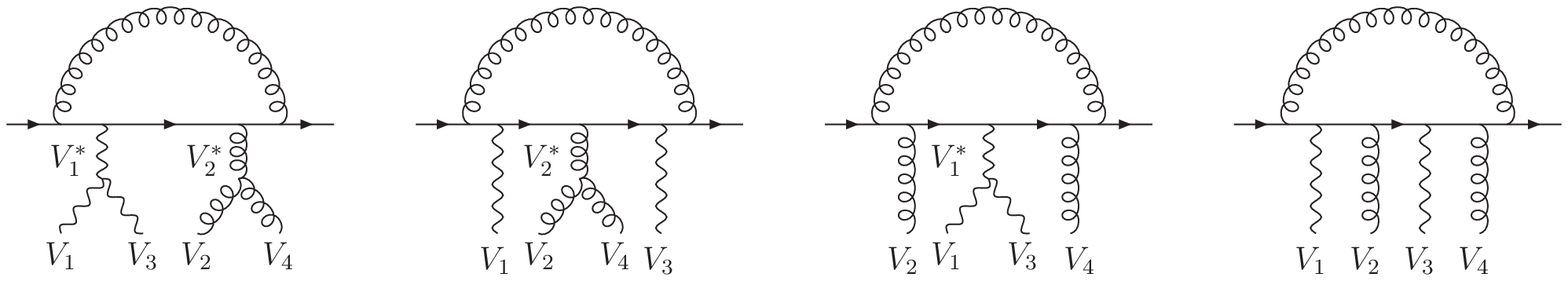}
\end{center}
\caption[]{\label{fig:LO}
{\figfont ``bosonic'' one loop QED-like topologies appearing in the
  calculation of the virtual contributions for $pp \to VVjj +X $ production, with V$
  \in ($W $^\pm$,Z,$\gamma)$.}  }
\end{figure}

In this section, all the ``bosonic'' QED-like loop
corrections along a quark line needed for $pp \to
VVjj+X$ production are considered. These contributions can be classified
by {\bf I)} virtual corrections along a quark line
with two vector bosons attached (first diagram of 
Figure~\ref{fig:LO}) {\bf II)} virtual corrections
with three vector bosons attached to the quark line (second and third
diagrams of Figure~\ref{fig:LO}) and {\bf III)} virtual corrections along a quark
line with four vector bosons attached (last diagram of
Figure~\ref{fig:LO}). 
%
The strategy followed is to search for a minimal set of universal
building blocks from which every amplitude can be constructed.
We use the effective current approach described and applied in
Refs.~\cite{Campanario:2009um,Campanario:2010hp,Campanario:2010hv,Campanario:2008yg,Bozzi:2009ig,Bozzi:2010sj,Bozzi:2011ww,Campanario:2010xn}.
As illustration, the first diagram of Fig.~\ref{fig:LO} is used. This can be
written as, 
\be
A_{V_1V_2V_3V_4,\tau}= J^{\mu_1}_{V_1^*} J^{\mu_2}_{V_2^*} {\cal
  M}_{\mu_1\mu_2,\tau}\equiv {\cal
  M}_{V_1^*V_2^*,\tau},
\ee
where the color indices have been omitted. Here, $J^{\mu_1}_{V_1^*} $
and $J^{\mu_2}_{V_2^*}$ represent effective polarization vectors
including finite width effects in the scheme of Refs.~\cite{Denner:1999gp,Oleari:2003tc} and
propagator factors, e.g., 
\be
J^{\mu_1}_{V_1^*}(q_1) = \frac{-I}{q_1^2-M_{V_1^*}^2-I \,M_{V_1^*}
  \Gamma_{V_1^*}} \left( g^{\mu_1}_{\mu} -
  \frac{q_1^{\mu_1}q_{1 \mu}}{q_1^2-M_{V_1^*}^2-I \,M_{V_1^*}\Gamma_{V_1^*}}\right) \Gamma^{\mu}_{V_1^*V_1 V_3}
\ee 
with $\Gamma_{V_1^*}$, the width of the $V_1^*$ vector boson, and
$\Gamma^{\mu}_{V_1^*V_1 V_3}$, the triple vertex which can also contain the leptonic decay of the EW vector bosons
including all off-shell effects or BSM physics.
In this manner, we can treat diagrams with triple vertexes as external
off-shell legs (in the sense that they will have the same IR
behavior). We can then concentrate in computing, instead of
$A_{V_1V_2V_3V_4,\tau}$, the virtual correction to
two massive vector bosons attached to the quark line, ${\cal  M}_{V_1^*V_2^*,\tau}$, or equivalently  ${\cal
  M}_{\mu_1\mu_2,\tau}$, where the polarization vectors or effective
currents have been factored out.
Thus, the strategy will be to combine in groups all the virtual corrections to a given
born-amplitude configuration,
independently of the effective currents  or polarization vectors attached to
the quark line. Furthermore, the order of the gauge bosons
are fixed and the full amplitude will be recovered by summing over the
physically-allowed permutations. Thus, three universal virtual
contributions are left, corrections to a born amplitude with two, three and four vector bosons
attached to the quark line for a given order permutation. 
%
%
%
%
%
%
%
%
%
%
%
%
%
%

\begin{figure}[t!]
\begin{center}
\includegraphics[scale=0.9]{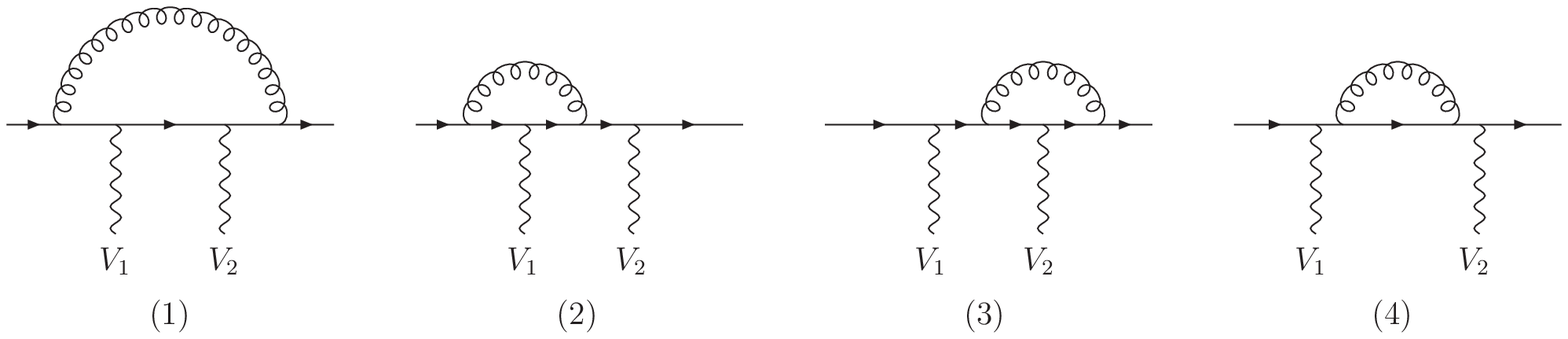}
\end{center}
\caption[]{\label{fig:box}
{\figfont Virtual corrections for a fermion line with two vector bosons
  attached, $V_1(k_1)$ and $V_2(k_2)$ in a given permutation. The sum of these graphs defines ${\cal 
M
}_{V_1V_2,\tau}$ in Eq.~(\ref{boxline}).}  }
\end{figure}

\noindent {\bf I)} The virtual corrections to the Feynman graph with two vector
bosons $V_1$ and $V_2$ attached to the quark line for a given order-permutation with
incoming momenta $k_1$ and $k_2$ are depicted in Figure~\ref{fig:box}
with kinematics,
\be
q(p_1) + \bar{q}(p_2) +V_1(k_1)+V_2(k_2) \to 0.
\ee
The sum of the four amplitudes with a given helicity is written in terms of,
\bea
\label{boxline}
{\cal 
M
}_{V_1V_2,\tau}& =&g^{V_1f}_{\tau}g^{V_2f}_{\tau}\frac{g_0^2}{(4\pi)^2}\sum_{n=1}^4 {\cal
  C}^{V_1 V_2}_{(n)}  {\cal M}^{(n)}_{V_1V_2,
\tau}~, \nonumber \\
{\cal
  M}^{(n)}_{V_1V_2,
\tau} &=& \widetilde{{\cal M}}^{(n)}_{V_1V_2,\tau}+
\widetilde{{\cal N}}^{(n)}_{V_1V_2,\tau} 
+ \frac{{\cal N}^{1,(n)}_{V_1V_2,\tau}}{\epsilon}
+\sum_{i=1}^2 \frac{{\cal M}^{i,(n)}_{V_1V_2,\tau}}{\epsilon^i},
\eea
where ${\cal 
M
}_{V_1V_2,\tau}$ is called from now on ``boxline'' contribution. Although, we are interested on off-shell vector bosons emitted
from the quark line~(effective polarization vectors), the divergent contributions and $\widetilde{{\cal N}}^{(n)}_{V_1V_2,\tau}
$ were also computed analytically for the different kinematic
configurations. These configurations are obtained by
considering the vector bosons massless or massive and all possible
ordering permutations, i.e.,
$(k_1^2=0,k_2^2=0),~(k_1^2=M_1,k_2^2=0),~(k_1^2=0,k_2^2=M_2),~(k_1^2=M_1,k_2^2=M_2)$. From the
analytical calculation, it is confirmed that the rational terms arise from
diagrams with ultraviolet divergences, thus, $\widetilde{{\cal
    N}}^{(1)}_{V_1V_2,\tau} $ is zero.  For configurations with
massless vector bosons, $k_i^2=0$, the transversality property of the bosons must be
used, $k_i \cdot \epsilon (k_i)=0$. A general proof of this statement
was presented in Ref.~\cite{Bredenstein:2008zb}. Moreover, ${\cal
  N}^{1,(n)}_{V_1V_2,\tau} $ is zero since there
are not $1/\epsilon^2$ UV poles. For the self-energy diagram, ${\cal
  M}^{2,(4)}_{V_1V_2,\tau}$=0 since the divergences are only
$1/\epsilon$. Finally, for ${\cal M}^{2,(2-3)}_{V_1V_2,\tau}$,  there are
not collinear and soft singularities simultaneously given rise to
$ 1/\epsilon^2$  poles, so that, ${\cal M}^{2,(2-3)}_{V_1V_2,\tau}$=0.

%
%
\begin{figure}[htb]
\begin{center}
\includegraphics[scale=0.9]{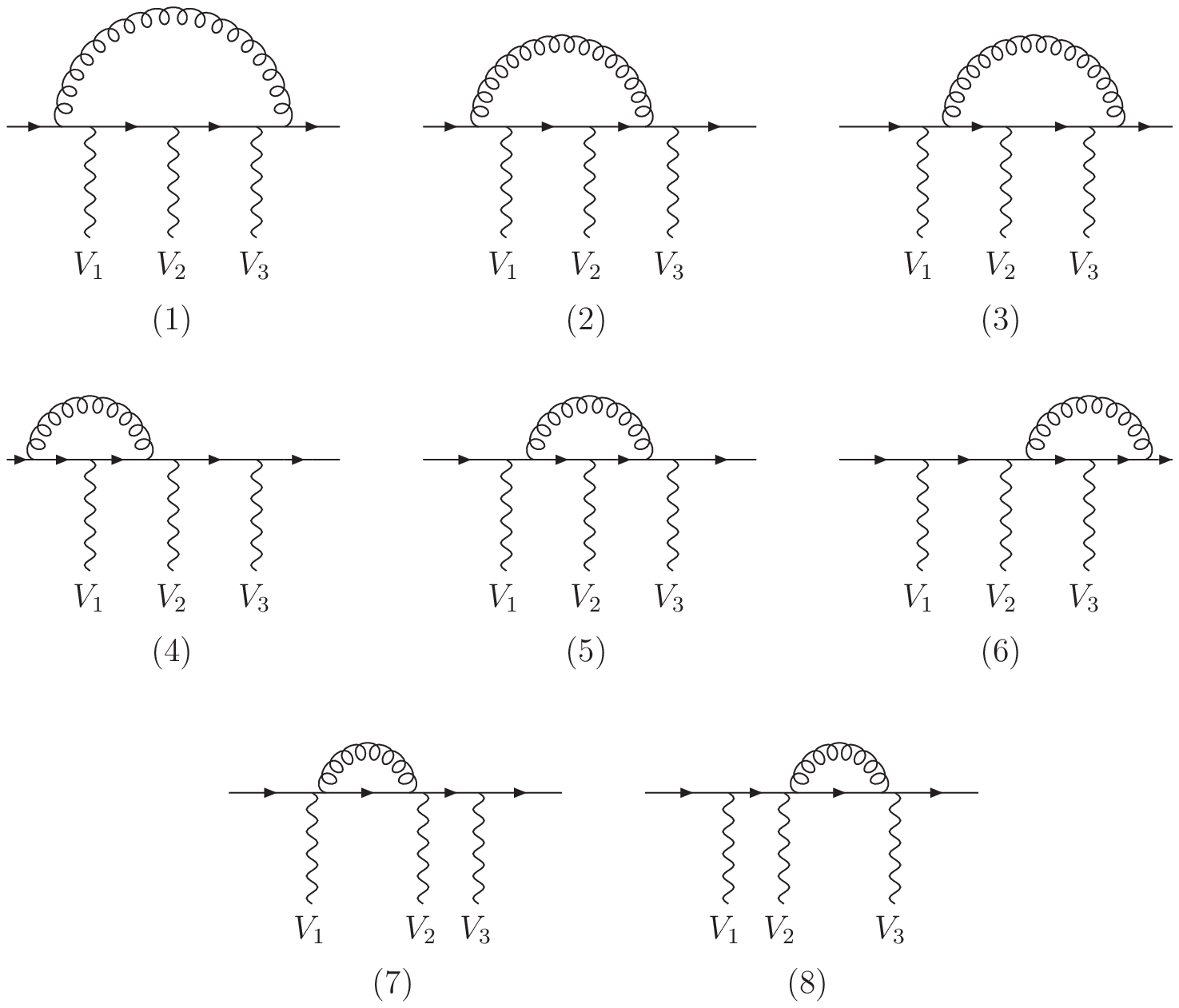}
\end{center}
\caption[]{\label{fig:pen}
{\figfont Virtual corrections for a fermion line with three vector
  bosons attached in a given permutation. The sum of these graphs defines ${\cal 
M
}_{V_1V_2V_3,\tau}$ in Eq.~(\ref{penline}).}  }
\end{figure}
%
%
%
%
%
%
%
%
%
%
%

{\bf II)} The virtual corrections to the Feynman graph with three vector
bosons $V_1$, $V_2$ and $V_3$ attached to
the quark line for a given permutation are depicted in Figure~\ref{fig:pen} with kinematics, 
\be
q(p_1) + \bar{q}(p_2) +V_1(k_1)+V_2(k_2)+V_3(k_3) \to 0.
\ee
The sum of the  eight diagrams with a given helicity is written in terms of,
\bea
\label{penline}
{\cal 
M
}_{V_1V_2V_3,\tau}& =&g^{V_1f}_{\tau}g^{V_2f}_{\tau}g^{V_3f}_{\tau}\frac{g_0^2}{(4 \pi)^2}\sum_{n=1}^8 {\cal
  C}^{V_1 V_2V_3}_{(n)}  {\cal M}^{(n)}_{V_1V_2V_3,\tau}~, \nonumber \\
 {\cal
  M}^{(n)}_{V_1V_2V_3,
\tau} &=& \widetilde{{\cal M}}^{(n)}_{V_1V_2V_3,\tau}+ \widetilde{{\cal N}}^{(n)}_{V_1V_2V_3,\tau} + \frac{{\cal N}^{1,(n)}_{V_1V_2V_3,\tau}}{\epsilon}+\sum_{i=1}^2
\frac{{\cal M}^{i,(n)}_{V_1V_2V_3,\tau}}{\epsilon^i},
\eea
where ${\cal 
M
}_{V_1V_2V_3,\tau}$ is called from now on ``penline'' contribution. 
We have computed also the divergent contributions and $\widetilde{{\cal N}}^{(n)}_{V_1V_2V_3,\tau}
$ analytically for the eight different kinematic configurations%
%
%
%
%
%
%
%
%
, $(k_1^2=0,k_2^2=0,k^2_3=0),~(k_1^2=M_1,k_2^2=0,k^2_3=0),~(k_1^2=M_1,k_2^2=M_2,k^2_3=0),\ldots$.
From the
analytical calculation, it is verified that once the transversality property
for massless on-shell particles is applied, the rational terms arise from
diagrams with ultraviolet divergences.
Therefore, $\widetilde{{\cal N}}^{(1,2,3)}_{V_1V_2V_3,\tau} $ and  ${\cal
  N}^{1,(n)}_{V_1V_2V_3,\tau} $ are zero. Analogously to the boxline, for self-energies, ${\cal
  M}^{2,(7,8)}_{V_1V_2V_3,\tau}$=0 as well as  ${\cal
  M}^{2,(2-6)}_{V_1V_2V_3,\tau}$=0.

%
%
%
%
%
%
%
%
%
%
%

\begin{figure}[h!]
\begin{center}
\includegraphics[scale=0.9]{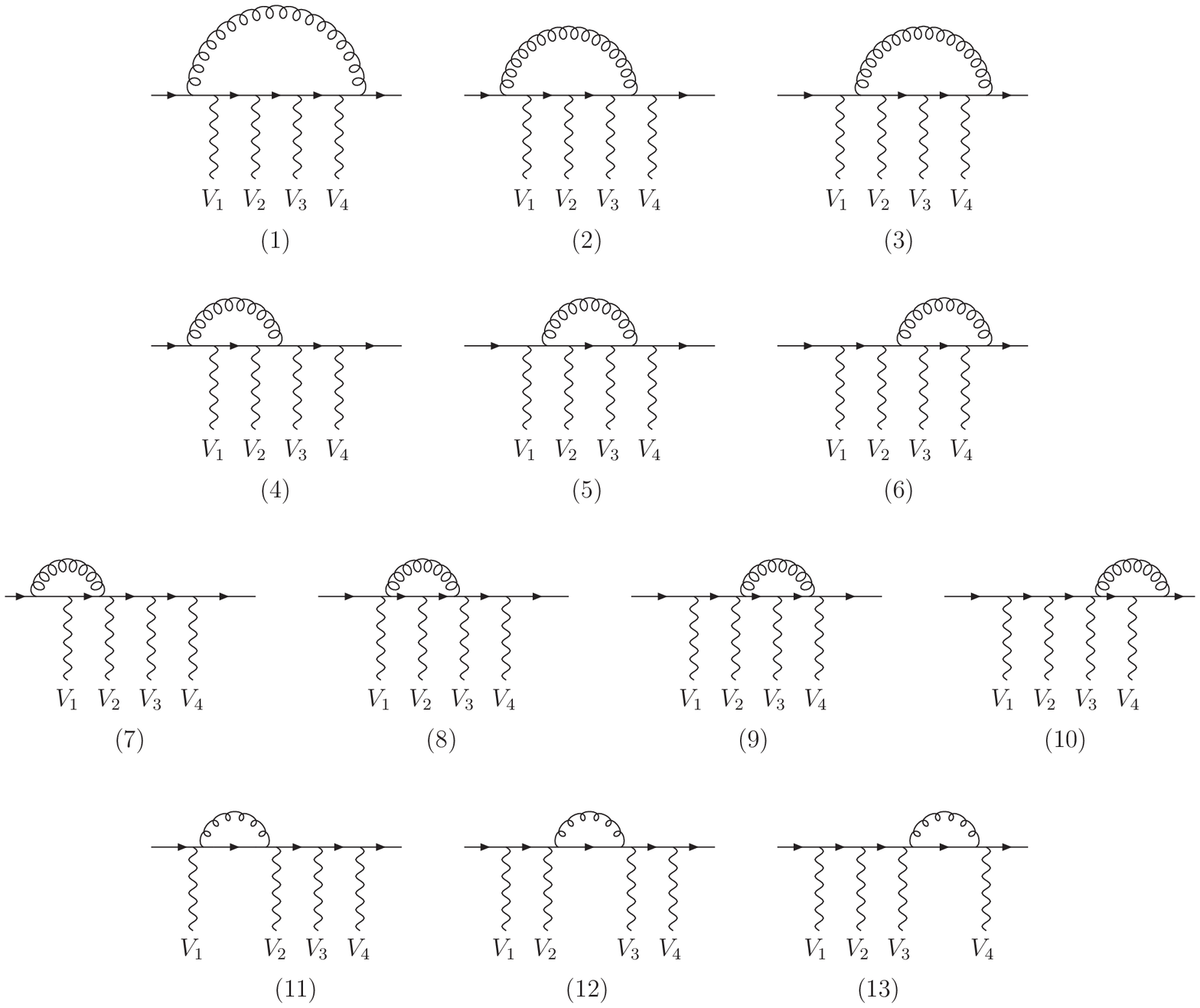}
\end{center}
\caption[]{\label{fig:hex}
{\figfont Virtual corrections for a fermion line with four vector
  bosons attached for a given order permutation. The sum of these graphs defines ${\cal 
M
}_{V_1V_2V_3V_4,\tau}$ in Eq.~(\ref{hexline}).}  }
\end{figure}
%
%
%
%
%
%
%
%
%
%
%
%
%
%
%
%

{\bf III)} The virtual corrections to the Feynman graph with four vector
bosons $V_1$, $V_2$, $V_3$ and $V_4$ attached to
the quark line for a given permutation are depicted in Figure~\ref{fig:hex}
with kinematics,
\be
q(p_1) + \bar{q}(p_2) +V_1(k_1)+V_2(k_2)+V_3(k_3)+V_4(k_4) \to 0.
\ee
The sum of the thirteen diagrams with a given helicity is written in terms of,
\bea
\label{hexline}
{\cal 
M
}_{V_1V_2V_3V_4,\tau}& =&g^{V_1f}_{\tau}g^{V_2f}_{\tau}g^{V_3f}_{\tau}g^{V_4f}_{\tau}\frac{g_0^2}{(4 \pi)^2}\sum_{n=1}^{13} {\cal
  C}^{V_1 V_2V_3V_4}_{(n)}  {\cal M}^{(n)}_{V_1V_2V_3V_4,\tau}~, \nonumber \\
  {\cal
  M}^{(n)}_{V_1V_2V_3V_4,
\tau} &=& \widetilde{{\cal M}}^{(n)}_{V_1V_2V_3V_4,\tau}+
\widetilde{{\cal N}}^{(n)}_{V_1V_2V_3V_4,\tau} + \frac{{\cal N}^{1,(n)}_{V_1V_2V_3V_4,\tau}}{\epsilon}+\sum_{i=1}^2
\frac{{\cal M}^{i,(n)}_{V_1V_2V_3V_4,\tau}}{\epsilon^i},
\eea
${\cal 
M
}_{V_1V_2V_3V_4,\tau}$ will be called ``hexline'' contribution in the
following.  For the hexline contribution, the divergences are not computed analytically. The explicit and naive recursive reduction and simplification of the tensor integrals for hexagons in terms of
$1/\epsilon$ poles exceeds the capacity of Mathematica with 4 GB of
memory RAM. Therefore, Eqs.~(\ref{D4terms},\ref{rational}) are
used to obtain numerically the different contributions both at the
Mathematica and FORTRAN level. %
We have checked with up to 30000 digits of precision in Mathematica for
the 16 different kinematic configurations, i.e.,
$(k_1^2=0,k_2^2=0,k_3^2=0,k_4^2=0),~(k_1^2=M_1,k_2^2=0,k_3^2=0,k_4^2=0),~(k_1^2=M_1,k_2^2=M_2,k_3^2=0,k_4^2=0),\ldots$
and several phase space points that  $\widetilde{{\cal
    N}}^{(1-6)}_{V_1V_2V_3V_4,\tau} $ is zero for all possible kinematic
configurations~(for massless particles the transversality property must
be used) as well as  ${\cal N}^{1,(n)}_{V_1V_2V_3V_4,\tau}= $ ${\cal
  M}^{2,(2-13)}_{V_1V_2V_3V_4,\tau}=0$. At the FORTRAN level for
non-singular points, the proof works at the working precision level.
\subsubsection{Checks}
In this section, the tests performed in order to ensure the
correctness of the calculation of the 1 loop diagrams/contributions are explained. We
consider electroweak  vector boson production ($V_1$,$V_2,V_3,V_4) \in $~(W$^\pm$,Z,$\gamma$). Then, the
color factor for all diagrams and contributions is proportional to
$C_F$. This is enough to test not only the individual diagrams but also the different
contributions. For electroweak vector bosons,  
the divergent terms for the boxline
and penline are known analytically, the general form generalizes for n
vector boson emission and 
reads,
\bea
\label{eq:hexAbe}
{\cal M}_{V_1\ldots V_n,\tau} &=&g^{V_1f}_{\tau} \ldots  g^{V_nf}_{\tau} C_F \frac{\as(\mu)}{4 \pi}
\left(\widetilde{{\cal M}}_V 
+  \right. \\ 
&&\left.
+\left( \frac{4 \pi \mu^2}{-s}
\right)^\epsilon \Gamma{(1 + \epsilon)} \left[ -\frac{2}{\epsilon^2} -
  \frac{10-D}{2\epsilon} 
\right] {\cal M}^B_{V_1 \ldots V_n,\tau} \right),~~~~~\,\,
(V_i)\in (\mbox{W}^\pm,\mbox{Z},\gamma)\nonumber 
\eea
where ${\cal M}^B_{V_1 \ldots V_n,\tau}$ is the corresponding Born amplitude
with helicity $\tau$ and $s$ the square of the
partonic center-of-mass energy.
If we use $c_\Gamma(-s)$ in Eq.~(\ref{cteFacC}) for the definition of the scalar and tensor
integrals and $D=4-2\epsilon$, then, the pre-factor $\left( \frac{4 \pi}{-s}
\right)^\epsilon \Gamma{(1 + \epsilon)}$ of Eq.~(\ref{eq:hexAbe}) is reproduced and each of the terms of Eq.~(\ref{eq:hexAbe}) can be identified, for example, for the hexline
contribution,~Eq.~(\ref{hexline}), with the different finite and pole
terms as,
\bea
\label{eq:identifyhex}
\sum_{n=1}^{13}\widetilde{{\cal M}}^{(n)}_{V_1V_2V_3V_4,\tau}= \widetilde{{\cal M}}_V,~ 
\sum_{n=1}^{13} {\cal M}^{1,(n)}_{V_1V_2V_3V_4,\tau}= -3 {\cal M}^B_{V_1V_2V_3V_4,\tau},~
\sum_{n=1}^{13} {\cal M}^{2,(n)}_{V_1V_2V_3V_4,\tau} =  -2 {\cal M}^B_{V_1V_2V_3V_4,\tau}, \nonumber \\
\sum_{n=1}^{13}\widetilde{{\cal N}}^{(n)}_{V_1V_2V_3V_4,\tau}= -{\cal M}^B_{V_1V_2V_3V_4,\tau}, \quad
\sum_{n=1}^{13}\widetilde{{\cal N}}^{1,(n)}_{V_1V_2V_3V_4,\tau}= 0.\hspace{7.15cm}
\eea 
%
%

These relations are very important for testing purposes. Note that the left
hand quantities for each of the above lines are obtained  numerically
from the
same analytical expression,~Eqs.~(\ref{D4terms},\ref{rational}). They
only differ in which terms of the $1/\epsilon^n$ expansion of the scalar
and tensor integrals are used. Therefore, the numerical check of the
factorization of the singularities and rational terms provides a strong
check of the correctness of the finite terms $\widetilde{{\cal
    M}}^{(n)}_{V_1V_2V_3v_4,\tau}$. 
We have checked the factorization formula analytically for the boxline
and penline contributions for all different kinematic configurations. To cast the singularities in this form for massless on-shell
vector bosons,~($\gamma$), the transversality property of the particle
must be used. Otherwise, additional singularities proportional to the
product  $k_i \cdot \epsilon (k_i)$ appear,
with $k_i$ the on-shell massless particle momentum. Thus, we have
checked that the presence of additional massless particles does not
introduce new singularities or new rational terms in agreement
with Ref.~\cite{Bredenstein:2008zb}.  For the hexline contribution, we have
checked numerically with high precision in Mathematica the factorization formula for the 16 different kinematic
configurations and for different phase space points . The coefficients
multiplying the poles of Eq.~(\ref{eq:hexAbe}) are obtained with up to 30000
digits of precision at least (at the FORTRAN level, for non-singular points,
the proof works at the working precision level). 
Nevertheless,
we will assume the result to be analytic due to the high precision
achieved. An analytical proof can be obtained using the
method described in Ref.~\cite{Dittmaier:2003bc} which is not automatized
within our approach. 

Concerning the origin of the rational terms in Eq.~(\ref{eq:hexAbe}), we note
that for electroweak boson production, 
the wave renormalization functions (WRF), which are zero in Dimensional
Regularization for massless on-shell quarks,  
together with $ {\cal M}_{V_1\ldots V_n,\tau} $ are UV finite. Therefore, all
the divergences in Eq.~(\ref{eq:hexAbe}), after adding the WRF, become
of IR origin including the terms containing the rational factor, 
$D/(2\epsilon)$, of UV origin. This is clear since the WRF
are zero due to the cancellation of IR and UV poles. Treating the UV and
IR divergences separately, one observes that the UV poles of the WRF cancel the rational terms of
Eq.~(\ref{eq:hexAbe}) and only IR divergences remain~(see also Ref.~\cite{Bredenstein:2008zb}). 

The factorization proof already provides an important check of the correctness
of the calculation.  
%
%
%
%
%
%
We can use the factorization
formula to perform an additional test. The factorization of
$(-s)^{-\epsilon}$ from the scalar integrals introduces an additional
dependence in $\widetilde{{\cal M}}_v$ on this variable.  If we consider (-s) as an independent scale
energy variable, $-s \equiv \mu_0$, and series expand $(\mu_0)^{-\epsilon}$ in $\epsilon$ in the second piece of
Eq.~(%
\ref{eq:hexAbe}), we obtain a new
finite term, $\widetilde{{\cal M}}^\prime_V$, of the form
%
%
%
%
\footnote{Note that
  $\widetilde{{\cal M}}_V(\mu_0)$ and  $\widetilde{{\cal M}}^\prime_V$,
  which is $\widetilde{{\cal M}}_V(\mu_0=1\text{GeV}) $, have
  the same analytical structure. They only differ in the input scalar
  integrals.},
%
%
%
%
%
\be
 \widetilde{{\cal M}}^\prime_V = \widetilde{{\cal M}}_V(\mu_0) + f(\mu_0)
 \cdot {\cal
   M}_B, \qquad  \text{with}\qquad  f(\mu_0)= -\left( \log ^2(\mu_0)-3 \log
   (\mu_0)     \right),
\label{mu-dep}
\ee
such that the $\log (\mu_0)$ terms compensate the $\mu_0$ dependence
propagated in all the scalar and 
tensor coefficient integrals in the complex $ \widetilde{{\cal
    M}}_V(\mu_0)$. Thus,  $\widetilde{{\cal M}}^\prime_V $ is $\mu_0$ independent.
This fact was satisfied for non-singular points in the three contributions at the working precision level
in the FORTRAN code~(for the hexline see Tab.~\ref{eq:Apprenhexline}) and at least with 30000 digits in the Mathematica
code.
Moreover, we have implemented Ward identity tests for the
virtual corrections in FORTRAN and Mathematica at different levels of complexity:
\begin{enumerate}
\item[{\bf 1)}] At the level of single diagrams.
\item[{\bf 2)}] For the  ${\cal 
M
}_{V_1V_2,\tau}$, ${\cal 
M
}_{V_1V_2V_3,\tau}$, ${\cal 
M
}_{V_1V_2V_3V_4,\tau}$ contributions.
\item[{\bf 3)}] At the level of gauge invariant quantities.
\item[{\bf 4)}] Subset of amplitudes invariant for a specific replacement ($\epsilon(p_k)=\epsilon_k\to p_k$).
\item[{\bf 5)}] Specific contractions that make the contributions to vanish.
\end{enumerate}
%
%
\begin{figure}[htb]
\begin{center}
\includegraphics[scale=1]{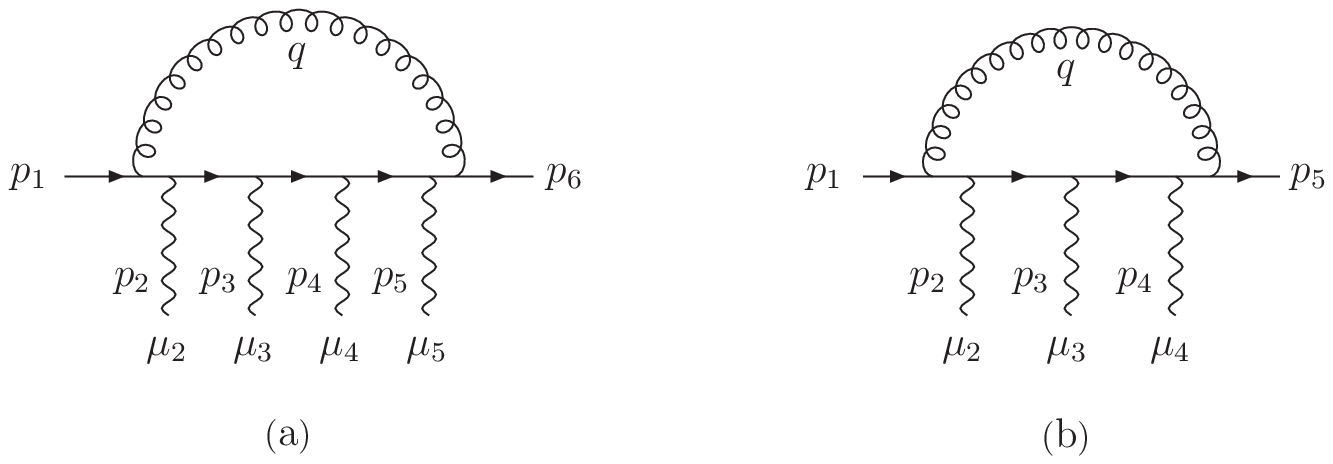}
\end{center}
\caption[]{\label{fig:gauge}
{\figfont ${\cal F}_{\mu_2\mu_3\mu_4\mu_5}(p_1,p_2,p_3,p_4,p_5)$ and 
${\cal E}_{\mu_2\mu_3\mu_4}(p_1,p_2,p_3,p_4) $ of Eq.~(\ref{eq:hex}) and
Eq.~(\ref{eq:hexgauge}).}  }
\end{figure}
%
%
%
%

\noindent {\bf 1)} Under the replacement of the polarization vector by its
momentum ($\epsilon(p_k)=\epsilon_k\to p_k$), relations among the
different diagrams/topologies can be found. 
As illustration, one can consider the hexagon of Fig.~\ref{fig:gauge},
\be
\label{eq:hex}
{\cal F}_{\mu_2\mu_3\mu_4\mu_5}(p_1,p_2,p_3,p_4,p_5)=\int
\frac{d^dq}{(2\pi)^d}
\frac{1}{q^2}
\gamma^\alpha%
\frac{1}{\slashed{q}+\slashed{p}_{15}}
\gamma_{\mu_5}\frac{1}{\slashed{q}+\slashed{p}_{14}}
\gamma_{\mu_4}\frac{1}{\slashed{q}+\slashed{p}_{13}}
\gamma_{\mu_3}\frac{1}{\slashed{q}+\slashed{p}_{12}}
\gamma_{\mu_2}\frac{1}{\slashed{q}+\slashed{p}_{1}}
\gamma_\alpha
\ee
where $p_{1j}=\sum_{k=1}^j p_k$. Contracting one of the open indices by
the corresponding momentum and expressing the contracted gamma matrix as
the difference of two adjacent fermionic propagators, the hexagon is
reduced to a difference of two pentagon integrals,
\bea
\label{eq:hexgauge}
p_2^{\mu_2}{\cal F}_{\mu_2\mu_3\mu_4\mu_5}(p_1,p_2,p_3,p_4,p_5)&=&
{\cal E}_{\mu_3\mu_4\mu_5}(p_1,p_2+p_3,p_4,p_5)
-{\cal E}_{\mu_3\mu_4\mu_5}(p_1+p_2,p_3,p_4,p_5), \nonumber \\
p_3^{\mu_3}{\cal F}_{\mu_2\mu_3\mu_4\mu_5}(p_1,p_2,p_3,p_4,p_5)&=&
{\cal E}_{\mu_2\mu_4\mu_5}(p_1,p_2,p_3+p_4,p_5)
-{\cal E}_{\mu_2\mu_4\mu_5}(p_1,p_2+p_3,p_4,p_5), \nonumber \\
p_4^{\mu_4}{\cal F}_{\mu_2\mu_3\mu_4\mu_5}(p_1,p_2,p_3,p_4,p_5)&=&
{\cal E}_{\mu_2\mu_3\mu_5}(p_1,p_2,p_3,p_4+p_5)
-{\cal E}_{\mu_2\mu_3\mu_5}(p_1,p_2,p_3+p_4,p_5), \nonumber \\
p_5^{\mu_5}{\cal F}_{\mu_2\mu_3\mu_4\mu_5}(p_1,p_2,p_3,p_4,p_5)&=&
{\cal E}_{\mu_2\mu_3\mu_4}(p_1,p_2,p_3,p_4)
-{\cal E}_{\mu_2\mu_3\mu_4}(p_1,p_2,p_3,p_4+p_5). 
\eea
where ${\cal E}_{\mu_i\mu_j\mu_k}$ represents the pentagon diagram
of Figure~\ref{fig:gauge}, defined similarly as
Eq.~(\ref{eq:hex}).
%
%
%
%
%

\noindent {\bf 2)} For the contributions, assuming EW vector boson
production, equivalent relations
can be obtained. First, we factor out 
the couplings and the polarization vectors,
\be
{\cal M}_{V_1 \ldots V_n,\tau}(p_1,\ldots,p_n)= g^{V_1f}_{\tau} \ldots  g^{V_nf}_{\tau}\epsilon_{V_1}^{\mu_2}(p_2) \ldots \epsilon_{V_n}^{\mu_n}(p_n)
{\cal M}_{\mu_2\ldots  \mu_n,\tau}(p_1,\ldots,p_n).
\ee
where ${\cal M}_{\mu_2\ldots
  \mu_n,\tau}(p_1,\ldots,p_n)$ represents the un-contracted 
contributions and we have explicitly written down the momentum's dependence following
the convention of Figure~\ref{fig:gauge}. Note that the color factors, $C_{m}^{V_1 \ldots V_n}$, are still present in ${\cal M}_{\mu_2\ldots
  \mu_n,\tau}(p_1,\ldots,p_n)$. Then,  Eq.~(\ref{eq:hexgauge}) holds for
the  hexline and penline contributions with the replacements, 
\bea
&{\cal
  F}_{\mu_2\mu_3\mu_4\mu_5} \to  {\cal M}_{\mu_2\mu_3\mu_4\mu_5,\tau},\nonumber\\
& {\cal E}_{\mu_i\mu_j\mu_l} \to {\cal M}_{\mu_i\mu_j\mu_l,\tau}.
\eea
%
Analogously, the penline contribution can be checked
against the boxline contribution obtaining similar relations. 
Additionally, to check the boxline contribution, we computed
the quark line with one vector boson attached, which we called vertline, ${\cal 
M
}_{\mu_2,\tau}$, and is formed by a simple vertex graph. These relations
are satisfied  independently both for the finite and the divergent
parts of each contribution.  \\

\begin{figure}[htb!]
\includegraphics[scale=1]{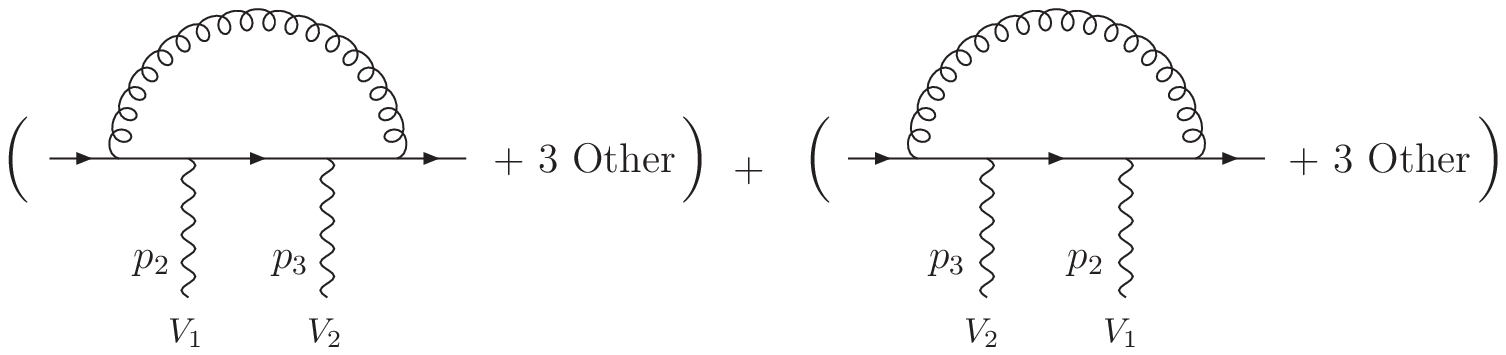}
\caption[]{\label{fig:cross}
{\figfont Direct and cross term diagrams which correspond to permuting
  the momenta and the polarization vectors in the boxline contributions
  yielding a gauge invariant subset for $V_i \in (\gamma,Z)$. }  }
\end{figure}

\noindent {\bf 3)} Vertline is gauge invariant under the replacement of
$\epsilon^{\mu_i}(p_i)\to p_i^{\mu_i}$ which means that the equivalent form of Eq.~(\ref{eq:hexgauge}) gives zero. 
To build gauge invariant quantities for the other un-contracted
contributions and for EW vector boson production, the
cross diagrams must be considered~\footnote{Note that for processes with
  $W$ vector
  bosons not all the permutations are physically allowed.} (See Fig.~\ref{fig:cross} as
illustration). 
%
Thus, the sum of direct and cross terms under the replacement
$\epsilon^{\mu_i}(p_i)\to p_i^{\mu_i}$ is gauge invariant: 
\be
\begin{array}{lll}
p_i^{\mu_i}\left({\cal M}_{\mu_2\mu_3,\tau}(p_1,p_2,p_3) +
{\cal M}_{\mu_3\mu_2,\tau}(p_1,p_3,p_2)\right) & =& 0,\\ 
p_i^{\mu_i}\left({\cal M}_{\mu_2\mu_3\mu_4,\tau}(p_1,p_2,p_3,p_4)+{\cal M}_{\mu_3\mu_2\mu_4,\tau}(p_1,p_3,p_2,p_4)
+ \mbox{4 other}\right) &=& 0,\\
p_i^{\mu_i}\left({\cal M}_{\mu_2\mu_3\mu_4\mu_5,\tau}(p_1,p_2,p_3,p_4,p_5)+{\cal M}_{\mu_3\mu_2\mu_4\mu_5,\tau}(p_1,p_3,p_2,p_4,p_5)+ \mbox{22 other}\right) &=& 0.\\
\end{array}
\ee
%
%
%
%

\noindent {\bf 4)} For the un-contracted penline and hexline and
considering EW vector boson production, there are
smaller subsets which give zero~(``gauge invariant'') for a given replacement ($\epsilon(p_k)=\epsilon_k\to p_k$).
For the  un-contracted penline $ {\cal M}_{\mu_2\mu_3\mu_4,\tau} (p_1,p_2,p_3,p_4)$, a ``gauge
invariant'' subset for a specific replacement, e.g.,
$\epsilon(p_2)^{\mu_2} \to p_2^{\mu_2}$  is obtained by permuting  the
position of the $p_2$ momentum and the corresponding Lorentz index and
keeping the relative order of the other vector bosons fixed, i.e.,
\be 
p_2^{\mu_2}\left( {\cal M}_{\mu_2\mu_3\mu_4,\tau} (p_1,p_2,p_3,p_4) +  
{\cal M}_{\mu_3\mu_2\mu_4,\tau} (p_1,p_3,p_2,p_4) +  
{\cal M}_{\mu_3\mu_4\mu_2,\tau}(p_1,p_3,p_4,p_2) \right)= 0.
\ee 
The same can be proved performing the
corresponding replacement~($\epsilon(p_k)=\epsilon_k\to p_k$) for the contracted contributions which include
the couplings $g^{V_nf}_{\tau}$, subject that $V_n\in (\gamma,Z)$.

\noindent {\bf 5)} In addition, we have checked analytically and later
on numerically that there are some combinations of replacements for the 
polarization vector which make for EW vector boson production the contributions to vanish. e.g.,
\be
\begin{array}{lll}
\left(p_2^{\mu_2} (p_2+p_3)^{\mu_3} || (p_2+p_3)^{\mu_2} p_3^{\mu_3}
\right) \,\, \left({\cal M}_{\mu_2\mu_3,\tau}(p_1,p_2,p_3)\right) & =& 0,\\ 
p_2^{\mu_2} (p_2+p_3+p_4)^{\mu_3} p_4^{\mu_4} \,\,\left({\cal M}_{\mu_2\mu_3\mu_4,\tau}(p_1,p_2,p_3,p_4)\right) &=& 0,\\
\left(p_2^{\mu_2} (p_2+p_3)^{\mu_3}(p_2+p_3+p_4+p_5)^{\mu_3}p_5^{\mu_5}
  || p_2^{\mu_2} (p_2+p_3+p_4+p_5)^{\mu_3}(p_4+p_5)^{\mu_3}p_5^{\mu_5} 
\right)\,\,  && \\ \qquad \qquad\qquad \qquad\qquad \qquad\qquad \qquad\qquad \qquad \qquad \left({\cal
    M}_{\mu_2\mu_3\mu_4\mu_5,\tau}(p_1,p_2,p_3,p_4,p_5)\right) &=& 0.\\
\end{array}
\ee

These tests have been implemented at the FORTRAN and Mathematica
level. At the Mathematica level a precision of 30000 digits was
achieved. At the FORTRAN level, for non-singular phase space points,
these tests are satisfied at the working precision level.
%
%
%
The finite boxline and penline contribution have essentially the same analytic
expressions found in the calculation of the NLO QCD  corrections in vector boson
fusion processes, $qq\to Vqq$ and $qq\to VVqq$, discussed in
Refs.~\cite{Oleari:2003tc} and~\cite{Jager:2006zc}, respectively. We have checked that our results agree at the
double precision level. 
The use of modular structure routines, as the above presented, has been
proved to be an advantage in the program {\it VBFNLO}~\cite{Arnold:2008rz} since once a
structure is computed and checked it can be reused for different
processes. For example, the boxline and penline contributions computed here have been
used to compute the NLQ QCD corrections to 
$W^\pm W^\pm Z$~\cite{Campanario:2008yg}, 
$W^+W^-\gamma~(ZZ\gamma)$~\cite{Bozzi:2009ig}, 
$W^\pm Z \gamma$~\cite{Bozzi:2010sj}, 
$W^\pm \gamma \gamma$~\cite{Bozzi:2011ww}, 
$W^{\pm} \gamma j$~\cite{Campanario:2009um,Campanario:2010hv}, 
$W^{\pm} Z j$~\cite{Campanario:2010hp,Campanario:2010xn} and also
$H \gamma jj$~\cite{Arnold:2010dx} production. They are publicly available as part of the
{\it VBFNLO} package together with the tensor reduction routines up
to the pentagon level, excluding the routines for small Gram determinants.
%
%
%
%
%
%
%
%
%
%
%
 \begin{figure}[h!]
\begin{center}
\includegraphics[scale=0.87]{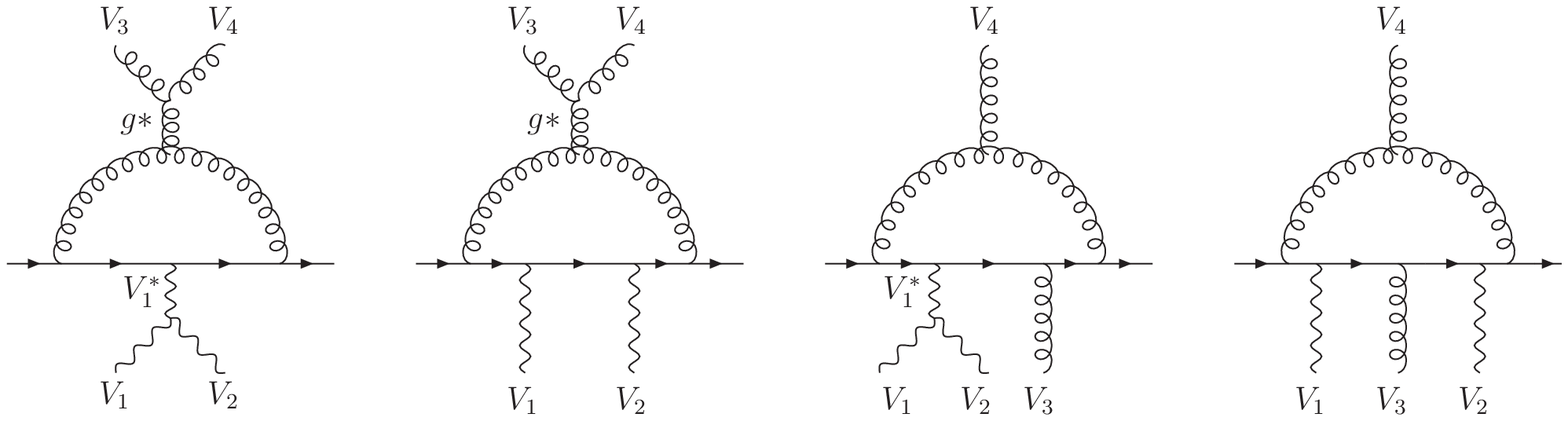}
\end{center}
\caption[]{\label{fig:NoAbegra}
{\figfont Topologies appearing in the
  calculation of the virtual contributions for $pp \to VVjj +X $
  production, with V$  \in ($W $^\pm$,Z,$\gamma)$ .}  }
\end{figure}
%
%
%
%
%
%
%
%
%
%
\subsection{``Bosonic'' contributions to $pp \to VVVj + X$} 
\label{sec:NOABE}
One loop non-abelian corrections involving one triple gluon vertex are
considered in this section,
Fig.~\ref{fig:NoAbegra}.  We follow the same strategy used in the
previous section 
and classify the virtual corrections to three
different groups depending whether they contribute to {\bf I)} $q
\bar{q} V_1^* g^*\to 0$, {\bf II)}  $q \bar{q} V_1 V_2 g^*\to 0$ or
$q \bar{q} V_1^* V_3 V_4\to
0$  or  {\bf III)} $q \bar{q}
V_1V_2 V_3 V_4\to 0$. In analogy to the previous section, for the last two, only diagrams with 
a specific permutation of the vector bosons are considered. Generally, the momentum
square for the off-shell legs $V_1^*$ and $g^*$, indicated with a star,
is not zero. This is important since it results in different IR
divergences for individual graphs. From now on, to avoid confusion, the star is only kept for the
gluon, but we remind the reader that at the programing
level all vector bosons are considered to be off-shell as described in
Section~\ref{sec:calc} since the divergences are reproduced numerically,
Eqs.~(\ref{D4terms},\ref{rational}). 

%
%
%
%
%
%
%
%
%
%
%
%
%
%
%
%
%
%

\begin{figure}[t!]
\begin{center}
\includegraphics[scale=1]{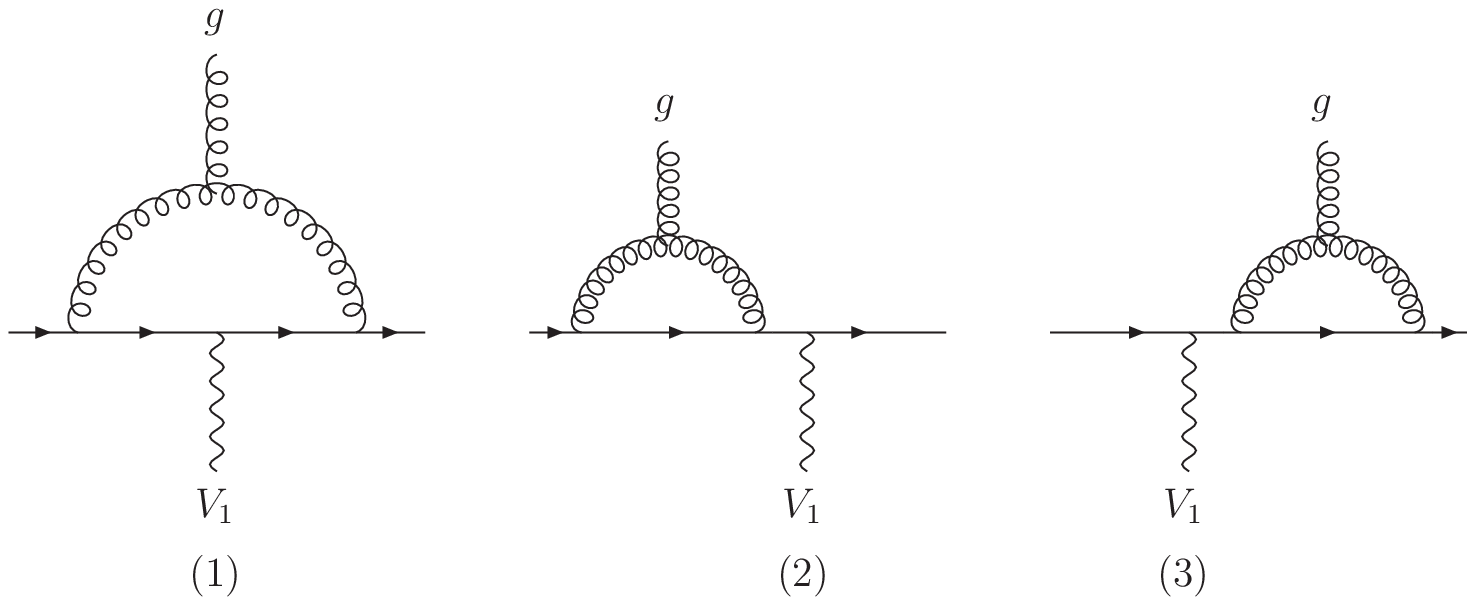}
\end{center}
\caption[]{\label{fig:boxNoAbe}
{\figfont Virtual corrections for a fermion line with one vector boson
  attached, $V_1(k_1)$. The sum of these graphs defines ${\cal 
M
}^{g^*}_{V_1,\tau}$ in Eq.~(\ref{boxNoAbe}). }  }
\end{figure}

{\bf I)} Non-abelian corrections to the Feynman graphs with one vector
boson $V_1$ attached to
the quark line are depicted in Figure~\ref{fig:boxNoAbe} with kinematics,
\be
q(p_1) + \bar{q}(p_2) +V_1(k_1)+g^*(k_2) \to 0.
\ee
%
%
%
The sum of the three graphs with a given helicity is written in terms of,
\bea
\label{boxNoAbe}
{\cal M}_{V_1,\tau}^{g^*}& =& g^{V_1f}_{\tau} g_0 \frac{g_0^2}{(4 \pi)^2}\sum_{n=1}^3 {\cal
  C}^{g,V_1}_{(n)}  {\cal M}^{g^*,(n)}_{V_1,\tau}, \nonumber \\
 {\cal
  M}^{g^*,(n)}_{V_1,\tau} &=& \widetilde{{\cal M}}_{V_1,\tau}^{g^*,(n)} + \widetilde{{\cal
    N}}_{V_1,\tau}^{g^*,(n)}
  + \frac{{\cal N}_{V_1,\tau}^{g^*,1,(n)}}{\epsilon}+\sum_{i=1}^2 \frac{{\cal M}_{V_1,\tau}^{g^*,i,(n)}}{\epsilon^i},
\eea
where $ {\cal M}_{V_1,\tau}^{g^*}$ is called from now on ``boxlineNoAbe''
contribution. We have computed the divergent contributions and
$\widetilde{{\cal  N}}_{V_1,\tau}^{g^*,(n)}$ analytically for the 4 different
kinematic configurations. From the analytical calculation using the
transversality property for massless particles, it is confirmed that $\widetilde{{\cal  N}}_{V_1,\tau}^{g^*,(1)}$ and  ${\cal
    N}_{V_1,\tau}^{g^*,1,(n)}$  are zero. 
\begin{figure}[htb]
\begin{center}
\includegraphics[scale=1]{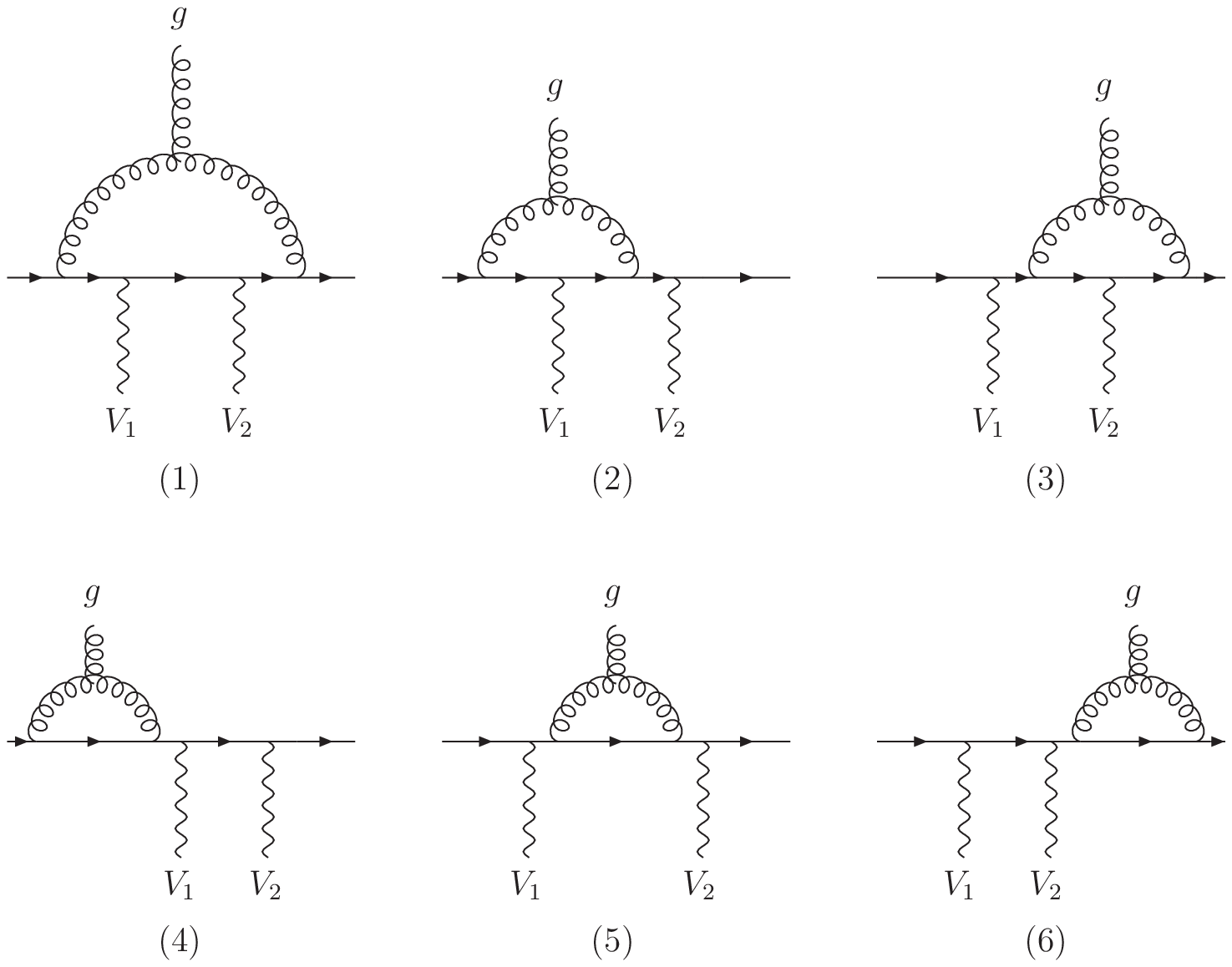}
\end{center}
\caption[]{\label{fig:penNoAbe}
{\figfont Virtual corrections for a fermion line with two vector bosons
  attached, $V_1(k_1)$ and $V_2(k_2)$ for a given order permutation. The sum of these graphs defines ${\cal 
M
}^{g^*}_{V_1V_2,\tau}$ in Eq.~(\ref{penNoAbe}).}  }
\end{figure}
%
%
%
%
%
%
%
%
%
%
%
%
%
%
%
%

{\bf II)} The virtual corrections to the Feynman graph with two vector
bosons $V_1$ and $V_2$  attached to
the quark line for a given permutation are depicted in
Figure~\ref{fig:penNoAbe} 
with kinematics,
\be
q(p_1) + \bar{q}(p_2) + V_1(k_1)+V_2(k_2)+g^*(k_3) \to 0.
\ee
The sum of the six graphs with a given helicity is written in terms of:
\bea
\label{penNoAbe}
{\cal 
M
}^{g^*}_{V_1V_2,\tau}& =&g^{V_1f}_{\tau}g^{V_2f}_{\tau}g_0\frac{g^2_0}{(4
  \pi)^2}\sum_{n=1}^6 {\cal
  C}^{g,V_1 V_2}_{n}  {\cal M}^{g^*,(n)}_{V_1V_2,\tau}, \nonumber \\
 {\cal
  M}^{g^*,(n)}_{V_1V_2,\tau} &=& \widetilde{{\cal M}}_{V_1V_2,\tau}^{g^*,(n)} + \widetilde{{\cal
    N}}_{V_1V_2,\tau}^{g^*,(n)}
  + \frac{{\cal N}_{V_1V_2,\tau}^{g^*,1,(n)}}{\epsilon}+\sum_{i=1}^2 \frac{{\cal M}_{V_1V_2,\tau}^{g^*,i,(n)}}{\epsilon^i},
\eea
where $ {\cal M}_{V_1V_2,\tau}^{g^*}$ is called  ``penlineNoAbe''
contribution in the following. 
We have computed the divergent contributions and
$\widetilde{{\cal  N}}_{V_1V_2,\tau}^{g^*,(n)}$ analytically for the 8 different
kinematic configurations. It is verified that $\widetilde{{\cal  N}}_{V_1V_2,\tau}^{g^*,(1-3)}$ and  $ {\cal
  N}_{V_1V_2,\tau}^{g^*,1,(n)}$ are zero (the transversality property
for massless particles must be used).
\begin{figure}[h!]
\begin{center}
\includegraphics[scale=0.9]{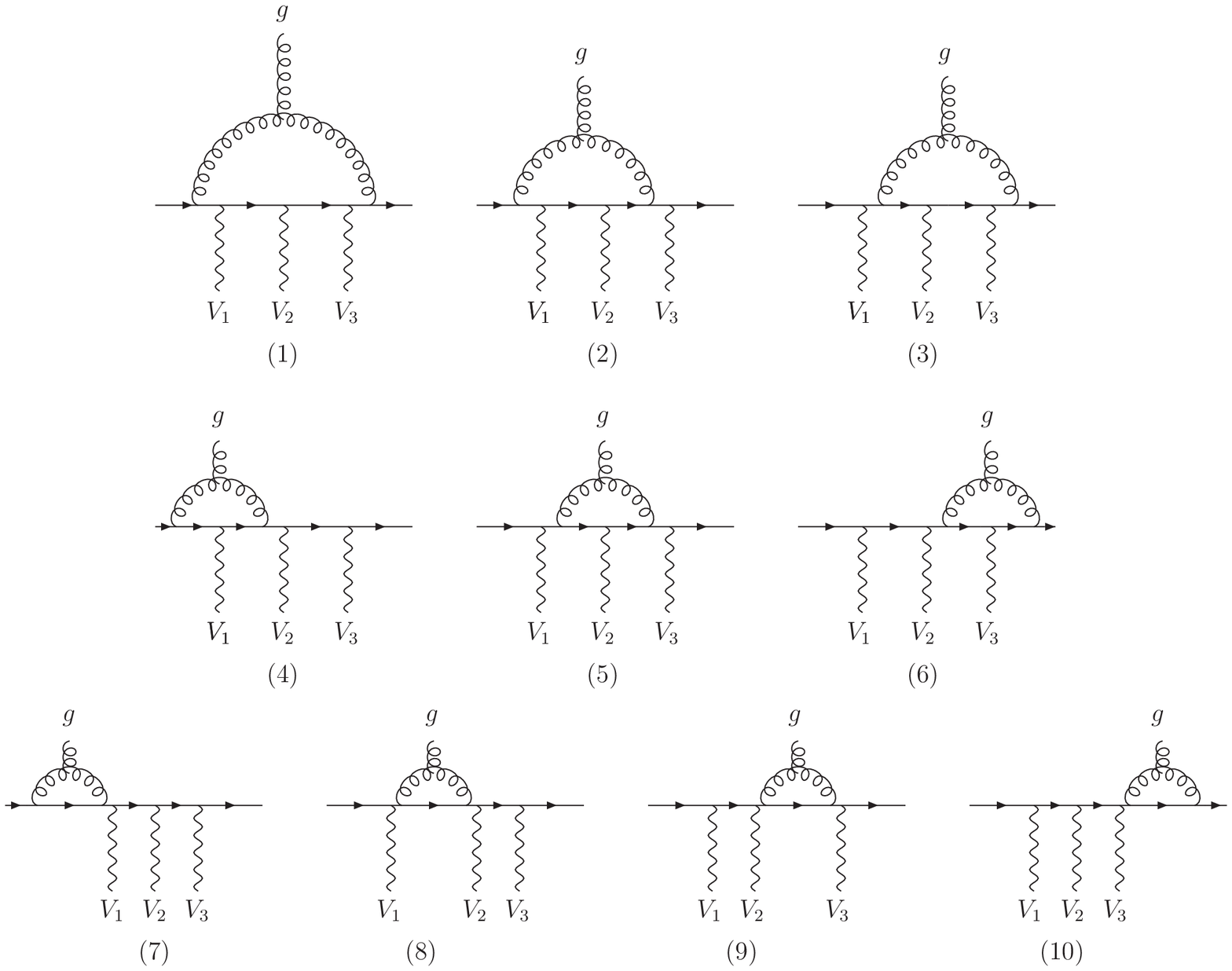}
\end{center}
\caption[]{\label{fig:hexNoAbe}
{\figfont Virtual corrections for a fermion line with three vector bosons
  attached, $V_1(k_1)$, $V_2(k_2)$ and $V_3(k_3)$ for one permutation. The sum of these graphs defines ${\cal 
M
}^{g^*}_{V_1V_2V_3,\tau}$ in Eq.~(\ref{hexNoAbe}).}  }
\end{figure}
%
%
%
%
%
%
%
%
%
%
%
%
%
%
%
%

{\bf III)} The virtual corrections to the Feynman graph with three vector
bosons $V_1$, $V_2$ and $V_3$ attached to
the quark line for a given permutation are depicted in
Figure~\ref{fig:hexNoAbe} 
with kinematics%
~\footnote{Note that we do
  not need to consider an off-shell gluon in this case since for $pp\to VV
  jj$ one of the $V_i$ must be also a gluon in Eq.~(\ref{hexNoAbe:ki}).},
%
%
%
\be
\label{hexNoAbe:ki}
q(p_1) + \bar{q}(p_2) + V_1(k_1)+V_2(k_2) +V_3(k_3)  +g^*(k_4)\to 0.
\ee
The sum of the ten diagrams with a given helicity is written in terms of:
\bea
\label{hexNoAbe}
{\cal 
M
}^{g^*}_{V_1V_2V_3,\tau}&
=&g^{V_1f}_{\tau}g^{V_2f}_{\tau}g^{V_3f}_{\tau}g_0\frac{g_0^2}{(4 \pi)^2}\sum_{n=1}^{10} {\cal
  C}^{g, V_1 V_2V_3}_{n}  {\cal M}^{g^*,(n)}_{V_1V_2V_3,\tau}~ , \nonumber \\
 {\cal
  M}^{g^*,(n)}_{V_1V_2V_3,\tau} &=& \widetilde{{\cal M}}_{V_1V_2V_3,\tau}^{g^*,(n)} + \widetilde{{\cal
    N}}_{V_1V_2V_3,\tau}^{g^*,(n)}
  + \frac{{\cal N}_{V_1V_2V_3,\tau}^{g^*,1,(n)}}{\epsilon}+\sum_{i=1}^2 \frac{{\cal M}_{V_1V_2V_3,\tau}^{g^*,i,(n)}}{\epsilon^i},
\eea
${\cal M}^{g^{*}}_{V_1V_2V_3,\tau}$ is called ``hexlineNoAbe'' contribution in the
following. We have checked  with high precision numerically in Mathematica
that once the transversality property for massless
particles is used, $\widetilde{{\cal N}}^{g^*(1-6)}_{V_1V_2V_3,\tau}$ and  ${\cal
  N}^{g^*1,(n)}_{V_1V_2V_3,\tau} $ are zero for the 16 different
kinematic configurations (at the FORTRAN level for non-singular points, this proof works at the working precision level).
\subsubsection{Checks}
\begin{figure}[htb]
\begin{center}
\includegraphics[scale=0.85]{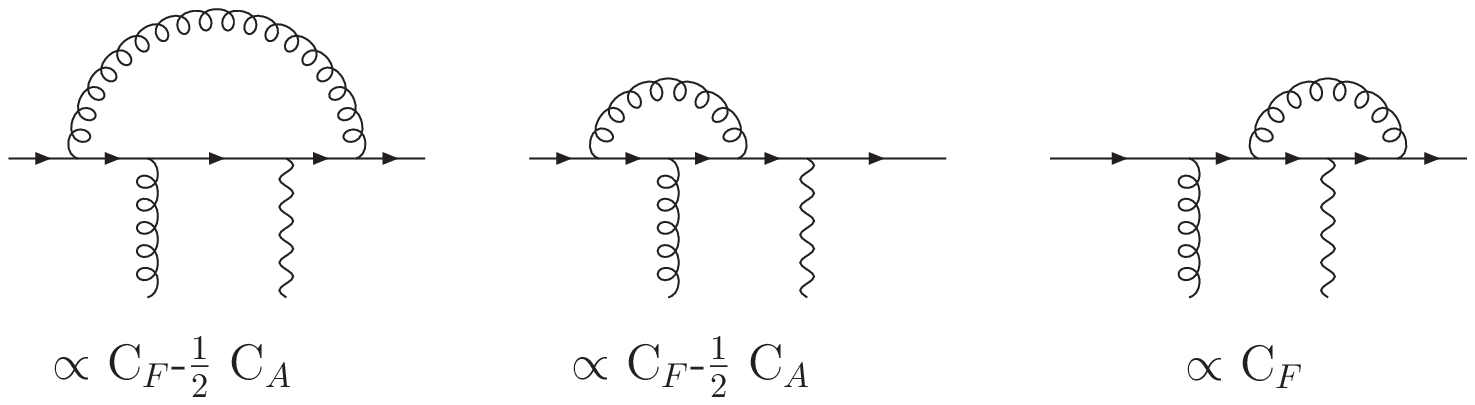}
\end{center}
\caption[]{\label{fig:color}
{\figfont Color factors that appear when one gluon is emitted from the quark line.}}
\end{figure}

The factorization of the infrared divergences for one
or two electroweak vector boson production in
association with one jet, $g u \to u V_1$ or $ g u \to u V_1 V_2$, is
known analytically~\cite{Figy:2007kv,Campanario:2009um}. We will make use of this fact to perform powerful tests
to our contributions. For one gluon emission, the color for all the non-abelian
contributions is proportional to $C_A$, and two color factors proportional to $C_F$ and to $C_F-1/2 C_A$~(see Fig.~\ref{fig:color}) appear in the abelian
contributions computed in the previous section. The
relevant color factors for this proof for all
diagrams of each abelian contribution are given in Appendix~\ref{appendixC}.
We will concentrate in the hexline contributions since the penline and
the boxline contributions are checked similarly. To verify the factorization of the divergences against the
born amplitude for
$(V_1,V_2,V_3)\in (W^\pm,Z,\gamma)$ and for an
on-shell gluon, it is enough to consider all the diagrams mixed under the
QCD gauge group for a given order combination of the vector bosons, $(V_1,V_2,V_3)$,
\be
\label{fac:hexNoAbe}
{\cal M }_{g V_1V_2V_3,\tau}+{\cal  M}_{V_1gV_2V_3,\tau}+{\cal 
M
}_{V_1V_2 g V_3 ,\tau}+{\cal 
M
}_{V_1V_2V_3 g,\tau}+
{\cal 
M
}^g_{V_1V_2V_3,\tau}.
\ee
\noindent The same is proved for the other six permutations ($V_1 \Leftrightarrow
V_2 \Leftrightarrow V_3$). This combination under the replacement
$\epsilon(k_4)\to k_4$, given the kinematics defined by
Eq.~(\ref{hexNoAbe:ki}) is ``gauge invariant''.  The virtual contributions
factorize on the Born amplitude generalizing the result of diboson plus
jet production.  The sum of the amplitudes, including
the counterterms~(CT), in Dimensional Regularization are given by,
%
\bea 
\label{eq:hexlineNoAbe}
&&{\cal M }_{g V_1V_2V_3,\tau}
+{\cal  M}_{V_1gV_2V_3,\tau}
+{\cal M}_{V_1V_2 g V_3 ,\tau}
+{\cal M}_{V_1V_2V_3 g,\tau}
+{\cal M}^g_{V_1V_2V_3,\tau}+CT
=
\nonumber \\
&&\mbox{~~~~~~~~~}={\cal M}^{B}_\tau \frac{\alpha_s(\mu^2)}{4 \pi} 
\Gamma(1+\epsilon)
\left \{ \frac{1}{2} 
\left(\left (\frac{4 \pi \mu^2}{-u }\right)^{\epsilon} +
\left (\frac{4 \pi \mu^2}{-t}\right)^{\epsilon} \right) 
(-\frac{C_A}{\epsilon^2}-\frac{\gamma_{g}}{\epsilon}) \right. \nonumber \\
&&\mbox{~~~~~~~~~}+ \frac{1}{2} \frac{C_A}{C_F} \left ( \left( \frac{4 \pi \mu^2}{-u} 
\right)^{\epsilon} +  
\left( \frac{4 \pi \mu^2}{-t} \right)^{\epsilon}-2  
\left( \frac{4 \pi \mu^2}{-s} \right)^{\epsilon} \right)
(-\frac{C_F}{\epsilon^2}-\frac{\gamma_q}{\epsilon}) \nonumber \\
&&\mbox{~~~~~~~~~}+ 2 \left (\frac{4 \pi \mu^2}{-s} \right)^{\epsilon}
\left. (-\frac{C_F}{\epsilon^2} -\frac{\gamma_q}{\epsilon} + \frac{C_F(D-4)}{4\epsilon}) 
\right \}  
+ \widetilde{{\cal M}}_{V},\quad (V_1,V_2,V_3)
\in(W^\pm,Z,\gamma),~ \quad ~~~~
\eea
with
\be
CT=-{\cal M}^{B}_\tau \frac{\alpha_s(\mu^2)}{4 \pi} 
\Gamma(1+\epsilon)(4\pi)^\epsilon\frac{\gamma_g}{\epsilon}%
, \quad \gamma_{q} = \frac{3}{2} C_{F}, \quad 
\gamma_{g} = \frac{11}{6} C_{A} - \frac{2}{3} T_{R} N_{f},
\ee
$T_{R}=1/2$, $C_{A}=N$, and $C_{F}=(N^2-1)/(2N)$ in $SU(N)$ 
gauge theory. The number of flavors is $N_{f}=5$,  $s$ is the square of the
partonic center-of-mass energy, $s=(p_1+p_2)^2$, $u=(p_1+k_4)^2$ and
$t=(p_2+k_4)^2$, given the kinematics defined by
Eq.~(\ref{hexNoAbe:ki}).
${\cal M}^B_\tau$
is the Born amplitude formed by four tree level graphs, the
corresponding born
amplitudes of each of the ``hexline'' contributions, ${\cal
  M}^B_\tau={\cal M}^B_{V_1V_2V_3g\tau}+{\cal
  M}^B_{V_1V_2gV_3\tau}+{\cal M}^B_{V_1gV_2V_3\tau}+{\cal
  M}^B_{gV_1V_2V_3\tau}$. Given our definition for the scalar and tensor
integrals, Eq.~(\ref{cteFacC}), $\widetilde{{\cal M}}_V$ is defined through,
\bea
\label{eq:identifyhexNoabef}
\sum_{n=1}^{13} \left(
{\cal C}_{(n)}^{V_1V_2V_3g}\widetilde{{\cal M}}^{(n)}_{V_1V_2V_3 g,\tau} 
+{\cal C}_{(n)}^{V_1V_2gV_3}\widetilde{{\cal M}}^{(n)}_{V_1V_2gV_3,\tau} 
+{\cal C}_{(n)}^{V_1gV_2V_3}\widetilde{{\cal M}}^{(n)}_{V_1gV_2V_3 ,\tau} 
+{\cal C}_{(n)}^{gV_1V_2V_3}\widetilde{{\cal M}}^{(n)}_{gV_1V_2V_3 ,\tau} \right)
+\nonumber\\
+\sum_{j=1}^{10}{\cal C}_{(j)}^{g,V_1V_2V_3}\widetilde{{\cal M}}^{g,(j)}_{V_1V_2V_3,\tau} 
= \widetilde{{\cal M}}_V + {\cal M}^B_\tau
f(u,s,t,\mu^2,C_F,C_A,\gamma_g),  \qquad \qquad \qquad 
\eea
%
%
%
%
%
%
%
%
%
%
where $f(u,s,t,\mu^2,C_F,C_A,\gamma_g)$ is given by the finite terms resulting
of the
epsilon expansion of Eq.~(\ref{eq:hexlineNoAbe}),
%
\bea
\label{eq:identifyf}
f(u,s,t,\mu^2,C_F,C_A,\gamma_g)=-\text{$C_F$} \left(\log \left(\frac{-s}{\mu^2 }\right)-\frac{2\gamma_q}{C_F}\right) \log \left(\frac{-s}{\mu^2
   }\right)+\frac{\gamma _g}{2} \left(\log \left(\frac{-t}{\mu^2
     }\right)+ \right. \nonumber\\
\left. +\log
   \left(\frac{-u}{\mu^2 }\right)\right)  
+\text{$C_A$} \left(\frac{1}{2}
   \left(\log ^2\left(\frac{-s}{\mu^2 }\right)-\log ^2\left(\frac{-t}{\mu^2
   }\right)-\log ^2\left(\frac{-u}{\mu^2 }\right)\right)+
\right. \nonumber
\\ 
+\left.
\frac{\gamma
 _q}{2 \text{$C_F$}}\left(-2 \log
   \left(\frac{-s}{\mu^2 }\right)+\log \left(\frac{-t}{\mu^2 }\right)+\log
   \left(\frac{-u}{\mu^2 }\right)\right) \right).
\qquad\qquad\qquad\quad
\eea
From the analytical form of the factorization formula, we can find
relations for the divergent and rational terms. For the terms
proportional to $C_A$, we get
%
%
%
\bea
\label{eq:identifyhexNoabeCA}
\frac{-1}{2}\left( 
\sum_{n=1,3,6,10} {\cal M}^{1,(n)}_{V_1 V_2
    V_3g,\tau}
+\sum_{n=1-3,5,6,10} {\cal M}^{1,(n)}_{V_1
     V_2gV_3,\tau}
+\sum_{n=1-5,8} {\cal M}^{1,(n)}_{V_1
    g V_2V_3,\tau}
+\sum_{n=1,2,4,7}{\cal M}^{1,(n)}_{gV_1V_2V_3,\tau} \right)&& 
\nonumber \\
+\sum_{j=1}^{10}{\cal
  M}^{g,1,(j)}_{V_1V_2V_3,\tau}={\cal M}^B_{\tau}\left ( -\log\left (\frac{-s}{\mu^2}\right)+ \log\left (\frac{-t}{\mu^2}\right)+\log
   \left(\frac{-u}{\mu^2}\right)\right),\text{~~~~~~~~~~~~~~~~~~~~~~~~~~~~~~~~~~~}
 &&\nonumber \\
\frac{-1}{2}\left( 
\sum_{n=1,3,6,10} {\cal M}^{2,(n)}_{V_1 V_2
    V_3g,\tau}
+\sum_{n=1-3,5,6,10} {\cal M}^{2,(n)}_{V_1
     V_2gV_3,\tau}
+\sum_{n=1-5,8} {\cal M}^{2,(n)}_{V_1
    g V_2V_3,\tau}
+\sum_{n=1,2,4,7}{\cal M}^{2,(n)}_{gV_1V_2V_3,\tau} \right)&& 
\nonumber \\
+\sum_{j=1}^{10}{\cal
  M}^{g,2,(j)}_{V_1V_2V_3,\tau}=-{\cal
  M}^B_\tau,\qquad\qquad\qquad\quad
\text{~~~~~~~~~~~~~~~~~~~~~~~~~~~~~~~~~~~~~~~~~~~~~~~~~~~~~~~~~~~~~~~~~~}&&
\nonumber\\ 
\frac{-1}{2}\left( 
\sum_{n=1,3,6,10} \widetilde{{\cal N}}^{(n)}_{V_1 V_2
    V_3g,\tau}
+\sum_{n=1-3,5,6,10} \widetilde{{\cal N}}^{(n)}_{V_1
     V_2gV_3,\tau}
+\sum_{n=1-5,8} \widetilde{{\cal N}}^{(n)}_{V_1
    g V_2V_3,\tau}
+\sum_{n=1,2,4,7}\widetilde{{\cal N}}^{(n)}_{gV_1V_2V_3,\tau} \right)\qquad&& 
\nonumber \\
+\sum_{j=1}^{10}\widetilde{{\cal
  N}}^{g,(j)}_{V_1V_2V_3,\tau}=0, \hspace{13cm}
 &&\nonumber \\
\frac{-1}{2}\left( 
\sum_{n=1,3,6,10} {\cal N}^{1,(n)}_{V_1 V_2
    V_3g,\tau}
+\sum_{n=1-3,5,6,10} {\cal N}^{1,(n)}_{V_1
     V_2gV_3,\tau}
+\sum_{n=1-5,8} {\cal N}^{1,(n)}_{V_1
    g V_2V_3,\tau}
+\sum_{n=1,2,4,7}{\cal N}^{1,(n)}_{gV_1V_2V_3,\tau} \right)\qquad&& 
\nonumber \\
+\sum_{j=1}^{10}{\cal
  N}^{g,2,(j)}_{V_1V_2V_3,\tau}=0,\hspace{13cm} &&
\eea
%
and for the $C_F$ terms,
\bea
\label{eq:identifyhexNoabeCF}
\sum_{n=1}^{13} \left(
{\cal M}^{1,(n)}_{V_1V_2V_3 g,\tau} 
+{\cal M}^{1,(n)}_{V_1V_2gV_3,\tau}
+{\cal M}^{1,(n)}_{V_1gV_2V_3,\tau} 
+{\cal M}^{1,(n)}_{gV_1V_2V_3,\tau} \right)
={\cal M}^B_\tau\left(-3 + 2 \log \left(\frac{-s}{\mu^2}\right)\right),
&&\nonumber \\
\sum_{n=1}^{13} \left(
{\cal M}^{2,(n)}_{V_1V_2V_3 g,\tau} 
+{\cal M}^{2,(n)}_{V_1V_2gV_3,\tau}
+{\cal M}^{2,(n)}_{V_1gV_2V_3,\tau} 
+{\cal M}^{2,(n)}_{gV_1V_2V_3,\tau} \right)
=-2{\cal M}^B_\tau,
\text{~~~~~~~~~~~~~~~~~~~~}\text{~~~~~~}&&
\nonumber \\
\sum_{n=1}^{13} \left(
\widetilde{{\cal N}}^{(n)}_{V_1V_2V_3 g,\tau} 
+\widetilde{{\cal N}}^{(n)}_{V_1V_2gV_3,\tau}
+\widetilde{{\cal N}}^{(n)}_{V_1gV_2V_3,\tau} 
+\widetilde{{\cal N}}^{(n)}_{gV_1V_2V_3,\tau} \right)
=-{\cal M}^B_\tau, \hspace{4.75cm}&&
\nonumber \\
\sum_{n=1}^{13} \left(
{\cal N}^{1,(n)}_{V_1V_2V_3 g,\tau} 
+{\cal N}^{1,(n)}_{V_1V_2gV_3,\tau}
+{\cal N}^{1,(n)}_{V_1gV_2V_3,\tau} 
+{\cal N}^{1,(n)}_{gV_1V_2V_3,\tau} \right)
=0.\hspace{5.35cm}
&&
\eea
%
%
We have checked numerically with high precision in Mathematica the factorization formula for the 8 different kinematic configurations of the
hexlines once $k_4^2=0$ is fixed, i.e.,
$(k_1^2=0,k_2^2=0,k_3^2=0,k_4^2=0),~(k_1^2=M_1,k_2^2=0,k_3^2=0,k_4^2=0),~(k_1^2=M_1,k_2^2=M_2,k_3^2=0,k_4^2=0),\ldots$. The coefficients
multiplying the poles of Eq.~(\ref{eq:hexlineNoAbe}) are obtained with 30000
digits of precision at least~(at the FORTRAN level for non-singular points,
the proof works at the working precision level). The transversality property for the on-shell massless 
particles must be applied also in this case. For the penline and boxline contributions,
similar relations are found. In addition, for them, we have checked the
factorization formula analytically for all possible kinematic
configurations, once the on-shellness of the gluon, $p_g^2=0$, is imposed.

As in the pure abelian case, the factorization of the divergences, for
the sum of the ${\cal M}^{(i)}$ terms, against
the born amplitude at the numerical and analytical level already
provides a strong check of the correctness of the result since
$\widetilde{{\cal M}}$ and ${\cal M}^{(i)}$ have the same analytical
structure, Eq.~(\ref{D4terms}).
The cancellation of the renormalization scale variable,
$\mu$, in $\widetilde{\cal M}_V$, Eq.~(\ref{eq:identifyhexNoabef}),
provides an additional strong check~(see
Table~\ref{eq:ApprenhexlineNoAbeC_FC_A} for hexagons). Additionally, similar to the
pure abelian case,  we have implemented Ward identity tests for the
virtual corrections in FORTRAN and Mathematica  at different
levels of complexity:

\begin{enumerate}
\item[{\bf 1)}] At the level of single diagrams.
\item[{\bf 2)}] For the different topologies ${\cal 
M
}^{g^*}_{V_1,\tau}$, ${\cal 
M
}^{g^*}_{V_1V_2,\tau}$ and ${\cal 
M
}^{g^*}_{V_1V_2V_3,\tau}$.
\item[{\bf 3)}] At the level of gauge invariant amplitudes for one gluon emission.
\item[{\bf 4)}] Subset of amplitudes invariant for a specific replacement ($\epsilon(p_k)=\epsilon_k\to p_k$).
\item[{\bf 5)}] Specific contractions that make the contributions to vanish.
\end{enumerate}
%
%
%
%
\begin{figure}[htb]
\begin{center}
\includegraphics[scale=1]{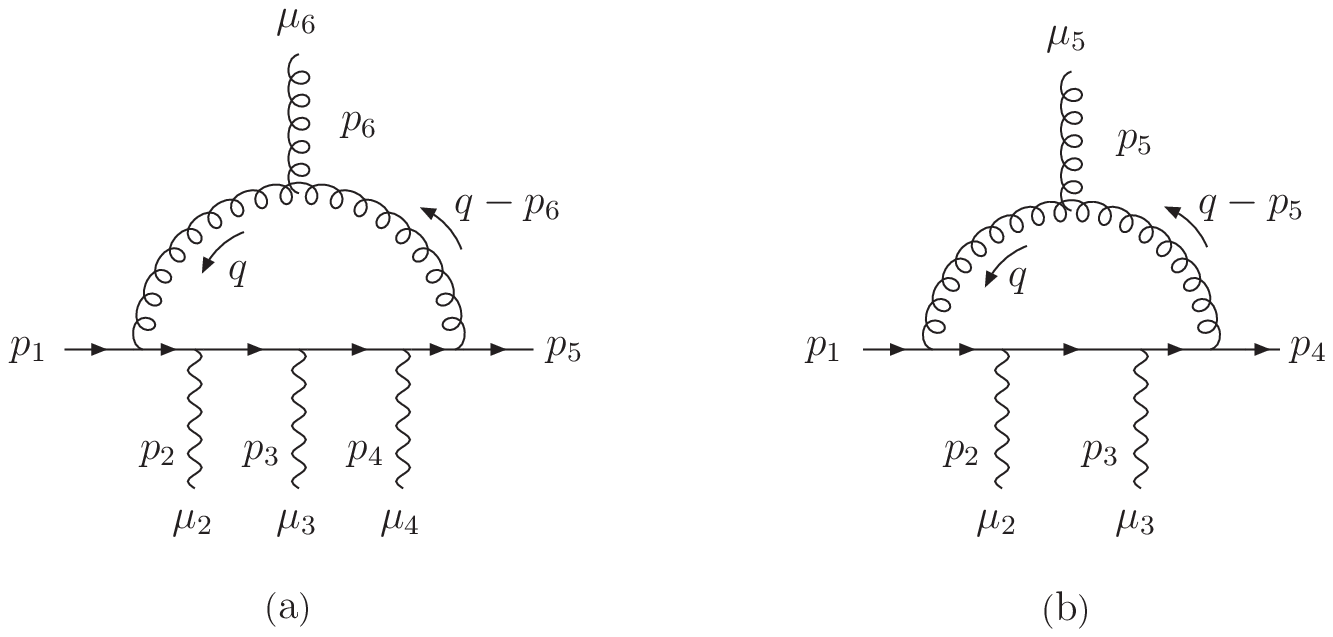}
\end{center}
\caption[]{\label{fig:gaugeNoAbe}
{\figfont ${\cal F}^{g^*}_{\mu_2\mu_3\mu_4\mu_6}(p_1,p_2,p_3,p_4,p_5,p_6)$ and 
${\cal E}^{g^*}_{\mu_2\mu_3\mu_5}(p_1,p_2,p_3,p_4,p_5) $ of Eq.~(\ref{eq:hexNoAbeg}) and
Eq.~(\ref{eq:hexgaugeNoAbe}).}  }
\end{figure}
%
%
%
%
%
%
%
%
%
%
%
%
%
%
{\bf 1)} and {\bf 2)}  under the replacement of the polarization vector by its
momentum ($\epsilon(p_k)=\epsilon_k\to p_k$) for the EW vector
bosons, relations among the different diagrams/contributions can be
obtained. Considering the hexagon of Figure~\ref{fig:gaugeNoAbe},
\bea
\label{eq:hexNoAbeg}
&&{\cal F}^{g^*}_{\mu_2\mu_3\mu_4\mu_6}(p_1,p_2,p_3,p_4,p_5,p_6)=\mbox{~~~~~~\hspace{9cm}~~~~~~} \\
&&~~~~=\int
\frac{d^dq}{(2\pi)^d}
\frac{1}{q^2}\frac{1}{(q-p_6)^2}
\gamma_\alpha%
\frac{1}{\slashed{q}+\slashed{p}_{14}}
\gamma_{\mu_4}\frac{1}{\slashed{q}+\slashed{p}_{13}}
\gamma_{\mu_3}\frac{1}{\slashed{q}+\slashed{p}_{12}}
\gamma_{\mu_2}\frac{1}{\slashed{q}+\slashed{p}_{1}}
\gamma_\beta V^{\alpha}_{\mu_{6}}(q-p_6,p_6,q),\nonumber
\eea
with $V^{\alpha}_{\mu_{6}}(q-p_6,p_6,q)$, the triple gluon vertex. Note
that we have kept the complete dependency of the momenta for clarity despite the fact that
we can use momentum conservation to eliminate one of them. Then, contracting one of the open indices by
the corresponding momentum and expressing the contracted gamma matrix as the difference of two adjacent fermionic propagators, we find
the following relations,
{\small
\bea
\label{eq:hexgaugeNoAbe}
p_2^{\mu_2}{\cal F}^{g^*}_{\mu_2\mu_3\mu_4\mu_6}(p_1,p_2,p_3,p_4,p_5,p_6)&=&
{\cal E}^{g^*}_{\mu_3\mu_4\mu_6}(p_1,p_2+p_3,p_4,p_5,p_6)
-{\cal E}^{g^*}_{\mu_3\mu_4\mu_6}(p_1+p_2,p_3,p_4,p_5,p_6) \nonumber \\
p_3^{\mu_3}{\cal F}^{g^*}_{\mu_2\mu_3\mu_4\mu_6}(p_1,p_2,p_3,p_4,p_5,p_6)&=&
{\cal E}^{g^*}_{\mu_2\mu_4\mu_6}(p_1,p_2,p_3+p_4,p_5,p_6)
-{\cal E}^{g^*}_{\mu_2\mu_4\mu_6}(p_1,p_2+p_3,p_4,p_5,p_6) \nonumber \\
p_4^{\mu_4}{\cal F}^{g^*}_{\mu_2\mu_3\mu_4\mu_6}(p_1,p_2,p_3,p_4,p_5,p_6)&=&
{\cal E}^{g^*}_{\mu_2\mu_3\mu_6}(p_1,p_2,p_3,p_4+p_5,p_6)
-{\cal E}^{g^*}_{\mu_2\mu_3\mu_6}(p_1,p_2,p_3+p_4,p_5,p_6), \nonumber \\
\eea
}\noindent
where ${\cal E}^{g^*}_{\mu_1\mu_2\mu_3}$ represents the pentagon diagram
of Figure~\ref{fig:gaugeNoAbe}, defined similarly as
Eq.~(\ref{eq:hexNoAbeg}).
For the contributions, $V_i \in (W^\pm,Z,\gamma)$ is considered such that
only one global color factor appears for all diagrams of
Figures~\ref{fig:boxNoAbe},~\ref{fig:penNoAbe} and
\ref{fig:hexNoAbe}. Then, factorizing out the polarization vectors
and couplings, 
\be
{\cal M}^{g^*}_{V_1 \ldots V_n,\tau}(p_1,\ldots,p_{n+2})=
g^{V_1f}_{\tau} \ldots  g^{V_nf}_{\tau} g_0\epsilon_{V_1}^{\mu_2}(p_2) \ldots \epsilon_{V_n}^{\mu_n}(p_n) \epsilon_{g}^{\mu_{n+2}}(p_{n+2})
{\cal M}^{g^*}_{\mu_2\ldots  \mu_n \mu_{n+2},\tau}(p_1,\ldots,p_{n+2}),
\ee
the same relation holds for
the  hexlineNoAbe and penlineNoAbe contributions under the replacements, 
\bea
&{\cal
  F}^{g^*}_{\mu_2\mu_3\mu_4\mu_6} \to  {\cal
  M}^{g^*}_{\mu_2\mu_3\mu_4\mu_6,\tau}, \nonumber \\
& {\cal E}^{g^*}_{\mu_i\mu_j\mu_l} \to {\cal M}^{g^*}_{\mu_i\mu_j\mu_l,\tau}.
\eea
Similar relations are obtained between the penlineNoAbe and boxlineNoAbe
contributions. Finally, the EW Ward Identity for the boxlineNoAbe routine returns
directly zero.

\begin{figure}[htb]
\begin{center}
\includegraphics[scale=0.85]{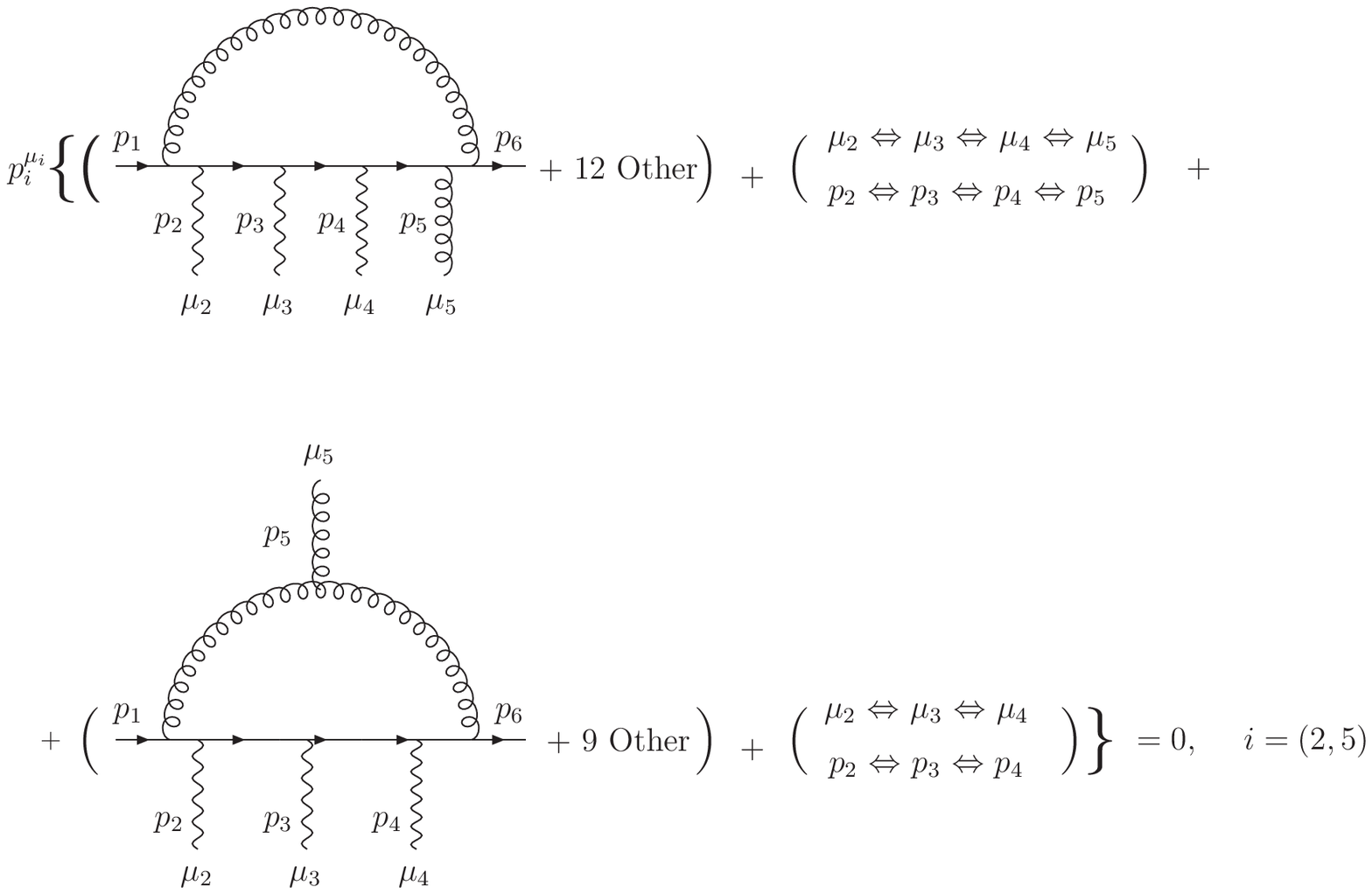}
\end{center}
\caption[]{\label{fig:crossNoAbehex}
{\figfont Direct and cross terms of Eq.~(\ref{hexline}) and
  Eq.~(\ref{hexNoAbe}) forming a gauge invariant quantity. 
}}
\end{figure}
%
%
%
%
%
%
%
\noindent {\bf 3)} To check gauge invariance for EW boson production
with an additional jet and due to the fact that
QCD mixes, for example, the
topologies of the Figures \ref{fig:hex} and \ref{fig:hexNoAbe}, we have to consider the
abelian type contributions of the previous section, including
the cross terms, along with the non-abelian type contributions~\footnote{Note that for processes with
  $W$ vector
  bosons not all the permutations are physically allowed.}~(see
Figure~\ref{fig:crossNoAbehex} for the convention of momenta and to illustration).
For the different topologies, it is satisfied that
{\small
\be
\label{gaugeNoAbe}
\begin{array}{lll}
p_i^{\mu_i}\left({\cal M}_{\mu_2\mu_3,\tau}(p_1,p_2,p_3) +
{\cal M}_{\mu_3\mu_2,\tau}(p_1,p_3,p_2) + {\cal M}^{g}_{\mu_2\mu_3,\tau}(p_1,p_2,p_4,p_3)\right) & =& 0\\ 
p_i^{\mu_i}\left({\cal
      M}_{\mu_2\mu_3\mu_4,\tau}(p_1,p_2,p_3,p_4)
+ \mbox{5 other} +{\cal M}^{g}_{\mu_2\mu_3\mu_4,\tau}(p_1,p_2,p_3,p_5,p_4)
+ \mbox{1 other}\right) &=& 0\\
p_i^{\mu_i}\left({\cal M}_{\mu_2\mu_3\mu_4\mu_5,\tau}(p_1,p_2,p_3,p_4,p_5)
+ \mbox{23 other}
 + {\cal M}^{g}_{\mu_3\mu_2\mu_4\mu_5,\tau}(p_1,p_3,p_2,p_4,p_6,p_5)+
  \mbox{5 other}  \right) &=& 0.\\
\end{array}
\ee
}\noindent
The abelian contributions with an emitted gluon give rise to two
different color structures, Fig.~\ref{fig:color}. The non-abelian type
contributions are proportional to C$_A$ and their contribution cancel in the gauge test Eq.~(\ref{gaugeNoAbe})
against the abelian type ones proportional to C$_A$. The C$_F$ terms
from the abelian contribution cancel against themselves as
discussed in the previous section, therefore, each one of the equalities
of Eq.~(\ref{gaugeNoAbe}) allow us to check on the one hand the abelian type
contributions and, on the other hand, the interplay between the abelian
and non-abelian contributions. 

\noindent {\bf 4) } For penlineNoAbe and hexlineNoAbe contributions,
there are as well ``gauge invariant'' subsets for an specific replacement. For example,
for the hexline and hexlineNoAbe contributions with gluon momentum, $p_5$, an invariant
subset under the replacement $\epsilon(p_5)=\epsilon_5\to p_5$ is obtained by permuting  the
position of the $p_5$ momentum and the corresponding Lorentz index and
keeping the relative order of the other vector bosons fixed, i.e.,
\bea
p_5^{\mu_5}\left({\cal M}_{\mu_2 \mu_3 \mu_4\mu_5,\tau}(p_1,p_2,p_3,p_4,p_5)+{\cal
  M}_{\mu_2 \mu_3 \mu_5\mu_4,\tau}(p_1,p_2,p_3,p_5,p_4) 
 \right. + \nonumber \\
\left. +{\cal M}_{\mu_2 \mu_5 \mu_3 \mu_4,\tau }(p_1,p_2,p_5,p_3,p_4)
+{\cal M}_{\mu_5 \mu_2 \mu_3 \mu_4,\tau }(p_1,p_5,p_2,p_3,p_4) + \right.\nonumber \\
\left.
+{\cal M}^g_{\mu_2 \mu_3 \mu_4\mu_5,\tau}(p_1,p_2,p_3,p_4,p_6,p_5) \right)=0.
\eea
Note also that the  divergences of this combination also factorize to the
born amplitude, Eq.~(\ref{fac:hexNoAbe}). \\

\noindent {\bf 5)} In addition, we have checked analytically and later
on numerically that there are some combinations of replacements for the 
polarization vectors which make the contributions to vanish.
\be
\begin{array}{lll}
\left(p_2^{\mu_2}\right) \,\, \left({\cal M}^g_{\mu_2\mu_4,\tau}(p_1,p_2,p_3,p_4)\right) & =& 0,\\ 
\left(p_2^{\mu_2} (p_2+p_3)^{\mu_3} || (p_2+p_3)^{\mu_2} p_3^{\mu_3}
\right) \,\, \left({\cal M}^g_{\mu_2\mu_3\mu_5,\tau}(p_1,p_2,p_3,p_4,p_5)\right) &=& 0,\\
p_2^{\mu_2} (p_2+p_3+p_4)^{\mu_3} p_4^{\mu_4}  \left({\cal
    M}^g_{\mu_2\mu_3\mu_4\mu_6,\tau}(p_1,p_2,p_3,p_4,p_5,p_6)\right) &=& 0.\\
\end{array}
\ee
Note that the first line represents the EW Ward identity for the
boxlineNoAbe, equivalent to Eq.~(\ref{eq:hexgaugeNoAbe}), which is zero
as mentioned previously.

\section{Numerical Instabilities and Timing}
\label{sec:ins}
To study the stability and timing of the contributions computed here, we
have used 5 $\cdot$ 10$^5$ cut-accepted points for the vector boson
fusion process  $qq
\to q q W^+ W^-$ at LO generated with {\it VBFNLO}~\cite{Arnold:2008rz}
applying the cuts, 
\be
\label{cut1}
p_{Tj}\ge 20\, \mbox{~Gev},~ |y_j| \le 4.5,~ \Delta y_{jj} > 4,
\ee
$p_{Tj}$ and  $|y_j|$ are the transverse momentum and rapidity
of the jets, respectively. $\Delta y_{jj}$ is the tagging jet
rapidity separation. Furthermore, we require the two tagging jets to
lay in opposite detector hemispheres,
\be
\label{cut2}
y_{j1} \times y_{j2} <0,
\ee
with an invariant mass of $M_{jj}> 600\mbox{GeV}$.
For the hexline(NoAbe) contribution, we assign the momenta of the two outgoing jets to two of the vector
bosons emitted from the quark line in Fig.~\ref{fig:hex}~(Fig.~\ref{fig:hexNoAbe}).
The corresponding polarization vectors are constructed
following the conventions of Ref.~\cite{Murayama:1992gi}~(Eq.~(A.11) of that paper). 
For the boxline(NoAbe) and penline(NoAbe) contributions, the momenta
of the external particles are combined to obtain $2 \to 3 $ and $2 \to 2$ kinematics. 
Different combinations and/or permutations yield similar results, we show here the more
unstable observed. 

The contributions are written automatically into FORTRAN modular routines in which some
flags are incorporated to take advantage of the structure of the result.
Note that the information of the spinor helicity is contained in the
matrix elements, SM$_{i,\tau}$~(Eq.~(\ref{eq:helmethod})), which are 
computed following the helicity method. %
Thus, calling the routines with a different spinor helicity only
requires to recalculate the reduced set of standard matrix elements
SM$_{i,\tau}$, which represents less than 1 $\%$ of
the total CPU time for the hexline(NoAbe) contributions. Additionally,
for changes on the vector bosons due to different helicities or to the
implementation of
gauge tests ($\epsilon(p_n) \to p_n $), only the
F1$_j$ and SM$_{i,\tau}$ of Eq.~(\ref{eq:helmethod}) should be
recalculated, but not the F$_k$ from Eq.~(\ref{eq:F1}).  
\begin{table}
\begin{center}
\begin{tabular}
{|l|c|c|c|}
\hline
Functions & ~~~~~~~~SM's~~~~~~~~ & ~~~~~~~~F1's~~~~~~~~ & ~~~~~~~~F's~~~~~~~~\\
\hline
Recalculate for&&&\\
different spinor&Yes&No&No\\
helicity&&& \\
\hline
Recalculate for &&& \\
different polarization&Yes&Yes &No\\
vector ($\epsilon(p_i)\to p_i,$ &&& \\
$\epsilon_+(p_i)\to \epsilon_-(p_i) $ &&&\\
\hline
\% CPU/Total &$\approx$ 1\%&$\approx$ 30\% &$\approx$ 70\%\\
\hline
\end{tabular}
\end{center}
\caption{CPU time to revaluate different parts of the code under
 changes of spinor helicities and polarization vectors for the hexline contribution.}
\label{reeva1}
\end{table}
%
%
In Table~\ref{reeva1}, as an example, the summary of the CPU time spent to
revaluate the different parts of the code, as well as the part of the code
that has to be revaluated under changes of spinor helicities and
polarization vectors, are shown for the hexline contribution.

The organization of the code in dependent/independent loop integral
parts is of advantage since additional callings of
the routines for making Ward identity gauge tests are obtained at a lower CPU time
cost.
To control the stability of our routines for the finite terms of each of
the contributions (boxline(NoAbe), penline(NoAbe), hexline(NoAbe)), for every
phase space point, all the possible Ward identity gauge tests
($\epsilon(p_j)\to p_j$) of
Eqs.~(\ref{eq:hexgauge},\ref{eq:hexgaugeNoAbe}) are performed. Except
for the boxlineNoAbe, which vanishes for
the EW  Ward identity gauge test of Eq.~(\ref{eq:hexgaugeNoAbe}), and
Eq.~(\ref{gaugeNoAbe}) is applied with the replacement $\epsilon(p_g)\to p_g$. These equations can be
rewritten in the more convenient form ``abs(a/b)-1=0'' such that we obtain normalized
results which allow us to test the accuracy of our numerical zeros,
subject that there is not any numerical cancellation in the determination
of the ``a'' and
``b'' quantity. In that case, we use ``a-b=0''.  In
practice, we take the worse normalized
value of the Ward identity gauge test  and compare it with a minimum accepted accuracy, i.e., abs(a/b)-1 $<$ ``accuracy''.
We define as unstable point the one that does not satisfy the Ward
identity gauge test 
at a given accuracy. The factorization proof of the
divergences to control the stability is not used since they need different input
integrals~(Eq.(\ref{eq.b0})) and the complete routine has to be
revaluated. The use of the Ward identities gauge test for the finite
contributions has the additional advantage that the
same integrals and tensor coefficients, stored in the $F_j$
functions~(Eq.(\ref{eq:F1})), needed to provide the finite result,
${\cal M}_v$, are used. Moreover,
since the rational terms and the poles will factorize against the born
amplitude, we  will not revaluate the
routines for the poles (Eqs.~(\ref{D4terms},\ref{rational})), but,
instead, in practical implementations, we will use the analytical form which only requires to evaluate
the born amplitude. 

The fraction of unstable points for the different
contributions depending on the 
accuracy of the Ward identity gauge tests can be seen in Tables~\ref{Accuracy1} and \ref{Accuracy2}. %
\begin{table}[h!]
\begin{center}
\begin{tabular}{|l|c|c|c|c|c|c|}
\hline
Test Accuracy& 10$^{-1}$ &10$^{-2}$ & 10$^{-3}$ & 10$^{-4}$  & 10$^{-5}$
& 10$^{-6}$ \\\hline
Failed points for Boxline &  0.002$\permil$ &0.008$\permil$
& 0.01 $\permil$ &  0.05$\permil $   & 0.1 $\permil$ &0.4$\permil$    \\\hline
Failed points for Penline & 0.1$\permil$ & 0.3$\permil$ &0.8$\permil$
& 2 $\permil$ & 0.9 $\% $   & 3.7 $\%$   \\\hline
Failed points for Hexline  &  2$\permil$ &  5$\permil$ & 1.1$\%$
& 2.8 $\%$ & 7.6 $\% $   & 18.1 $\%$   \\ \hline
\end{tabular}
\end{center}
\caption{Fraction of unstable points, out of the sample of $5\cdot 10^5$
  events, depending on the accuracy of the
  gauge test for the abelian contributions Eq.~(\ref{boxline}), 
Eq.~(\ref{penline}) and Eq.~(\ref{hexline}).}
\label{Accuracy1}
\end{table}
%
%
\begin{table}[h!]
\begin{center}
\begin{tabular}{|l|c|c|c|c|c|c|}
\hline
Test Accuracy& 10$^{-1}$ &10$^{-2}$ & 10$^{-3}$ & 10$^{-4}$  & 10$^{-5}$
& 10$^{-6}$ \\\hline
Failed points for BoxlineNoAbe & 0.008$\permil$ & 0.03$\permil$ &0.07$\permil$
& 0.2 $\permil$ & 0.6 $\permil $   & 1.8 $\permil$   \\\hline
Failed points for PenlineNoAbe & none & 0.004$\permil$ &0.01$\permil$
& 0.3 $\permil$ & 4.6 $\permil $   & 3.6 $\%$   \\\hline
Failed points for HexlineNoAbe  &  0.5$\permil$ & 1.2$\permil$ & 4.1$\permil$
& 1.4 $\%$ & 4.4 $\% $   & 12$\% $   \\ \hline
\end{tabular}
\end{center}
\caption{Fraction of unstable points, out of the sample of $5\cdot 10^5$
  events, depending on the accuracy of the
  gauge test for the non-abelian contributions Eq.~(\ref{boxNoAbe}), Eq.~(\ref{penNoAbe})
 and Eq.~(\ref{hexNoAbe}).}
\label{Accuracy2}
\end{table}
%
For the determination of the tensor coefficients, we have applied the
Passarino-Veltman tensor decomposition method following Eqs.~(\ref{masPV}) and (\ref{masPV00}) up to the box level, and the Denner-Dittmaier method for
pentagons and hexagons applying Eqs.~(\ref{Emas}) and (\ref{E00mas}). 
To deal with these unstable points, some knowledge about the
origin of these instabilities is required. It is known that small Gram
determinants appearing in $C$ and $D$ functions and small Cayley determinants
for $E$ and $F$ functions result in a loss of
precision in the determination of the  tensor coefficient integrals, therefore,
in the amplitudes. To solve this problem within double precision
accuracy, we try to improve the determination of the tensor
coefficients up to the box level. First, using a fast implementation of
the LU decomposition method which avoids the explicit calculation of
inverse Gram determinants of Eq.~(\ref{masPV}) by solving numerically
a system of linear equations, Eq.~(\ref{masPVLU}). It turns out that this procedure is more
stable close to singular regions and reduces considerably the number of identified
instabilities without CPU penalty, Tabs.~\ref{LUAbe}
and~\ref{LUNoAbe}. 
\begin{table}[t!]
\begin{center}
\begin{tabular}{|l|c|c|c|c|c|c|}
\hline
Test Accuracy& 10$^{-1}$ &10$^{-2}$ & 10$^{-3}$ & 10$^{-4}$  & 10$^{-5}$
& 10$^{-6}$ \\\hline
\multicolumn{7}{|c|}{Boxline}\\
\hline
Failed points with Dble  & 0.002$\permil$ &  0.008$\permil$ & 0.01$\permil$
& 0.05 $\permil$ & 0.1 $\permil $   & 0.4 $\permil$   \\ \hline
Failed pts with LU  &  none & none & none
& none & none   &  none   \\ \hline
Failed points with LU and Gram &  none & none & none
& none & none   &  none   \\ \hline
\multicolumn{7}{|c|}{Penline}\\
\hline
Failed points with Dble  & 0.1$\permil$ &  0.3$\permil$ & 0.8$\permil$
& 2 $\permil$ & 0.9 $\% $   & 3.7$\%$   \\ \hline
Failed pts with LU  &  0.002$\permil$ & 0.008 $\permil$ & 0.04$\permil$
& 0.09 $\permil$ & 0.3 $\permil $   &  1 $\permil$   \\ \hline
Failed points with LU and Gram  &  none & none & none
& 0.002 $\permil$ & 0.07 $\permil $   &  0.2$\permil$   \\ \hline
\multicolumn{7}{|c|}{Hexline}\\
\hline
Failed points with Dble  & 2$\permil$ &  5$\permil$ & 1.1$\%$
& 2.8 $\%$ & 7.6 $\% $   & 18.1 $\%$   \\ \hline
Failed pts with LU  &  0.1$\permil$ & 0.3 $\permil$ & 1$\permil$
&  5 $\permil$ & 1.7 $\% $   &  5.5$\%$   \\ \hline
Failed points with LU and Gram  &  0.08$\permil$ & 0.3 $\permil$ & 1$\permil$
& 4 $\permil$ & 1.4 $\%$   &  4.8$\%$   \\ \hline
\end{tabular}
\end{center}
\caption{Fraction of unstable points, out of the sample of $5\cdot 10^5$
  events, depending on the accuracy of the
  gauge test for the abelian contributions.}
\label{LUAbe}
\end{table}
%
\begin{table}[h!]
\begin{center}
\begin{tabular}{|l|c|c|c|c|c|c|}
\hline
Test Accuracy & 10$^{-1}$ &10$^{-2}$ & 10$^{-3}$ & 10$^{-4}$  & 10$^{-5}$
& 10$^{-6}$ \\\hline
\multicolumn{7}{|c|}{BoxlineNoAbe}\\
\hline
Failed points with Dble  & 0.008$\permil$ &  0.03$\permil$ & 0.07$\permil$
& 0.2 $\permil$ & 0.6 $\permil $   & 1.8 $\permil$   \\ \hline
Failed pts with LU  &  none & 0.01 $\permil$ & 0.04$\permil$
& 0.1 $\permil$ & 0.3 $\permil $   &  1$\permil$   \\ \hline
Failed points with LU and Gram  &  none & 0.008 $\permil$ & 0.03$\permil$
& 0.08 $\permil$ & 0.3 $\permil $   &  0.8$\permil$   \\ \hline
\multicolumn{7}{|c|}{PenlineNoAbe}\\
\hline
Failed points with Dble  & none &  0.004$\permil$ & 0.01$\permil$
& 0.3 $\permil$ & 4.6 $\% $   & 3.6$\%$   \\ \hline
Failed pts with LU  &  none & 0.002 $\permil$ & 0.006$\permil$
& 0.1 $\permil$ & 1.3 $\permil $   &  1.2$\%$   \\ \hline
Failed points with LU and Gram  &  none & none & 0.004$\permil$
& 0.08 $\permil$ & 0.7 $\permil $   &  7$\permil$   \\ \hline
\multicolumn{7}{|c|}{HexlineNoAbe}\\
\hline
Failed points with Dble  &  0.5$\permil$ & 1.2$\permil$ & 4.1$\permil$
& 1.4 $\%$ & 4.4 $\% $   & 12$\% $   \\ \hline
Failed pts with LU &  0.03 $\permil$ & 0.08$\permil$ & 0.3$\permil$
& 1.6 $\permil$ & 8 $\permil $   &  3.2 $\% $   \\ \hline
Failed pts with LU and Gram&  0.02 $\permil$ & 0.04$\permil$ & 0.2$\permil$
& 0.9 $\permil$ & 4 $\permil $   &  1.9 $\% $   \\ \hline
\end{tabular}
\end{center}
\caption{Fraction of unstable points, out of the sample of $5\cdot 10^5$
  events, depending on the accuracy of the
  gauge test for the non-abelian contributions.}
\label{LUNoAbe}
\end{table}%
In addition, we apply also special tensor reduction routines
for small Gram determinants in analogy to Ref.~\cite{Denner:2005nn} using
Eq.~(\ref{massGram}). These routines are switched on whenever
a cancellation in the Gram determinant defined by,
\be
\Delta= \frac{2 \sum_{i_1\ldots i_n=1}^n \epsilon_{i_1\ldots i_n} p_1\cdot p_{i_1}\ldots
  p_n\cdot p_{i_n}}{2 \sum_{i_1\ldots i_n=1}^n |\epsilon_{i_1\ldots i_n} p_1\cdot p_{i_1}\ldots p_n\cdot p_{i_n} |},
\ee  
is larger than a given cut
off.  The milder the cancellation the larger the number of
terms in the expansion that has to be included to provide a good accuracy.
For $C_{ij}$~($D_{ij}$) functions, we include tensor integrals up to rank 9~(7) 
which guarantees approximately an error of ${\cal O}( \Delta^7)$ (${\cal O}(\Delta^4)$) for
the $C_{ij}$~($D_{ij}$) tensor
coefficients of rank 2~(rank 3)~\footnote{For $D$ functions,
  cancellations in more that one sub-determinant can appear. Nevertheless,
  the estimate seems to work.}. Such that, e.g. the cancellation of two
digits in the Gram determinant typically results in a precision of 14
(8) digits for the $C_{ij}~(D_{ij})$ tensor coefficients of rank 2~(rank
3). Cancellations of two to three digits, $\Delta=10^{-2,-3}$, in the Gram determinant of the
$D$ functions represent the borderline, in our present set up, for which
the special tensor reduction routines for $D_{ij}$ functions provide better results than the
ones obtained with the LU decomposition. Practically, below the cut off, we
compare the precision provided by the two methods using sum rules and
apply the most precise one.
For up to penline(NoAbe), the combined
method works well leaving the remaining instabilities below the per
mill level for an accuracy of the gauge test of $10^{-4}$. Note, however, that for the boxlineNoAbe and
the penlineNoAbe routines the behavior is worse than for the abelian
contributions. In these cases, cancellations in the calculation of Cayley determinants
in $C$ and/or $D$ functions also take place and other special routines have to be
applied, Ref.~\cite{Denner:2005nn}. 

For the hexagons, the remaining
instabilities are still considerable and, generically, cancellations in Cayley
determinants of the $E$ and $F$ functions are also present. 
At this stage, the simplest and fastest way of rescuing the remaining
instabilities is calling the contributions  with 
quadruple~(QUAD) precision. 
However, quadruple precision 
is 20 times slower than 
double~(Dble) precision, thus, revaluating the identified 
unstable points, 5$\%$ of the points for the hexline contribution (the most
time consuming and most unstable object) for an accuracy of the Ward
identity gauge test of $10^{-6}$,
results in an addition of 100$\%$ CPU time. 
A better approach which reduces the slowing factor of QUAD precision
consists in  applying QUAD precision to compute the input 
scalar and the tensor coefficient integrals. 
\begin{table}[t!]
\begin{center}
\begin{tabular}{|l|c|c|c|c|c|c|}
\hline
Test Accuracy& 10$^{-1}$ &10$^{-2}$ & 10$^{-3}$ & 10$^{-4}$  & 10$^{-5}$
& 10$^{-6}$ \\\hline
\multicolumn{7}{|c|}{Boxline}\\
\hline
Failed points with LU & none &  none & none
& none & none   & none   \\ \hline
Failed pts with QUAD  &  none & none & none
& none & none   &  none   \\ \hline
Failed points with Full QUAD  &  none & none & none
& none & none   &  none   \\ \hline
Failed points with LU and Gram & none &  none & none
& none & none   & none   \\ \hline
Failed pts with QUAD  &  none & none & none
& none & none   &  none   \\ \hline
Failed points with Full QUAD  &  none & none & none
& none & none   &  none   \\ \hline
\multicolumn{7}{|c|}{Penline}\\
\hline
Failed points with LU &  0.002$\permil$ & 0.008$\permil$
& 0.04 $\permil$ & 0.09$\permil $   & 0.3$\permil$  & 0.1 $\%$ \\ \hline
Failed pts with QUAD  &  none & none & none
& none & none   &  0.002$\permil$   \\ \hline
Failed pts with full QUAD  &  none & none & none
& none & none   &  0.002$\permil$   \\ \hline
Failed points with LU and Gram & none &  none & none
& 0.002 $\permil$ & 0.07$\permil $   & 0.2$\permil$   \\ \hline
Failed pts with QUAD  &  none & none & none
& none & none   &  none   \\ \hline
Failed points with Full QUAD  &  none & none & none
& none & none   &  none   \\ \hline
\multicolumn{7}{|c|}{Hexline}\\
\hline
Failed points with LU   & 0.1$\permil$ &  0.3$\permil$ & 1$\permil$
& 5 $\permil$ & 1.6 $\% $   & 5.5 $\%$   \\ \hline
Failed pts with QUAD  &  none & none & none
& none & 0.02 $\permil $   &  0.19$\permil$   \\ \hline
Failed points with Full QUAD  &  none & none & none
& none & 0.004 $\permil $   &  0.07$\permil$   \\ \hline
Failed points with LU and Gram  & 0.08$\permil$ &  0.3$\permil$ & 1$\permil$
& 4 $\permil$ & 1.4 $\% $   & 4.8 $\%$   \\ \hline
Failed pts with QUAD  &  none & none & none
& none & 0.018 $\permil $   &  0.15$\permil$   \\ \hline
Failed points with Full QUAD  &  none & none & none
& none & 0.002 $\permil $   &  0.03$\permil$   \\ \hline
\end{tabular}
\end{center}
\caption{Fraction of unstable points, out of the sample of $5\cdot 10^5$
  events, depending on the accuracy of the
  gauge test for Dble and QUAD precision for the abelian contributions.}
\label{Accuracy3}
\end{table}
%
The tensor
decomposition routines and scalar integrals amount a fraction of the
total CPU time of the hexline(NoAbe) routines and evaluating
these in QUAD precision 
results in routines only a factor 4
slower than the original one.
Therefore, for instabilities at the 5$\%$
level, only an additional 20$\%$ CPU time is added, in contrast to the 100$\%$ when applying quadruple precision 
to the whole routine.
%
After this procedure, the instabilities are reduced for an accuracy of 10$^{-6}$ approximately a
 300 factor for the hexagon contributions, leaving the instabilities at the 
$\approx 0.15 \permil$  level, confirming that the lost of accuracy is at the
level of the tensor integral coefficient determination due to
cancellations in the Gram and/or Cayley determinants. 
%
\begin{table}[h!]
\begin{center}
\begin{tabular}{|l|c|c|c|c|c|c|}
\hline
Test Accuracy& 10$^{-1}$ &10$^{-2}$ & 10$^{-3}$ & 10$^{-4}$  & 10$^{-5}$
& 10$^{-6}$ \\\hline
\multicolumn{7}{|c|}{BoxlineNoAbe}\\
\hline
Failed points with LU  & none &  0.01$\permil$ & 0.04$\permil$
& 0.1 $\permil$ & 0.3 $\permil $   & 1 $\permil$   \\ \hline
Failed pts with QUAD  &  none & none & none
& none & 0.002 $\permil $   &  0.01$\permil$   \\ \hline
Failed pts with Full QUAD  &  none & none & none
& none & none   &  0.002$\permil$   \\ \hline
Failed points with LU and Gram  & none &  0.008$\permil$ & 0.03$\permil$
& 0.08 $\permil$ & 0.3 $\permil $   & 0.8 $\permil$   \\ \hline
Failed pts with QUAD  &  none & none & none
& none & none   &  0.008$\permil$   \\ \hline
Failed points with Full QUAD  &  none & none & none
& none & none   &  none   \\ \hline
\multicolumn{7}{|c|}{PenlineNoAbe}\\
\hline
Failed points with LU & none &  0.002$\permil$ & 0.006$\permil$
& 0.1 $\permil$ & 1 $\permil $   & 1.2$\%$   \\ \hline
Failed pts with QUAD  &  none & none & none
& none & none   &  none   \\ \hline
 Failed points with Full QUAD &  none & none & none
& none & none   &  0.002$\permil$   \\ \hline
Failed points with LU and Gram & none &  none & 0.004$\permil$
& 0.08 $\permil$ & 0.7 $\permil $   & 0.76$\%$   \\ \hline
Failed pts with QUAD  &  none & none & none
& none & none   &  none   \\ \hline
Failed points with Full QUAD  &  none & none & none
& none & none   &  none   \\ \hline
\multicolumn{7}{|c|}{HexlineNoAbe}\\
\hline
Failed points with LU  &  0.03$\permil$ & 0.08$\permil$ & 0.3$\permil$
& 1.6 $\permil$ & 0.8$\% $   & 3.3$\% $   \\ \hline
Failed pts with QUAD&  none & none & none
& none & 0.05 $\permil $   &  0.4 $\permil $   \\ \hline
Failed pts with Full QUAD &  none & none & none
& none & 0.006 $\permil $   &  0.008 $\permil $   \\ \hline
Failed points with LU and Gram  &  0.02$\permil$ & 0.04$\permil$ & 0.2$\permil$
& 0.9 $\permil$ & 0.4$\% $   & 1.9$\% $   \\ \hline
Failed pts with QUAD&  none & none & none
& none & 0.002 $\permil $   &  0.15 $\permil $   \\ \hline
Failed pts with Full QUAD&  none & none & none
& none & none   &  0.002 $\permil $   \\ \hline
\end{tabular}
\end{center}
\caption{Fraction of unstable points, out of the sample of $5\cdot 10^5$
  events, depending on the accuracy of the
  gauge test for Dble and QUAD precision for the non-abelian contributions.}
\label{Accuracy4}
\end{table}
%
For this reduced set of points, $ \approx 0.15 \permil $, Tab.~\ref{Accuracy3}, we can call the
routines with full QUAD precision with only an additional 5 $\permil $
CPU time, reducing the instabilities by an order of magnitude.
In
Tables~\ref{Accuracy3} and \ref{Accuracy4}, the fraction of unstable
points after these steps depending on the
accuracy demanded for the Ward identity gauge test are shown for the abelian and the
non-abelian contributions, respectively, for two setups, with and without
previously applying the special routines for small Gram determinants.
The non-existence of instabilities for an accuracy of 
the Ward identity gauge test of
$10^{-4}$, both for the hexline and for the hexlineNoAbe routines, without
using the special routines for small Gram determinants in
double precision, suggests that an increase of
precision would be enough to rescue these points, and the Landau
singularities are not present for cut accepted points or they are the
sub-million level for the kinematics of EW $pp\to W^+W^-jj+X$ production and given
the cuts of Eqs.~(\ref{cut1},\ref{cut2}). Once, we are forced to use QUAD precision for the
hexagons, the use of the special routines does not bring too much of an
improvement concerning the final number of instabilities but it
introduces a delaying factor. To reduce this factor, we only switch on
the special routines whenever the cancellation of the Gram determinant
are severe, larger than 3 digits. This procedure is used to compute the timing
of the routines in Tables~\ref{timing} and \ref{timing2}.
%
%
\begin{table}[h!]
\begin{center}
\begin{tabular}{|l|c|c|c|}
\hline
& boxline & penline & hexline  \\\hline
CPU time  & 8$\mu$ s & 70$\mu$ s & 2 ms \\ \hline
additional CPU time for Gauge Test& 8$\%$ & 30$\%$ & 28$\%$ \\ \hline
n$^o$ Gauge Test &   2 &   3  &  4  \\\hline
Total additional CPU time for Gauge Test& 16$\%$ & 90 $\%$ & 112 $\%$ \\
\hline
additional CPU time for spinor helicity& 12$\%$ & 5$\%$ & 1$\%$ \\ \hline
n$^o$ Helicity(Worse Case) &   2 &   2  &  2  \\\hline
Total additional CPU time for Helicity& 24$\%$ & 10$\%$ & 2$\%$ \\ \hline
Average failed Gauge test(Accuracy 10$^{-6}$)& none & 0.3$\permil$ & 5$\%$
\\ \hline
Average additional CPU time for bad point & 
\multirow{2}{*}{0} &
\multirow{2}{*}{4.6$\permil$} &
\multirow{2}{*}{20$\%$} 
\\ 
~~~~~~~~~~QUAD precision &&&\\ \hline
Additional CPU time for revaluate gauge tests & 0 & 4.1$\permil$ &
5.6$\%$ \\ \hline
Total CPU time (See test)  & 11.2$\mu$ s & 141$\mu$ s & 4.8(3) ms \\ \hline
Final Instabilities& none & none & 0.07$\permil$
\\ \hline
\end{tabular}
\end{center}
\caption{Average CPU time per point for the abelian contributions
  including additional time spent performing gauge tests and rescuing unstable points.}
\label{timing}
\end{table}
In these Tables, one finds the average CPU time
for the evaluation of the abelian and
 non-abelian type contributions, respectively, including a detailed
description of how this time is distributed for the calculation of
the Ward identity gauge tests, different helicities and the revaluation of unstable points 
with QUAD precision using a single core of an Intel(R) Core(TM)2 Quad CPU Q6600 @ 2.40GHz
computer with the Intel FORTRAN compiler. 
\begin{table}[t!]
\begin{center}
\begin{tabular}{|l|c|c|c|}
\hline
&{\small boxlineNoAbe} & {\small penlineNoAbe} & {\small hexlineNoAbe}  \\\hline
CPU time  & 6$\mu$ s & 60$\mu$ s & 2 ms \\ \hline
additional CPU time for Gauge Test& 10$\%$ & 20$\%$ & 40$\%$ \\ \hline
n$^o$ Gauge Test &   1 &   2  &  3  \\\hline
Total additional CPU time for Gauge Test& 10$\%$ & 40$\%$ & 120$\%$ \\
\hline
additional CPU time for spinor helicity& 14$\%$ & 8.5$\%$ & 1$\%$ \\ \hline
n$^o$ Helicity(Worse Case) &   2 &   2 &  2  \\\hline
Total additional CPU time for Helicity& 28$\%$ & 19$\%$ & 2$\%$ \\ \hline
Average failed Gauge test(Accuracy 10$^{-6}$)& 0.1$\permil$ & 1.2$\%$ & 3$\%$
\\ \hline
~~~~Average additional CPU time for bad &
\multirow{2}{*}{1.5$\permil$}& 
\multirow{2}{*}{27$\%$} &
\multirow{2}{*}{12$\%$} \\ 
~~~~~~~~~~~~~~~~point QUAD precision& && \\ \hline
~~~~Additional CPU time for revaluate  &
\multirow{2}{*}{0.01$\permil$} & 
\multirow{2}{*}{0.5$\%$} &
\multirow{2}{*}{3.6$\%$} \\
~~~~~~~~~~~~~~~~~~~~gauge tests &&&
\\ \hline
Total CPU time (See test)  & 8$\mu$ s & 111$\mu$ s & 4.8(3) ms \\ \hline
Final Instabilities& 0.002$\permil$ & 0.002$\permil$ & 0.02$\permil$
\\ \hline
\end{tabular}
\end{center}
\caption{Average CPU time per point for the non-abelian contributions
  including additional time spent performing gauge tests and rescuing unstable points.}
\label{timing2}
\end{table}
We observe that most of the time of the hexline(NoAbe) routine is spent in
applying the Ward identity gauge tests despite the fact that not all of
the routine should be revaluated, Tab.~\ref{reeva1}. We can reduce this slowing factor without
affecting statistically the identified number of instabilities given in
Tables~\ref{Accuracy3} and \ref{Accuracy4} by applying instead of the
four~(three) gauge tests for the hexline(NoAbe) contributions
only one selected randomly. The total CPU time in parenthesis in
Tables~\ref{timing} and \ref{timing2} is computed applying this procedure. Moreover, the
time shown is referred to the calculation of the finite pieces
$\widetilde{{\cal M}}_v$ since we will assume that the divergent pieces and
$\widetilde{{\cal N}}_v$ for
complete processes are known analytically, such that, we do
not have to revaluate $\widetilde{{\cal M}}_v$ and  $\widetilde{{\cal
    N}}_v$  for the divergent part
of the input functions.

\begin{center}
\begin{figure}[h!]
\hspace*{-1cm}
\includegraphics[scale=1.17,angle=0]{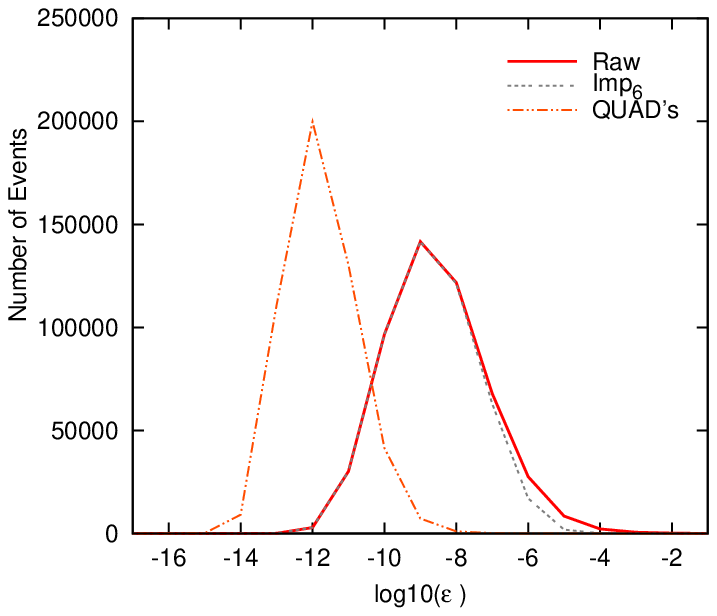}
\includegraphics[scale=1.17,angle=0]{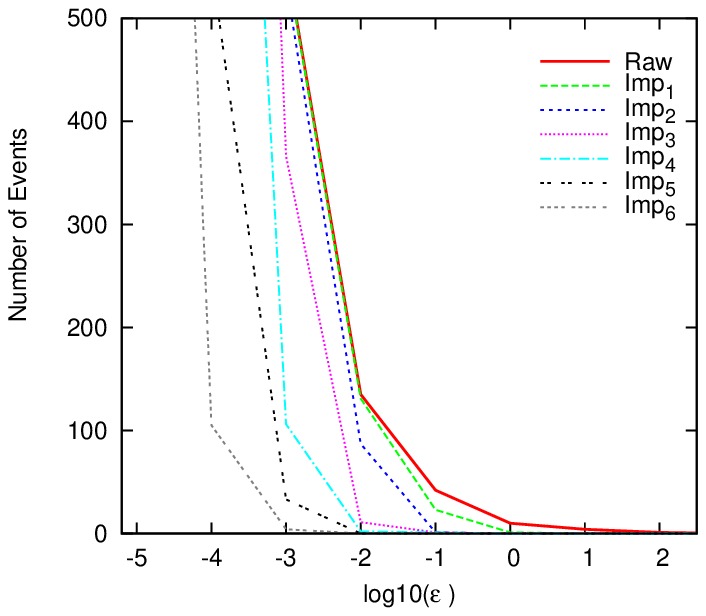}\\
\hspace*{-1cm}
\includegraphics[scale=1.17,angle=0]{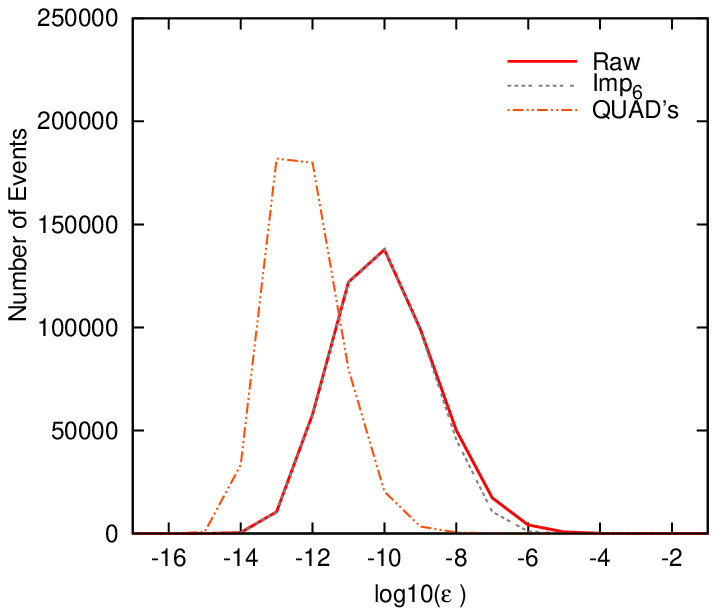}
\includegraphics[scale=1.17,angle=0]{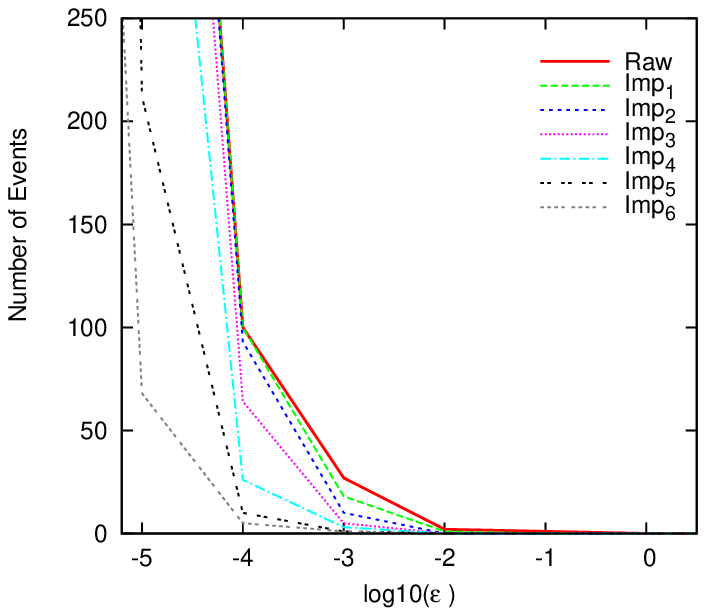}
\caption{The normalized difference between Double precision and Quadruple
  precision for the hexline~(top) and the hexlineNoAbe~(bottom)
  contributions is represented by the ``Raw'' line. ``QUAD's''
  represents the difference between the amplitudes evaluated with 
  QUAD precision and with QUAD only applied to the basis of scalar and tensor integral routines. ``Imp$_X$'' represents the effect of the rescue system
  activated for an accuracy of the Ward identity gauge test of $10^{-X}$. The right
  panels describe the critical region where double precision is not
  accurate and the effect of the rescue system for different values of
  the Ward identity gauge test.}
\label{Difference2}
\end{figure}
\end{center}

In Fig.~\ref{Difference2},  the relative
accuracy 
of the double precision
result, $\epsilon_0$, defined as the absolute value of the difference
between the double and quadruple precision divided by the quadruple
precision result is plotted. We also plot represented by the ``QUAD's'' label, the difference between the amplitudes evaluated with 
  full QUAD precision and with QUAD only applied to the basis of scalar integrals
  and tensor integral routines 
  dotted-dashed line. The effect of the rescue system, i.e., the double result is set to the QUAD precision when the Ward identity gauge test
for the amplitudes fails at the accuracy 10$^{-(X)}$ are described by
the ``Imp$_X$'' lines. In the right
  panels, one can see the critical region, where double precision is not
  accurate, and the effect of the rescue system for different values of
  the Ward identity gauge test. One observes that the choice
10$^{-4}$ (Imp$_4$) gives 3 digits of precision at the per mill level with
non-identified left instabilities at this level of accuracy, Tables~\ref{Accuracy3} and~\ref{Accuracy4}.  From the plot,  we can conclude
that the Ward identity gauge test together with the use of QUAD
precision is an efficient system to control the accuracy of our results.
%
%
%
%
%
%
%
%
%
%
%
%
%
%
%

%
%
%
\section{Conclusions}
\label{sec:concl}
In this paper, some of the NLO QCD one-loop amplitudes contributing to
the process $pp \to VV jj+X$ have been computed. It has been shown that the numerical
instabilities due to the presence of small Gram and Cayley determinants
 are under control. The use of the LU
 decomposition method for $C$ and $D$ functions reduces considerably the
 fraction of unstable points at the double precision level (without CPU penalty). This improvement
 together with the use of special routines for small Gram determinants
is enough to reduce the instabilities for pentagons well bellow the per
mill level for an accuracy of the Ward Identity gauge test of $10^{-4}$.
Using  quadruple precision in two steps, the
instabilities for the hexagons for a precision of $10^{-6}$ in the Ward identity gauge
tests are reduced up to the 0.03$\permil$ level. To decrease further this number the
use of higher precision can be safely applied since for an accuracy of $10^{-4}$ the identified
instabilities  disappear even without previously using special routines
for small Gram determinants. This suggests that the presence of Landau singularities which
translate into exactly vanishing Gram determinants, for
accepted cut points, is small for the kinematics of EW $pp\to WWjj+X$ production, given
the cuts of Eqs.~(\ref{cut1}) and (\ref{cut2}).

It has been shown factorization proofs of the infrared
divergences of the bosonic contributions of $VVVV$ and $VVVj$ production
with $V \in (W,Z,\gamma)$. We have also given the master equations for
the evaluation of the tensor coefficient integrals in the external
convention including those needed for small Gram determinants.


The CPU time of the contributions involving hexagons, which
rounds the milliseconds, are competitive with other more sophisticated
methods. We plan to use these tools not only to compute diboson plus two
jet production at hadron colliders at NLO QCD, but also other interesting $2\to
4$ processes like $W\gamma\gamma + j$~\cite{Campanario:2011}.


%
%
%
%
%
%
%
%
%
%
%
%
%
%
%
%
\section*{Acknowledgments}
F.C thanks M.~Rauch for pointing out the LU method, D.~Zeppenfeld
for encouraging discussions and S.~Khodayar for reading the manuscript. F.C.~acknowledges partial support of a postdoctoral fellowship of the 
Generalitat Valenciana, Spain (Beca Postdoctoral d`Excel$\cdot$l\`encia), by European FEDER and Spanish MICINN
under grant FPA2008-02878 and by the Deutsche Forschungsgemeinschaft 
via the Sonderforschungsbereich/Transregio  SFB/TR-9 ``Computational Particle
Physics''.
The Feynman diagrams in this paper were drawn using Axodraw~\cite{Axodraw}.
\appendix
\section{Tensor Decomposition Master Equations}
\label{appendixa}
For the derivation of the Master Equations presented in this Appendix, we
have followed closely Ref.~\cite{Denner:2005nn}. The difference resides in the
convention used to derive them. Meanwhile Ref.~\cite{Denner:2005nn} uses 
the internal propagator
momenta, $q_k$, we use the external momenta, $p_k$. The conversion from one notation to the other is trivial. Nevertheless,
we have derived the recursion relations in this notation, first, to avoid
loss of accuracy in the conversion for critical points,
second, the conversion for pentagons and hexagons of high rank starts to be lengthy and time consuming.

The n-point one loop tensor/scalar integrals can be written as
\be
\label{A:1}
I_N^{\mu_1\ldots
  \mu_P}(p_1,\ldots,p_{N-1},m_0,\ldots,m_{N-1})=c_\Gamma(\mu^2)
\int d^D l
\frac{l^{\mu_1} \ldots l^{\mu_p}}{N_0\ldots  N_{N-1}},\quad \quad P \ge 0
\ee
with 
\be
\label{cteFacC}
c_\Gamma(\mu^2)= \frac{-I (4\pi)^{D/2} (\mu^2)^{(4-D)/2} }{(2\pi)^{D}\Gamma(3-D/2)},
\ee
and
\be
N_k=(l+q_k)^2-m_k^2+i\epsilon=\left(l+\sum_{j=0}^k p_j\right)^2-m_k^2+i\epsilon,~~~~k
=0, \ldots , N-1,~~~~~q_0=p_0=0.
\ee
Although the introduction of external momenta may seem an additional
complication, if one carelessly uses the internal propagator notation
but uses external momenta in their expressions, the number of terms grow
factorially. 
As an example, we present the tensor integral of rank 1 for
a box integral.
With the external momenta convention, the tensor integral is written in terms of their
tensor integral coefficients as,
\be
I_4^{\mu_1}(p_1,p_2,p_3)= D_1 p_1^{\mu_1} +D_2 p_2^{\mu_1}+ D_3 p_3^{\mu_1}.
\ee
If we use the internal momenta notation, IM, for the same integral and
translate it to the external momenta notation to apply
equation of motions or transversality properties, the number of terms increases
from 3 to 6:
\be
\label{boxIM}
I_4^{\mu_1}(q_1,q_2,q_3)= D^{IM}_1 q_1 +D^{IM}_2 q_2+ D^{IM}_3 q_3 =D^{IM}_1 p_1^{\mu_1}+D^{IM}_2 (p_1+p_2)^{\mu_1}+ D^{IM}_3 (p_1+p_2+p_3)^{\mu_1}.
\ee
For Hexagons of rank 5, this means passing from 4500 to 200000
terms. 
Although, this problem can be solved trivially, e.g., 
applying conservation of momentum, some care has to be
taken to avoid exceeding the memory capacities of the computer at
intermediate stages.

Generally, the tensor integrals are written in terms of its tensor
integral coefficients using Lorentz-covariant structures,
\be
\label{ref:tens1}
I_N^{\mu_1 \ldots \mu_P}=
\sum_{n=0}^{[P/2]}\sum_{i_{2n+1},\ldots,i_P=1}^{N-1} \{\underbrace{g
    \ldots g}_n p\ldots p\}^{\mu_1 \ldots \mu_P}_{i_{2n+1}\ldots i_P}
I^N_{\underbrace{0 \ldots 0}_{2 n} i_{2 n +1}\ldots i_P}
\ee
where the curly braces are defined similarly as in Ref.~\cite{Denner:2005nn}.

Following Ref.~\cite{Denner:2005nn}, we find for the external momenta
convention up to the pentagon level, the following Master equations for the tensor coefficient integrals:
\bea
\label{masPV}
I^N_{i_1 \ldots i_P }& =&
\sum_{n=1}^{N-1}(Z^{(N-1)})^{(-1)}_{i_1n}\left(S^{(N-1)}_{n i_2 \ldots i_P}
  -2\sum_{r=2}^{P}\delta_{n i_r} I_{00i_2\ldots \hat{i}_r \ldots
    i_P}^N \right),\nonumber  \\
 &&  \quad  N \le 5, \quad 1 \le i_n\le N-1
 \eea
%
\bea
\label{masPV00}
I^N_{00i_3 \ldots i_P }& = &\frac{1}{2(D+P-N)}\left(
  (-1)^{(\delta_{0(i_3)_{0}}+\ldots+\delta_{0(i_P)_{0}})} I^{(N-1)}_{i3 \ldots
    i_P}(0) +2 m_0^2 I^N_{i_3\ldots i_P} + \right. \\
 && \left.+
  \sum_{n=1}^{N-1}(r_n-r_{n-1})I^N_{n i3 \ldots i_P}\right), \quad N \le
5, i_3 \ldots i_P=0,\ldots N-1 \nonumber
\eea
$(Z^{(N-1)})^{(-1)}$ is the inverse Gram determinant which is built with 
external momenta, $Z_{ij}=2(p_i \cdot p_j)$ and $r_n=\left(\sum_{j=1}^n p_j\right)^2-m_n^2$. $S^{N+1}_{j i_2 \ldots i_P}$
is defined by
\bea
S^{N+1}_{j i_2 \ldots i_P}&=&\bar{\delta}_{N(i_2)_j}\ldots\bar{\delta}_{N(i_P)_j}
I^{N}_{(i_2)_j\ldots(i_P)_j}(j)-(-1)^{(\delta_{0(i_2)_{j-1}}+\ldots+\delta_{0(i_P)_{j-1}})} I^N_{(i_2)_{j-1}\ldots(i_P)_{j-1}}(j-1)-\nonumber\\
&&-(r^2_{j}-r_{j-1}^2)I^{N+1}_{i_2\ldots i_p}\,\,\,\,\,\,\,\,\, ,
\eea
$\bar{\delta}_{ij}=1-\delta_{ij}$, $(i_1)_j$ is given by,
\be
(i_1)_j=\left(\begin{array}{cc} 
i_1 \le j & i_1\\
i_1>j & i_1-1 
\end{array} \right. ,
\ee
and j in $I^{(N)}_{i_1\ldots}(j)$ means that the $N_j$ propagator has been
canceled, that is, the tensor coefficients for the tensor integral with
the $N_j$ propagator removed from $I^{(N+1)}_{i_1\ldots}$. For the LU
implementation, the inversion of the Matrix in Eq.~(\ref{masPV}) is not
done analytically, instead, a system of equations for the tensor
integrals given by,
\bea
\label{masPVLU}
(Z^{(N-1)})_{i_1n} I^N_{i_1 \ldots i_P }& =&
S^{(N-1)}_{n i_2 \ldots i_P}
  -2\sum_{r=2}^{P}\delta_{n i_r} I_{00i_2\ldots \hat{i}_r \ldots i_P}^N,
\eea
is solved using a fast implementation of partial pivoting of the LU
decomposition method. 
For small Gram determinants, we define in analogy to Ref.~\cite{Denner:2005nn},
\bea
\hat{S}^{N+1}_{j i_2 \ldots i_P}&=&\bar{\delta}_{N(i_2)_j}\ldots\bar{\delta}_{N(i_P)_j}
I^{N}_{(i_2)_j\ldots(i_P)_j}(j)-(-1)^{(\delta_{0(i_2)_{j-1}}+\ldots+\delta_{0(i_P)_{j-1}})}
I^N_{(i_2)_{j-1}\ldots(i_P)_{j-1}}(j-1)\nonumber\\
S^{N+1}_{0 0 i_3 \ldots i_P}&=&2 (-1)^{(\delta_{0(i_3)_{0}}+\ldots+\delta_{0(i_P)_{0}})}
I^N_{(i_3)_{0}\ldots(i_P)_{0}}(0) + 2\,m_0\,I^{N+1}_{(i_3)_0\ldots(i_P)_0},\nonumber\\
\label{massGram}
\eea
and the recursion relations are obtained directly from Eqs.~(5.41-5.48)
of Ref.~\cite{Denner:2005nn} using external momenta in all
(sub)determinants and the replacement $f_n \to (r_n-r_{n-1})$. We have
implemented these recursion relations up to rank 9 for the $C_{ij}$,
and up to rank 7 for the $D_{ij}$ functions~\footnote{The UV part of the
integrals are not given explicitly here. They can be obtained by
contacting the author.}. 

For the pentagons, we have followed closely section 6 of
Ref.~\cite{Denner:2005nn} and derived their master formulas
Eqs.~(6.12,6.13) for the external momenta convention. Although, the cancellation
that takes place to get their simple equations are different, the
final master integrals are quite similar:
\bea
\label{Emas}
&&det(X^{(4)})\bar{E}_{k i_1 \ldots i_P}=\sum_{n=1}^{4}
X_{kn}^{(4)}\left(\bar{\delta}_{i_1 n}\ldots \bar{\delta}_{i_P
    n}D_{(i_1)_n \ldots (i_P)_n}(n) - \right.  \\ 
&&\left.
-(-1)^{(\delta_{0(i_1)_{j-1}}+\ldots+\delta_{0(i_P)_{j-1}})} D_{(i_1)_{n-1}
  \ldots (i_P)_{n-1}}(n-1)\right) - X_{k0}^{(4)}(-1)^{(\delta_{0(i_1)_{0}}+\ldots+\delta_{0(i_P)_{0}})}  D_{(i_1)_0-
  \ldots (i_P)_0}(0)\nonumber \\
&&- 2 \sum_{n=1}^{4}\sum_{r=1}^{P}
X^{4}_{(kn)(0i_r)}\left((\bar{\delta}_{i_1 n}\ldots\bar{\delta}_{i_{r-1}
    n}\bar{\delta}_{i_{r+1} n}\ldots\bar{\delta}_{i_Pn} D_{00(i_1)_n
    \ldots\widehat{(i_r)_n} \ldots (i_P)_n}(n) - \right. \nonumber \\
&& \left. -
  (-1)^{(\delta_{0(i_2)_{j-1}}+\ldots+\delta_{0(i_{r-1})_{j-1}}+\delta_{0(i_{r+1})_{j-1}}+\ldots+\delta_{0(i_P)_{j-1}})} D_{00(i_1)_{n-1}  \ldots \widehat{(i_{r})_{n-1}} \ldots (i_P)_{n-1}}(n-1)\right)\nonumber\\
&& ~~~~k=1,\ldots,4,\mbox{~~~~~~~~~~~~~~~~~~~~~~~~~~~~} P<4, \nonumber
\eea
%
\bea
\label{E00mas}
&& det(X^{(4)})\bar{E}_{00i_2 \ldots i_P}=\sum_{n=1}^{4}
X_{n0}^{(4)}\left(\bar{\delta}_{i_2 n}\ldots \bar{\delta}_{i_P
    n}D_{00(i_2)_n \ldots (i_P)_n}(n) - \right. \nonumber \\ 
&&\left.
-(-1)^{(\delta_{0(i_2)_{j-1}}+\ldots+\delta_{0(i_P)_{j-1}})} D_{00(i_2)_{n-1}
  \ldots (i_P)_{n-1}}(n-1)\right),\mbox{~~~~~~~~~~~~~~} P<4,
\eea
where $X^{(4)}$ and related quantities are defined/obtained from Eq.~(2.23) in Ref.\cite{Denner:2005nn}
making the replacement $f_n \to (r_n-r_{n-1})$ and using directly
external momenta, $q_k \to p_k$.

For the hexagons, we have followed section 7 of Ref.~\cite{Denner:2005nn}. The
master equations with external momenta convention are obtained from their
Eqs.~(7.13,7.16-7.18) by making the replacement:
\be
E_{i_2i_3\ldots i_P}(0) \to
(-1)^{(\delta_{0(i_2)_{n-1}}+\ldots+\delta_{0(i_{P})_{n-1}})}E_{(i_2)_{n-1}\ldots
  (i_P)_{n-1}}(n-1),
\ee
 using directly
external momenta, $q_k \to p_k$. We have checked the derivation of all the tensor coefficients up to
Hexagons of rank 5 by contracting the tensor integrals with external
momenta. Thus, we can relate all the tensor coefficients of order $N$ and
rank $P$ to tensor integrals of order $N$ and $N-1$ and rank $P-1$:
\bea
&&p_{i \mu_1} p_{j \mu_2}\ldots p_{k\mu_P} I_N^{\mu_1 \ldots \mu_P}=  \sum_{i_1 \ldots
  i_P}^{N-1} Z_{ii_1} Z_{ji_2}\ldots Z_{k i_P} I_{i_1\ldots i_P}^N =  \nonumber
\\
&& p_{j \mu_2}\ldots p_{k \mu_P} (I_{N-1}^{\mu_2 \ldots \mu_P}(i)
-I_{N-1}^{\mu_2 \ldots \mu_P}(i-1) -(r_i-r_{i-1})I_N^{\mu_2 \ldots \mu_P}).
\eea
For hexagons of rank 5, up to 166 different combinations~(not
all of them independent since only four momenta out of six are linearly
independent) where built
and checked at the FORTRAN and Mathematica level. The implementation of
these routines together with the massless input integrals in Mathematica have
proved to be quite useful since we can overcome the loss of accuracy due to
small Gram determinants by increasing the working precision, and check accurately
the special routines for small Gram determinants, the tensor
reductions for pentagons a la Passarino Veltman vs a la Denner-Dittmaier
as well as the complete numerical result of the contributions.
\section{Benchmark numbers for the hexagon contributions}
\label{appendixb}
In this section, we provide numbers for the hexline and the hexlineNoAbe contributions. The set of
momenta used for $q(p_1)\bar{q}(p_2) V_1 V_2 V_3 g \to 0$ is chosen to be:
%
%
\bea
   p_1 &=&  (   2210.591640000411,      0,      0,   2210.591640000411),\nonumber \\
   p_2 &=&  (    410.6465697388802,      0,      0,   -410.6465697388802),\nonumber \\
p_{V_1} &=& ( -1644.598252136518,   -186.7167811992445,     14.28499809343437,  -1633.902137026130),\nonumber \\
p_{V_2} &=& (    -266.9005208707261,    -59.67643702980636,   -176.1526035584612,   -173.8364669053697),\nonumber \\
p_{V_3} &=& (    -402.5905961231937,    166.3844370751274,    218.6277243718783,   -283.2688206929862),\nonumber \\
 p_{g} &=&  (   -307.1488406088535,     80.00878115392345,    -56.76011890685155,    291.0623543629552),\nonumber \\
 \eea
%
with the notation $p=(p^0,p^1,p^2,p^3)$ and all the components given in
GeV. This represents a cut-accepted point for the EW $qq\to W^+W^- qq$
process with a ``mild'' behavior, with ``mild'' meaning that there are
not small Gram/Cayley determinants appearing up to the hexagon level, therefore the double
precision routines provide accurate results. The corresponding
polarization vectors are constructed
following the conventions of Ref.~\cite{Murayama:1992gi}, Eq.~(A.11). First, we give results
for each of the diagrams that constitute the hexline contribution for a particular permutation of the
electroweak vector bosons, ${\cal M}_{V_1V_2V_3g,\tau}$, with $\tau=-1$,
in Table~\ref{appen:numhex} and Table~\ref{appen:numhex_d}, using the decomposition,
\bea
\label{eq:hexnum}
2 \frac{\text{Re} \left( {\cal M}^B_{\tau} \cdot {\cal M}_{V_1V_2V_3g,\tau}^*\right)}
{| {\cal M}^B_{\tau} {\cal M}_{\tau}^{B*}|} 
=
\frac{g_0^2}{(4\pi)^2}
(4\pi)^\epsilon
\Gamma (1+\epsilon)(\mu_0)^{-2\epsilon}
\sum_{n=1}^{13}\left(\frac{{\cal C}^{V_1V_2V_3g}_n}{T^a}\right) \times \hspace{2.3cm}
 \\
 \left(
 c^{(n)}_{(0),V_1V_2V_3g} 
+\frac{c^{(n)}_{(1),V_1V_2V_3 g}}{\epsilon}
+\frac{c^{(n)}_{(2),V_1V_2V_3 g}}{\epsilon^2}
+\frac{(D-4)}{-2\epsilon} d^{(n)}_{(0),V_1V_2V_3g} 
+\frac{(D-4)}{-2\epsilon} \frac{d^{(n)}_{(1),V_1V_2V_3 g}}{\epsilon}
\right), \nonumber
\eea
correspondingly to Eq.~(\ref{hexline}). 
${\cal M}^B_{\tau}= {\cal M}^B_{V_1V_2V_3g\tau}$, 
i.e. the born
amplitude for the specific order of vector bosons. The
$(4\pi)^\epsilon \Gamma (1+\epsilon)(\mu_0)^{-2\epsilon} $ factor comes
from the definition for the scalar and tensor integrals, Eq.~(\ref{A:1}).
%
\begin{table}[h!]
\begin{tabular}{|c|c|c|c|}
\hline
&\multirow{2}{*}{ $ c^{(n)}_{(0),V_1V_2V_3g} $} 
&\multirow{2}{*}{$c^{(n)}_{(1),V_1V_2V_3 g} $}
&\multirow{2}{*}{ $c^{(n)}_{(2),V_1V_2V_3g} $}
\\
&
& 
&
\\ \hline
n= 1& -0.8495224242548815E+02& -0.2718191945175511E+02& -0.3999999999999874E+01\nonumber \\ \hline
n= 2&  0.7373715226545116E+02&  0.1475364608414150E+02& -0.4823858365463354E-13\nonumber \\ \hline
n= 3&  0.7720679379271682E+02&  0.1032635756877249E+02&  0.4778146726503880E-13\nonumber \\ \hline
n= 4& -0.7023321057368833E+01& -0.2247107183707837E+01& -0.2411275830791782E-14\nonumber \\ \hline
n= 5&  0.5075981735087911E+01&  0.0000000000000000E+00&  0.0000000000000000E+00\nonumber \\ \hline
n= 6& -0.1241513230258542E+03& -0.1483523289520358E+02&  0.3441661498683095E-13\nonumber \\ \hline
n= 7&  0.1268199899143818E+02&  0.2000000000000156E+01&  0.1502284155171117E-26\nonumber \\ \hline
n= 8&  0.9572211011177236E+01&  0.2000000000000001E+01&  0.0000000000000000E+00\nonumber \\ \hline
n= 9&  0.2215410796479108E+02&  0.2000000000000001E+01&  0.0000000000000000E+00\nonumber \\ \hline
n= 10&  0.1060058346584089E+03&  0.1318425587774655E+02& -0.1013265897019504E-13\nonumber \\ \hline
n=11& -0.1268199899143716E+02& -0.2000000000000000E+01&  0.0000000000000000E+00\nonumber \\ \hline
n=12& -0.1024742028732465E+02& -0.2000000000000000E+01&  0.0000000000000000E+00\nonumber \\ \hline
n=13& -0.1523236618746480E+02& -0.2000000000000000E+01&  0.0000000000000000E+00\nonumber \\ \hline
\end{tabular}
\caption{Contributions of the hexline diagrams. $
  c^{(n)}_{(j),V_1V_2V_3g}$ are defined in Eq.~(\ref{eq:hexnum}).}
\label{appen:numhex}
\end{table}
%
\begin{table}[t!]
\begin{tabular}{|c|c|c|}
\hline
&\multirow{2}{*}{ $d^{(n)}_{(0),V_1V_2V_3g,\tau} $} 
&\multirow{2}{*}{$ d^{1,(n)}_{(1),V_1V_2V_3g ,\tau} $}
\\
&
& 
\\ \hline
n= 1&  0.8702382558218599E-12& -0.8613060482805703E-13\nonumber \\ \hline
n= 2&  0.1709040047358825E-13& -0.8545281855364259E-15\nonumber \\ \hline
n= 3&  0.3736943504519546E-13& -0.4397635645750366E-14\nonumber \\ \hline
n= 4& -0.2396002150441086E-13&  0.1363187596482680E-15\nonumber \\ \hline
n= 5&  0.0000000000000000E+00&  0.0000000000000000E+00\nonumber \\ \hline
n= 6& -0.3591236576358064E-12& -0.4690957329075479E-14\nonumber \\ \hline
n= 7& -0.2000000000000079E+01&  0.0000000000000000E+00\nonumber \\ \hline
n= 8& -0.2000000000000001E+01&  0.0000000000000000E+00\nonumber \\ \hline
n= 9& -0.2000000000000001E+01&  0.0000000000000000E+00\nonumber \\ \hline
n= 10& -0.2000000000000000E+01& -0.6325727406404221E-14\nonumber \\ \hline
n= 11&  0.2000000000000001E+01&  0.0000000000000000E+00\nonumber \\ \hline
n= 12&  0.2000000000000001E+01&  0.0000000000000000E+00\nonumber \\ \hline
n= 13&  0.2000000000000001E+01&  0.0000000000000000E+00\nonumber \\ \hline
\end{tabular}
\caption{Contributions of the hexline diagrams. $
  d^{(n)}_{(j),V_1V_2V_3g}$ are defined in Eq.~(\ref{eq:hexnum}).}
\label{appen:numhex_d}
\end{table}
%
\begin{table}[h!]
\begin{tabular}{|c|c|c|}
\hline
\multirow{2}{*}{ $ $} 
&\multirow{2}{*}{$\mu_0=s $}
&\multirow{2}{*}{ $\mu_0=1\text{GeV} $}
\\ 
&
&
\\\hline
$\sum c^{(n)}_{(0),V_1V_2V_3g} $
&0.5214540844413348E+02
&-0.3135489678574997E+03
\\ \hline
$2\text{Re}(f(s,t,u,\mu_0,1,0,0)^*)$
&0.1973920880217872E+02
&-0.3459551674994638E+03
\\ \hline
Eq.~(\ref{eq:Appfmu})
&0.3240619964195476E+02
&0.3240619964196406E+02
\\ \hline
ratio-1&
\multicolumn{2}{c|}{ -0.2868816295631405E-12}
\\
\hline
\end{tabular}
\caption{Check of factorization scale energy independence through Eq.~(\ref{eq:Appfmu}) for two
different sets of factorization energy scales.}
\label{eq:Apprenhexline}
\end{table}
We can implement some tests to these numbers. Assuming a general common
factor, $C_F$, we can
check the value of the sum of $c^{(n)}_{(0),V_1V_2V_3g}$  by testing the factorized scale energy independence, 
similarly as Eq.~(\ref{eq:identifyhexNoabef})
through, 
\be
\sum c^{(n)}_{(0),V_1V_2V_3g}(\mu_0)-2\text{Re}\left( f(s,t,u,\mu_0,1,0,0)^*\right)\not\equiv  F(\mu_0)
\label{eq:Appfmu}
\ee
 with $f(s,u,t,\mu_0,1,0,0)$ given by Eq.~(\ref{eq:identifyf}) with
 $s=(p_1+p_2)^2, t=(p_1+p_g)^2$ and $u=(p_2+p_g)^2$. %
In Table~\ref{eq:Apprenhexline}, we show the values of the sum of
 $c^{(n)}_{(0),V_1V_2V_3g}$  and  $f(s,u,t,\mu_0,1,0,0)$ for two
 different sets of factorization energy scales, for $\mu_0=s$,
 corresponding 
to the values given in
 Table~\ref{appen:numhex}, and for $\mu_0=1$GeV. One can see that
 Eq.~(\ref{eq:Appfmu}) is satisfied with an accuracy of 12 digits.
Additionally, for this specific phase space point, the Ward identities of Eq.~(\ref{eq:hexgauge}) for the
complete contributions are satisfied at the 12 digit
level. For the divergent contributions and the rational terms, 
the factorization against the born
amplitude is shown in Table~\ref{App:fachex} through,
\be
\sum c^{(n)}_{(1),V_1V_2V_3g}= -6\quad , \quad
\sum c^{(n)}_{(1),V_1V_2V_3g}= -4\quad , \quad
\sum d^{(n)}_{(0),V_1V_2V_3g}= -2\quad , \quad
\sum d^{(n)}_{(1),V_1V_2V_3g}= 0,
\label{eqApp:fachex}
\ee
%
correspondingly to Eq.~(\ref{eq:identifyhexNoabeCF}). Note that the
numerical values for $ d^{(n)}_{(1),V_1V_2V_3g}$ 
in Table~\ref{appen:numhex_d} are below $10^{(-13)}$.
%
\begin{table}[t!]
\begin{tabular}{|c|c|c|c|}
\hline
\multirow{2}{*}{ $ $} 
&\multirow{2}{*}{$  c^{(n)}_{(1),V_1V_2V_3g}$}
&\multirow{2}{*}{ $  c^{(n)}_{(2),V_1V_2V_3g}$}
&\multirow{2}{*}{ $  d^{(n)}_{(0),V_1V_2V_3g}$}
\\ 
&
&
&
\\\hline
$\sum$ &-0.6000000000005826E+01 
&-0.3999999999999853E+01
&-0.1999999999999537E+01
\\\hline
Exact&-0.6000000000000000E+01
&-0.4000000000000000E+01
&-0.2000000000000000E+01
\\\hline
ratio-1
&0.9710010573371619E-12
&-0.3674838211509268E-13
&-0.4524345419647128E-12
\\
\hline
\end{tabular}
\caption{Factorization of the divergent contributions through 
Eq.~(\ref{eqApp:fachex}) against the born
  amplitude and accuracy.}
\label{App:fachex}
\end{table}
%
%
%
%
%
%
%
%
%
%
%
%
%
%
%
%
\begin{table}[h!]
\begin{tabular}{|c|c|c|c|}
\hline
&\multirow{2}{*}{ $c^{g,(n)}_{(0),V_1V_2V_3} $} 
&\multirow{2}{*}{$c^{g,(n)}_{(1),V_1V_2V_3} $}
&\multirow{2}{*}{ $c^{g,(n)}_{(2),V_1V_2V_3} $}
\\
&
&
& 
\\ \hline
n= 1& -0.3970812374704999E+02& -0.1265721313949629E+02& -0.1813569591709497E+01\nonumber \\ \hline
n= 2& -0.1350281050903405E+01& -0.7519564890398996E+00& -0.2022904490007958E+00\nonumber \\ \hline
n= 3&  0.2728377087017317E+02&  0.2055535050768235E+01& -0.5549746075488889E+00\nonumber \\ \hline
n= 4& -0.3264136921557698E+00& -0.1238348703073990E+00& -0.1416214405745540E-01\nonumber \\ \hline
n= 5&  0.1723360559577099E+01&  0.3328092079219656E+00&  0.0000000000000000E+00\nonumber \\ \hline
n= 6& -0.5023976788785847E+02& -0.1779284983273073E+02& -0.2954984559380555E+01\nonumber \\ \hline
n= 7&  0.1253736490780382E-02& -0.2491704328619187E-03& -0.7861847392712649E-03\nonumber \\ \hline
n= 8&  0.4621898462078639E+00&  0.1193463736617884E+00&  0.0000000000000000E+00\nonumber \\ \hline
n= 9&  0.1418038518281030E+01&  0.4012945693937532E+00&  0.0000000000000000E+00\nonumber \\ \hline
n= 10&  0.1937597641023809E+02&  0.8653232589507967E+01&  0.1540767536436781E+01\nonumber \\ \hline
\end{tabular}
\caption{Contributions of the hexlineNoAbe diagrams. $
  c^{g,(n)}_{(j),V_1V_2V_3}$ are defined in Eq.~(\ref{eq:hexNoAbenum}).}
\label{appen:numhexNo}
\end{table}

In the following, we give results for the hexlineNoAbe contribution for $\tau=-1$.
We follow the decomposition of Eq.~(\ref{eq:hexNoAbenum}) assuming EW
production with an additional jet, thus, the color factor is $C_A$ for all
the diagrams,
\bea
\label{eq:hexNoAbenum}
2 \frac{\text{Re} \left( {\cal M}^B_{\tau} \cdot {\cal
      M}_{V_1V_2V_3,\tau}^{g*}\right)}
{| {\cal M}^B_{\tau} {\cal M}^{B*}_{\tau}|} 
=
\frac{g_0^2}{(4\pi)^2}
(4\pi)^\epsilon
\Gamma (1+\epsilon)(\mu_0)^{-2\epsilon}
\sum_{n=1}^{10} C_A\times \hspace{2.3cm}
 \\
\left(
 c^{g,(n)}_{(0),V_1V_2V_3} 
+\frac{c^{g,(n)}_{(1),V_1V_2V_3 }}{\epsilon}
+\frac{c^{g,(n)}_{(2),V_1V_2V_3 }}{\epsilon^2}
+\frac{(D-4)}{-2\epsilon} d^{g,(n)}_{(0),V_1V_2V_3} 
+ \frac{(D-4)}{-2\epsilon}\frac{d^{g,(n)}_{(1),V_1V_2V_3 }}{\epsilon}
\right) \nonumber
\eea
correspondingly to Eq.~(\ref{hexNoAbe}), with, 
\begin{equation}
{\cal M}^B_{\tau}=
{\cal M}^B_{V_1V_2V_3g,\tau}
+{\cal M}^B_{V_1V_2gV_3,\tau}
+{\cal M}^B_{V_1gV_2V_3,\tau}
+{\cal M}^B_{gV_1,V_2V_3,\tau},
\label{eq:b_con}
\end{equation}
%
\noindent that is, the sum of the four Born amplitudes for a given order of the EW
vector bosons. The individual contributions are given in
Table~\ref{appen:numhexNo} and Table~\ref{appen:numhexNo_d}. %
%
\begin{table}[t!]
\begin{tabular}{|c|c|c|}
\hline
&\multirow{2}{*}{ $ d^{g,(n)}_{(0),V_1V_2V_3} $} 
&\multirow{2}{*}{$d^{g,(n)}_{(1),V_1V_2V_3} $}
\\
&
& 
\\ \hline
n= 1& -0.1436483954568111E-11&  0.3386511581831592E-13\nonumber \\ \hline
n= 2& -0.6922590837003719E-18&  0.1939122411552382E-16\nonumber \\ \hline
n= 3&  0.5899544848540121E-13&  0.1263944582753722E-13\nonumber \\ \hline
n= 4& -0.5274108820645013E-16& -0.1071274072706729E-17\nonumber \\ \hline
n= 5&  0.3633510145273396E-19&  0.0000000000000000E+00\nonumber \\ \hline
n= 6&  0.5975463630574398E-13&  0.1078373100666381E-14\nonumber \\ \hline
n= 7& -0.7861847392712663E-03&  0.0000000000000000E+00\nonumber \\ \hline
n= 8& -0.1416214405745516E-01&  0.0000000000000000E+00\nonumber \\ \hline
n= 9& -0.1371873399923628E+00&  0.0000000000000000E+00\nonumber \\ \hline
n= 10& -0.8478643312109194E+00&  0.1159003560547673E-14\nonumber \\ \hline
\end{tabular}
\caption{Contributions of the hexlineNoAbe diagrams. $
  d^{(n)}_{(j),V_1V_2V_3g}$ are defined in Eq.~(\ref{eq:hexNoAbenum}).}
\label{appen:numhexNo_d}
\end{table}
%
\begin{table}[h!]
\begin{tabular}{|c|c|c|}
\hline
&\multirow{2}{*}{ $C_F $} 
&\multirow{2}{*}{$C_F -1/2 C_A $}
\\
&
& 
\\ \hline
${ c}_{(0),gV_1V_2V_3}$&  0.1186972860337361E+00  & -0.2513048326546336E+00  \nonumber \\ \hline
${ c}_{(0),V_1gV_2V_3}$& -0.3260632362384948E+00  &  0.1175575424874967E+01  \nonumber \\ \hline
${ c}_{(0),V_1V_2gV_3}$& -0.3587963452897030E+01  &  0.6010092883764584E+01  \nonumber \\ \hline
${ c}_{(0),V_1V_2V_3g}$&  0.6748418542481737E+02  & -0.2188150305629134E+02  \nonumber \\ \hline
${ c}_{(1),gV_1V_2V_3}$&  0.3435253401752263E-01  & -0.2242562120604109E-01  \nonumber \\ \hline
${ c}_{(1),V_1gV_2V_3}$& -0.1745714679327207E+00  &  0.9440497420994520E-01  \nonumber \\ \hline
${ c}_{(1),V_1V_2gV_3}$& -0.9768988228536071E+00  & -0.1710884136252724E+00  \nonumber \\ \hline
${ c}_{(1),V_1V_2V_3g}$&  0.1091241185529769E+02  & -0.1569618503789184E+02  \nonumber \\ \hline
${ c}_{(2),gV_1V_2V_3}$&  0.0000000000000000E+00  & -0.3144738956157922E-02  \nonumber \\ \hline
${ c}_{(2),V_1gV_2V_3}$&  0.0000000000000000E+00  & -0.5664857622981814E-01  \nonumber \\ \hline
${ c}_{(2),V_1V_2gV_3}$&  0.6661338147750939E-15  & -0.5487493599694454E+00  \nonumber \\ \hline
${ c}_{(2),V_1V_2V_3g}$& -0.4263256414560601E-13  & -0.3391457324843469E+01  \nonumber \\ \hline
${ d}_{(0),gV_1V_2V_3}$& -0.2765908162227770E-14  &  0.7861847503483757E-03  \nonumber \\ \hline
${ d}_{(0),V_1gV_2V_3}$&  0.4302114220422482E-15  &  0.1416214405720360E-01  \nonumber \\ \hline
${ d}_{(0),V_1V_2gV_3}$&  0.4385380947269368E-14  &  0.1371873399918132E+00  \nonumber \\ \hline
${ d}_{(0),V_1V_2V_3g}$&  0.3652633751016765E-13  &  0.8478643312106809E+00  \nonumber \\ \hline
\end{tabular}
\caption{Contributions of the hexline diagrams for the different
  colors and for the four permutations for a fixed order of
  the EW vector bosons.}
\label{appen:numhex_con}
\end{table}
%
In addition, in Table~\ref{appen:numhex_con} we show the hexline
results for the four contributions that are mixed under the QCD
group for a given permutation of EW vector bosons since they will help
us to construct useful checks for the hexlineNoAbe contributions. We use
the notation of Eq.~(\ref{eq:hexnum}) with ${\cal M}^B_{\tau}$ given by
Eq.~(\ref{eq:b_con}) and provide separate numbers for the sum of the two
different color structures that appear in this case. 
We can check the finite pieces, ${c}^{(n)}_{(0)}$, by testing the
factorized energy scale, $\mu_0$, independence for 
the two color factors, $C_F$ and $C_A$ through,

$C_F$
\bea
c^{'}_{(0),V_1V_2V_3g}(\mu_0) +
c^{'}_{(0),V_1V_2gV_3}(\mu_0) +
c^{'}_{(0),V_1gV_2V_3g}(\mu_0) +
c^{'}_{(0),gV_1V_2V_3g}(\mu_0)) -\hspace{1cm}\nonumber\\
-2\text{Re}\left( f(s,t,u,\mu_0,1,0,0)^*\right)\not\equiv  F(\mu_0) 
\label{eq:AppfmuNoAbe1}
\eea

$C_A$
\bea
\frac{-1}{2}\left( 
c^{}_{(0),V_1 V_2V_3g}+
c^{}_{(0),V_1 V_2gV_3}+
c^{}_{(0),V_1g V_2V_3}+
c^{}_{(0),gV_1V_2V_3} \right)%
+
\sum_{j=1}^{10} c^{g,(j)}_{(0),V_1V_2V_3}-\hspace{2cm}\nonumber\\
- 2\text{Re}\left( f(s,t,u,\mu_0,0,1,0)^*\right)\not\equiv  F(\mu_0)
\label{eq:AppfmuNoAbe2}
\eea
similarly to Eq.~(\ref{eq:identifyhexNoabef}). The prime quantities are
built by adding the two numbers of each line in
Table~\ref{appen:numhex_con}. For Eq.~(\ref{eq:AppfmuNoAbe2}),
$c_{(0)}$ is obtained by taking the number of the second column of each
line 
in Table~\ref{appen:numhex_con}.  We give the results for two different sets of
factorization scale energies in Table~\ref{eq:ApprenhexlineNoAbeC_FC_A} and the
accuracy of it. We obtain 11 digits of precision for the specific phase
space point.
\begin{table}[h!]
\begin{tabular}{|c|c|c|}
\hline
\multirow{2}{*}{ $ $} 
&\multirow{2}{*}{$\mu_0=s $}
&\multirow{2}{*}{ $\mu_0=1 $}
\\ 
&
&
\\\hline
&\multicolumn{2}{c|}{$C_F$ color factor}\\\hline
First line of in Eq.~(\ref{eq:AppfmuNoAbe1})&0.4874171644140917E+02
&-0.3169526598603159E+03
\\ \hline
$2\text{Re}(f(s,t,u,\mu_0,1,0,0)^*)$&0.1973920880217872E+02
&-0.3459551674994638E+03
\\ \hline
Eq.~(\ref{eq:AppfmuNoAbe1})&0.2900250763923045E+02
&0.2900250763914789E+02
\\ \hline
ratio-1&
\multicolumn{2}{c|}{ 0.2846611835138901E-11}
\\
\hline
&\multicolumn{2}{c|}{$C_A$ color factor}\\\hline
First line of Eq.~(\ref{eq:AppfmuNoAbe2}) &-0.4135999643699960E+02
&-0.1991502907112327E+03
\\\hline
$2\text{Re}(f(s,t,u,\mu_0,0,1,0)^*)$&0.7711028150622136E+02
&0.1260325556690930E+03
\\ \hline
Eq.~(\ref{eq:AppfmuNoAbe2})&0.4322385485937497E+02
&0.4322385485945347E+02
\\ \hline
ratio-1&
\multicolumn{2}{c|}{-0.1816102823681831E-11 }
\\
\hline
\end{tabular}
\caption{Check of factorization scale energy independence through 
Eqs.~(\ref{eq:AppfmuNoAbe1},\ref{eq:AppfmuNoAbe2}) for two
different sets of factorization energy scales.}
\label{eq:ApprenhexlineNoAbeC_FC_A}
\end{table}
%
Finally, proceeding in a similar manner from
Eqs.~(\ref{eq:identifyhexNoabeCA},\ref{eq:identifyhexNoabeCF}), the factorization of the divergence contributions 
can be checked for the two color factors, the results are shown in
Table~\ref{App:fachexNoAbe}. The value for all of the  $d_{(1)}$
quantities is always
below $10^{-13}$ (not shown). 
``Exact'' for $c_{(1)}$ is given by the 16 first digits of~(See Eqs.(\ref{eq:identifyhexNoabeCA},\ref{eq:hexnum},\ref{eq:hexNoAbenum})):
\be
2 \cdot \mbox{Re} \big[\left ( -\log\left (\frac{-s}{\mu^2}\right)+ \log\left (\frac{-t}{\mu^2}\right)+\log
   \left(\frac{-u}{\mu^2}\right)\right)^* ].
\ee
\begin{table}[h!]
\begin{tabular}{|c|c|c|c|}
\hline
\multirow{2}{*}{ $ $} 
&\multirow{2}{*}{$c_{(1)} $}
&\multirow{2}{*}{ $c_{(2)} $}
&\multirow{2}{*}{ $d_{(0)} $}
\\ 
&
&
&
\\\hline
&\multicolumn{3}{c|}{$C_F$ color factor}\\\hline
$\sum $&-5.99999999998432
&-3.99999999999893
&-1.00000000001008
\\\hline
Exact&-6.00000000000000
&-4.00000000000000 
&-1.00000000000000
\\\hline
ratio-1
&-2.612909888455306E-012
&-2.668976151198876E-013
&1.008459982188015E-011
\\
\hline
&\multicolumn{3}{c|}{$C_A$ color factor}\\\hline
$\sum$ &-11.8662386614969
&-2.00000000000024
& $<$ E-13
\\\hline
``Exact''&-11.8662386614914
&-2.00000000000000 
& 0
\\\hline
ratio-1
&4.600764214046649E-013
&1.187938636348917E-013
&
\\
\hline
\end{tabular}
\caption{Factorization of the divergent contributions against the born
  amplitude and accuracy for the two color factors.}
\label{App:fachexNoAbe}
\end{table}
\section{Color factors}
\label{appendixC}
The color factors for the two permutations of the boxline for one gluon
emission are given by:
\bea
&\,& {\cal C}_{(1)}^{gV}={\cal C}_{(1)}^{Vg}={\cal C}_{(2)}^{gV}={\cal C}_{(3)}^{Vg}=T^a (C_F-1/2C_A) \nonumber \\
&\,& {\cal C}_{(2)}^{Vg}={\cal C}_{(3)}^{gV}={\cal C}_{(4)}^{gV}={\cal C}_{(4)}^{Vg} =T^a C_F.
\eea
\\
The color factors for the penline routines for one gluon emission are given by:

\noindent $(g V_1V_2) $
%
%
\bea
&\,&{\cal C}_{(1)}^{gV_1V_2}={\cal C}_{(2)}^{gV_1V_2}={\cal C}_{(4)}^{g V_1V_2} =T^a (C_F-1/2C_A), \nonumber \\
&\,&{\cal C}_{(3)}^{gV_1V_2}={\cal C}_{(5)}^{gV_1V_2}={\cal C}_{(6)}^{gV_1V_2}={\cal C}_{(7)}^{gV_1V_2}={\cal C}_{(8)}^{gV_1V_2} =T^a C_F,
\eea
%
%
$(V_1g  V_2) $
%
%
\bea
&\,&{\cal C}_{(1)}^{V_1 g V_2}={\cal C}_{(2)}^{V_1 g V_2}={\cal C}_{(3)}^{ V_1 g V_2}={\cal C}_{(5)}^{ V_1 g V_2} =T^a (C_F-1/2C_A), \nonumber \\
&\,&{\cal C}_{(4)}^{V_1 g V_2}={\cal C}_{(6)}^{V_1 g V_2}={\cal C}_{(7)}^{V_1 g V_2}={\cal C}_{(8)}^{V_1gV_2}
=T^a C_F,
\eea
%
%
$(V_1  V_2 g) $
%
%
\bea
&\,&{\cal C}_{(1)}^{V_1V_2g}={\cal C}_{(3)}^{ V_1V_2g} ={\cal C}_{(6)}^{ V_1V_2g} =T^a (C_F-1/2C_A), \nonumber \\
&\,&{\cal C}_{(2)}^{V_1V_2g}={\cal C}_{(4)}^{V_1V_2g}={\cal C}_{(5)}^{V_1V_2g}={\cal C}_{(7)}^{V_1V_2g}={\cal C}_{(8)}^{V_1V_2g}
=T^a C_F,
\eea
the same color factors are obtained for the permutations corresponding to
$V_1 \leftrightarrow V_2$ with $V_i$ $\in
~($W$^\pm$,Z,$\gamma)$.

Finally, the color factors for the one gluon emission case for the
hexline routine are given by:

\noindent $(g V_1V_2V_3) $
%
%
\bea
&\,&{\cal C}_{(1-2)}^{gV_1V_2V_3}={\cal C}_{(4)}^{g V_1V_2V_3}={\cal C}_{(7)}^{g V_1V_2V_3} =T^a (C_F-1/2C_A) \nonumber \\
&\,&{\cal C}_{(3)}^{gV_1V_2V_3}={\cal C}_{(5-6)}^{gV_1V_2V_3}={\cal C}_{(8-13)}^{gV_1V_2V_3}=T^a C_F
\eea
$(V_1g  V_2V_3) $
%
%
\bea
&\,&{\cal C}_{(1-5)}^{V_1 g V_2V_3}={\cal C}_{(8)}^{V_1 g V_2V_3}=T^a (C_F-1/2C_A) \nonumber \\
&\,&{\cal C}_{(6-7)}^{V_1 g V_2V_3}={\cal C}_{(9-13)}^{V_1gV_2V_3}
=T^a C_F
\eea
%
%
$(V_1  V_2 gV_3) $
%
%
\bea
&\,&{\cal C}_{(1-3)}^{V_1V_2gV_3}={\cal C}_{(5-6)}^{V_1V_2gV_3}={\cal C}_{(9)}^{ V_1V_2gV_3} =T^a (C_F-1/2C_A) \nonumber \\
&\,&{\cal C}_{(4)}^{V_1V_2gV_3}={\cal C}_{(7-8)}^{V_1V_2gV_3}={\cal C}_{(10-13)}^{V_1V_2gV_3}
=T^a C_F
\eea
%
%
$(V_1  V_2 V_3g) $
%
%
\bea
&\,&{\cal C}_{(1)}^{V_1V_2V_3g}={\cal C}_{(3)}^{V_1V_2V_3g}={\cal C}_{(6)}^{ V_1V_2V_3g} ={\cal C}_{(10)}^{ V_1V_2V_3g} =T^a (C_F-1/2C_A) \nonumber \\
&\,&{\cal C}_{(2)}^{V_1V_2V_3g}={\cal C}_{(4-5)}^{V_1V_2V_3g}={\cal C}_{(7-9)}^{V_1V_2V_3g}={\cal C}_{(11-13)}^{V_1V_2V_3g}
=T^a C_F,
\eea
the same color factors are obtained for the six other permutations of the
vector bosons $V_1, V_2$ and $V_3$ with $V_i\in ($W$^\pm, Z,\gamma$). 
%
%
%
%
%
%
%
  
%
%

%
%
%
%
\end{document}